\documentclass[11pt]{article}
\pdfoutput=1 % if your are submitting a pdflatex (i.e. if you have
\usepackage{jheppub_qedNfenergy} % for details on the use of the package, please
\usepackage[T1]{fontenc} % if needed

\usepackage{graphicx}
\usepackage{bm}%	bold math
\usepackage{amsfonts}
\usepackage{booktabs}%   for \cmidrule

\usepackage{xcolor}

\usepackage{dsfont}

\def\b{{\bm b}}

\def\p{{\bm p}}

\def\B{{\bm B}}
\def\C{{\bm C}}
\def\P{{\bm P}}

\def\eps{\epsilon}

\def\Nc{N_{\rm c}}
\def\Nf{N_{\rm f}}

\def\alphas{\alpha_{\rm s}}

\def\Re{\operatorname{Re}}

\def\grad{{\bm\nabla}}

\def\yfrak{{\mathfrak y}}
\def\yfrakE{\yfrak_\ssE}
\def\yfrakEbar{\yfrak_\ssEbar}

\def\xe{x_e}
\def\ye{y_e}

\def\yfrake{{\mathfrak y}_e}

\def\MSbar{\overline{\mbox{MS}}}

\def\qhatA{\hat q_{\rm A}}

\def\qhat{\hat q}
\def\lstop{\ell_{\rm stop}}
\def\LO{{\rm LO}}
\def\NLO{{\rm NLO}}

\def\E{{\rm E}}

\def\ssE{{\scriptscriptstyle{\E}}}
\def\Ebar{{\bar \E}}

\def\ssEbar{{\scriptscriptstyle{\Ebar}}}
\def\xE{x_\ssE}

\def\xEbar{x_\ssEbar}

\def\net{{\rm net}}
\def\uee{{\underline{e\to e}}}
\def\ee{{e\to e}}

\def\eeEEsub{{e\to e\ssE\ssEbar}}
\def\ueE{{\underline{e\to\E}}}
\def\ueEbar{{\underline{e\to\Ebar}}}
\def\ueg{{\underline{e\to\gamma}}}
\def\ugE{{\underline{\gamma\to\E}}}
\def\ugEbar{{\underline{\gamma\to\Ebar}}}
\def\uij{{\underline{i\to j}}}

\def\indep{{\rm indep}}
\def\super{{\rm super}}
\def\snet{{\substack{\super \\ \net}}}
\def\inet{{\substack{\indep \\ \net}}}

\def\fac{{\rm fac}}
\def\form{{\rm form}}
\def\Avg{\operatorname{Avg}}
\def\dAvg{\operatorname{\delta Avg}}

\begin {document}

%%%%%%%%%%%%%%%%%%%%%%%%%%%%%%%%%%%%%%%%%%%%%%%%%%%%%%%%%%%%%%%%%%%%%%%%%%%%%%%

%%%%%%%%%%%%%%%%%%%%%%%%%%%%%%%%%%%%%%%%%%%%%%%%%%%%%%%%%%%%%%%%%%%%%%%%%%%%%%%

\title{Strongly vs.\ weakly coupled in-medium showers:
       \\energy stopping in large-$\Nf$ QED}

\author[a]{Peter Arnold,}
\author[b,c,a]{Omar Elgedawy,}
\author[d,e]{Shahin Iqbal}

% The "\note" macro will give a warning: "Ignoring empty anchor..."
% you can safely ignore it.

%% update Omar's affiliation
\affiliation[a]{Department of Physics, University of Virginia,
  P.O.\ Box 400714, 
  Charlottesville, VA 22904, U.S.A.}
\affiliation[b]{
  Key Laboratory of Atomic and Subatomic Structure and Quantum Control (MOE),
  Guangdong Basic Research Center of Excellence for Structure and Fundamental
    Interactions of Matter,
  Institute of Quantum Matter,
  South China Normal University, Guangzhou 510006, China}
\affiliation[c]{
  Guangdong-Hong Kong Joint Laboratory of Quantum Matter,
  Guangdong Provincial Key Laboratory of Nuclear Science,
  Southern Nuclear Science Computing Center,
  South China Normal University, Guangzhou 510006, China}
\affiliation[d]{National Centre for Physics,
  Quaid-i-Azam University Campus,
  Islamabad, 45320 Pakistan}
\affiliation[e]{Theoretical Physics Department, CERN,
  CH-1211 Geneva 23, Switzerland}

% e-mail addresses: one for each author, in the same order as the authors
\emailAdd{parnold@virginia.edu}
\emailAdd{oae2ft@virginia.edu}
\emailAdd{smi6nd@virginia.edu}

\begin {abstract}%
{%
Inside a medium, showers originating from a very high-energy particle may
develop
via medium-induced
splitting processes such as hard bremsstrahlung or pair production.
During shower development, two consecutive splittings
sometimes overlap quantum mechanically, so that
they cannot be treated independently.
Some of these effects can be absorbed into an effective value of a medium
parameter known as $\qhat$.
Previous calculations (with certain simplifying assumptions)
have found that, after adjusting the value of $\qhat$, the
leftover effect of overlapping splittings is quite
small for purely gluonic large-$\Nc$ showers but is very much larger
for large-$\Nf$ QED showers, at comparable values of $N\alpha$.
Those works did not quite make for apples-to-apples comparisons:
the gluon shower work investigated energy deposition from a
gluon-initiated shower, whereas the QED work investigated charge-deposition
from an electron-initiated shower.  As a first step to tighten up
the comparison, this paper investigates energy deposition in the QED case.
Along the way, we develop a framework that should be useful in the future
to explore whether the very small effect of overlapping splitting
in purely gluonic showers is an artifact of having ignored quarks.
}%
\end {abstract}

\maketitle
\thispagestyle {empty}

%{\def\boldmath{}\tableofcontents}
\newpage

%%%%%%%%%%%%%%%%%%%%%%%%%%%%%%%%%%%%%%%%%%%%%%%%%%%%%%%%%%%%%%%%%%%%%%%%%%%%%%%

\section{Introduction and Results}
\label{sec:intro}

\subsection {Introduction}

When passing through matter, high energy particles lose energy by
showering, via the splitting processes of hard bremsstrahlung and pair
production.  At very high energy, the quantum mechanical duration of
each splitting process, known as the formation time, exceeds the mean
free time for collisions with the medium, leading to a significant
reduction in the splitting rate known as the Landau-Pomeranchuk-Migdal
(LPM) effect.
The LPM effect was originally worked out for QED in the 1950's
\cite{LP1,LP2,Migdal}%
\footnote{
  The papers of Landau and Pomeranchuk \cite{LP1,LP2} are also available in
  English translation \cite{LPenglish}.
}
and then later generalized to QCD in the 1990s by
Baier, Dokshitzer, Mueller, Peigne, and Schiff \cite{BDMPS1,BDMPS2,BDMPS3}
and by Zakharov \cite{Zakharov1,Zakharov2} (BDMPS-Z).

Modeling of the development of
high-energy in-medium showers typically treats each splitting
as an independent dice roll, with probabilities set by calculations
of single-splitting rates that take into account the LPM effect.
The question then arises whether consecutive splittings in a shower
can really be treated as probabilistically independent, or whether
there is any significant chance that the formation times of
splittings could overlap so that there are significant quantum interference
effects entangling one splitting with the next.
A number of years ago, several authors \cite{Blaizot,Iancu,Wu}
showed, in a leading-log calculation,
that the effects of overlapping formation times in QCD showers
could become large when one of the two overlapping splittings
is parametrically softer than the other.
They also showed that those large leading logarithms could be absorbed into
a redefinition of the medium parameter $\hat q$, which parametrizes
the effectiveness with which the medium deflects high-energy particles.%
\footnote{
  Specifically, the typical total transverse momentum change $p_\perp$ to
  a high-energy particle after
  traveling through a length $L$ of the medium behaves like a random
  walk, $\langle p_\perp^2 \rangle = \qhat L$.
}
A refined question arose: How large are overlapping formation
time effects that {\it cannot}\/ be absorbed into a redefinition of
$\qhat$?

To provide a simpler arena than QCD for developing methods and
calculational tools to answer this question,
ref.\ \cite{qedNfstop} first studied it in large-$\Nf$ QED
(where $\Nf$ is the number of electron flavors).%
\footnote{
  The advantage of
  the large-$\Nf$ limit was mainly that it reduced the number of
  medium-averaged interference diagrams that had to be calculated.
}
That paper used a thought experiment to determine how important overlap
effects could be.  Consider a shower initiated by a high-energy electron
moving in the $z$ direction, starting at $z=0$.  Imagine for simplicity
that the medium is static, homogeneous, and of infinite extent.
The shower will create
more and more electrons, positrons, and photons, of lower and lower
energy, eventually depositing various $+$ and $-$ charges
into the medium at various positions.
Let $\rho(z)$ be the distribution in $z$ of net charge deposited
in the medium, statistically averaged over many such showers.
Define the charge stopping length $\lstop^Q$
to be the first moment of that distribution,
$\lstop^Q \equiv \langle z \rangle_\rho
 \equiv Q^{-1} \int dz \> z \, \rho(z)$,
where $Q$ is the charge of the initial electron (and so is the total charge
of the shower).  Let $\sigma^Q$ be the width of the distribution $\rho(z)$.
Ignoring overlap effects, both $\lstop^Q$ and $\sigma^Q$ scale with
$\qhat$, coupling constant, and the energy $E_0$ of the initial electron as
\begin {equation}
  \lstop \sim \sigma \sim \frac{1}{\alpha} \sqrt{ \frac{E_0}{\qhat} } .
\label {eq:scale}
\end {equation}
The value of $\qhat$ then cancels in the ratio $\sigma/\lstop$.
Any effect that can be absorbed into $\qhat$ would not affect the
value of $\sigma/\lstop$, and so that ratio could be used to
test how large are overlapping formation time effects that cannot be
absorbed into $\qhat$.
To leading order in $\alpha$, ref.\ \cite{qedNfstop} found that the
{\it relative} size of overlap effects was
\begin {equation}
  \mbox{overlap correction} = -87\% \times \Nf\alpha
  \qquad
  \mbox{(large-$\Nf$ QED charge stopping $\sigma/\lstop$).}
\label {eq:chiqed}
\end {equation}

Later, when we were doing a related calculation \cite{finale,finale2}
for large-$\Nc$
QCD, we fully expected to
find an answer of the same order of magnitude, with $\Nc\alphas$
playing the role of $\Nf\alpha$.
So far, that calculation has only been completed for purely-gluonic
showers.  Since gluons have no charge, we studied the {\it energy}
deposition distribution $\eps(z)$ instead of a charge deposition
distribution.  We similarly define an energy stopping distance
$\lstop^E$ and width $\sigma^E$, which also scale like
(\ref{eq:scale}).
We may again
look to the ratio $\sigma/\lstop$ as a vehicle for measuring overlap
effects that
cannot be absorbed into $\qhat$.  In the case of QCD, the question of
$\qhat$ insensitivity of $\sigma/\lstop$ is quite a bit more subtle
than in QED because of enhanced soft emissions in the
QCD version of the LPM effect.
Those subtleties do not matter for the QED analysis we will carry
out in the present paper, and so we will not review them here.
(See refs.\ \cite{finale,finale2} for details.)
To our great surprise, the result found for gluon showers was%
\footnote{
   This is the result quoted in eq.\ (11) of ref.\ \cite{finale}
   for the choice $\Lambda_{\rm fac}=x(1{-}x)E$ of factorization scale.
   As discussed in ref.\ \cite{finale}, the qualitative conclusion
   that overlap effects are at most a few percent times $\Nc\alphas$ is
   insensitive to any reasonable variation of factorization scale.
}
\begin {equation}
  \mbox{overlap correction} = -2\% \times \Nc\alphas
  \qquad
  \mbox{(large-$\Nc$ pure-gluon energy stopping $\sigma/\lstop$).}
\label {eq:chiqcd}
\end {equation}
For similar values of $N\alpha$, this is a {\it drastically} smaller overlap
effect than the corresponding QED result
(\ref{eq:chiqed}).

Refs.\ \cite{finale,finale2} also looked at the shape
$S_\eps(Z)$ of $\eps(z)$, defined by
$S_\eps(Z) \equiv \lstop^E \, \eps(Z\lstop^E)/E_0$ where
$Z$ represents distance measured in units of $\lstop$.
The width of $S_\eps(Z)$ is the ratio $\sigma^E/\lstop^E$ just discussed.
More generally,
overlap effects
on the full function $S_\eps(Z)$ were found to be very small for QCD.

There remains the open question of {\it why} the QED and QCD results are
so very different!  Perhaps the tiny result (\ref{eq:chiqcd}) is
merely a coincidence,
arising from an accidental cancellation for the special case
of large-$\Nc$ purely gluonic showers.
Perhaps showers involving fermions behave differently from those that don't.
Or perhaps the shape of energy deposition, in any theory,
is for some reason less
sensitive to changes (such as from overlap effects) than the shape
of charge deposition.

In this paper, we take a first look at the last possibility by
calculating the relative size of overlapping formation time effects
on the value of $\sigma/\lstop$ for {\it energy} deposition in large-$\Nf$
QED.  An equally important goal is that developing the tools to
better analyze overlap effects for the $e^\pm$/photon showers will
prepare us in later work to add quarks to our QCD showers and so
eventually address the other possible explanations as well.

% ----------------------------------------------------------------------------

\subsection {Results}

Our main results for large-$\Nf$ QED are summarized in
table \ref{tab:chi2}.  We will discuss later the different choices of
renormalization scale shown in the table.  That's a detail that does
not impact the qualitative conclusion, which is that the relatively large size
of the QED result (\ref{eq:chiqed}) compared to the gluon shower
result (\ref{eq:chiqcd}) is not due to any qualitative
difference between charge deposition and energy deposition
in the QED case.
For large-$\Nf$ QED, the overlap effects on energy deposition are comparable
in size to the ones for charge deposition.

\begin {table}[tp]

\setlength{\tabcolsep}{7pt}
\begin {center}
\begin{tabular}{ccccc}
\hline
\hline
  && \multicolumn{3}{c}{overlap correction to $\sigma/\lstop$}
\\
\cline{3-5}
  \parbox{0.7in}{deposition\\distribution}
  & \parbox{0.7in}{initiating\\particle}
  & $\mu\propto(\qhat E_0)^{1/4}$
  & $\mu\propto(\qhat E)^{1/4}$
 & $\begin {matrix}
       \mu_{e\to e\gamma}\propto(\xe \qhat E/(1{-}\xe))^{1/4} \\
       \mu_{\gamma\to e\bar e}\propto\bigl(\xe(1{-}\xe) \qhat E\bigr)^{1/4}
    \end{matrix}$
\\
\hline
  charge & $e$ & $-87\%\times\Nf\alpha$
               & $~{-}85\%\times\Nf\alpha$
               & $~{-}80\%\times\Nf\alpha$
\\
  energy & $e$ &
               & $+113\%\times\Nf\alpha$
               & $+113\%\times\Nf\alpha$
\\
  energy & $\gamma$ &
               & $~{+}99\%\times\Nf\alpha$
               & $~{+}98\%\times\Nf\alpha$
\\
\hline
\hline
\end{tabular}
\end {center}
\caption{%
\label{tab:chi2}%
  The relative size of corrections to the ratio $\sigma/\lstop$ of width to
  stopping distance in
  large-$\Nf$ QED for the cases of (i) charge deposition
  of electron-initiated showers, (ii) energy deposition of
  electron-initiated showers, and (iii) energy deposition of
  photon-initiated showers.  The last three columns correspond to
  three different prescriptions for the choice of
  renormalization scale, of which the last two will be used in this
  paper.  The $\mu\propto(\qhat E_0)^{1/4}$ entry for charge deposition of
  electron-initiated showers is provided merely to make contact with
  the value (\ref{eq:chiqed})
  of the QED result previously calculated in ref.\ \cite{qedNfstop}.
  The exact proportionality constants in
  $\mu \propto (\qhat E_0)^{1/4}$ and $\mu \propto (\qhat E)^{1/4}$
  do not matter; only the energy and $\xe$
  dependence of $\mu$ affect the results.
  That's also true of the specifications of $\mu$ in the last
  column provided (i) the proportionality constants are chosen
  the same for $\mu_{e\to e\gamma}$ and $\mu_{\gamma\to e\bar e}$ or (ii)
  one is looking at the charge deposition (which depends only on
  $\mu_{e\to e\gamma}$ at this order).
}
\end{table}

% ----------------------------------------------------------------------------

\subsection {Outline}

In the remainder of this introduction, we summarize the assumptions
made in this paper.
Like refs.\ \cite{finale,finale2}, our philosophy is to perform a
complete calculation of overlap effects in the simplest possible
theoretical situation.

In section \ref{sec:review},
we review diagrams and our notation for (i)
LPM/BDMPS-Z in-medium splitting rates [which we call ``leading order''
rates] and (ii) the corrections to those rates due to overlapping formation
times, which we call next-to-leading-order (NLO) corrections.
Complicated formulas for the NLO rate corrections may be found in
ref.\ \cite{qedNf} for large-$\Nf$ QED, but we will not review those
NLO formulas explicitly.

In section \ref{sec:netrates}, we review the concept of net rates
$[d\Gamma/dx]_{\rm net}$ used by refs.\ \cite{finale,finale2,qcd}
(i) to simplify shower evolution equations in cases where there are
effective $1{\to}3$ splittings (due to overlap effects) in addition to
just $1{\to}2$ splittings and (ii) to provide a convenient way to package
numerical results for rates, which can then be fit by analytic functions
that are more efficient to evaluate.  The previous analysis
of refs.\ \cite{finale,finale2,qcd} only considered gluons, where all
particles are identical, and here we adapt that discussion to
the case of distinguishable particles.
Some analytic results are also presented,
for logarithmic dependence of the net rates when one daughter of an
overlapping splitting is soft, with details left to an appendix.

Section \ref{sec:mu} discusses sensible choices of
ultraviolet (UV) renormalization scale
for this problem.

Section \ref{sec:charge} reviews the formalism used by ref.\ \cite{qedNfstop}
to find the earlier overlap correction (\ref{eq:chiqed})
for $\sigma^Q/\lstop^Q$, which is
the width of the shape function $S_\rho(Z)$ for the charge deposition
distribution $\rho(z)$.  Results for other moments of the shape
are also presented for completeness.
Section \ref{sec:energy} then generalizes that discussion to the energy
deposition distribution $\eps(z)$.  Both of these sections provide the
values presented in table \ref{tab:chi2}.

A very brief conclusion is offered in section \ref{sec:conclusion}.
%where we discuss how adaptation of this work to QCD may shed further
%light on the drastic difference in the size of overlap
%effects (\ref{eq:chiqcd}) and (\ref{eq:chiqed}) for comparable values
%of $N\alpha$.

% -------------------------------------------------------------------------

\subsection {Assumptions}

In this paper, we make use of formulas for overlap corrections to
splitting rates that were computed for large-$\Nf$ QED in ref.\ \cite{qedNf}
and applied to $\sigma/\lstop$ for charge deposition in ref.\ \cite{qedNfstop}.
We make the same simplifying assumptions as those papers, similar
to those later made in the gluon shower analysis of
refs.\ \cite{qcd,finale,finale2}. 
For the splitting rate calculations, we assume a static, homogeneous
medium that is large enough to contain (i) formation times in the case
of splitting rate calculations and (ii) the entire development of the
shower for calculation of overlap corrections to $\sigma/\lstop$.
We will ignore the mass (vacuum and medium-induced)
of all high-energy particles.
We take the multiple-scattering ($\hat q$) approximation for transverse
momentum transfer from the medium.  (This is equivalent to Migdal's
large Coulomb logarithm approximation \cite{Migdal} in the case of QED.)
We will in particular approximate the bare value $\qhat_{(0)}$ of $\qhat$
as constant, ignoring any
logarithmic energy dependence of $\qhat_{(0)}$.
(Here $\qhat_{(0)}$ represents
the value from scattering of the high-energy particle with the medium
{\it without} any high-energy splitting.)
We assume that the particle initiating the shower
can be approximated as on-shell.  Taking the large-$\Nf$ limit
reduced the number of diagrams that had to be computed in ref.\ \cite{qedNf},
somewhat simplified the structure of equations for charge deposition in
ref.\ \cite{qedNfstop}, and will somewhat simplify the structure of
equations for energy deposition in this paper.
The overlap corrections \cite{qedNf}
to splitting rates have so far only been computed
for $p_\perp$-integrated rates because integration over $\p_\perp$
makes the
calculations much simpler.  In any case, $p_\perp$-integrated rates are all that
we need to study features of
charge and energy deposition distributions $\rho(z)$
and $\eps(z)$ since we will not keep track of the (parametrically small)
spread of the deposition in directions transverse to $z$.

Throughout this paper, we formally treat $\Nf\alpha(\mu)$ as small,
where $\alpha(\mu)$ is the coupling associated with high-energy splitting.
Like in the QCD discussion of ref.\ \cite{finale}, the relevant scale
$\mu$ for the running coupling scales with $\qhat$ and energy $E$ as
roughly $(\qhat E)^{1/4}$.  [We'll discuss detailed choices of $\mu$ later.]
Unlike QCD, the running coupling in QED gets larger with increasing energy.
That means that, in the QED case, the value of $\Nf\alpha$ at medium scales
would necessarily be small as well.  We will not take advantage
of that; we summarize all medium effects by the value of $\qhat$
in order to (i) simplify the calculation and (ii) make
everything as closely
parallel to the QCD calculations of refs.\ \cite{finale,finale2,qcd} as
possible.

% ...........................................................................

\subsection{Examples of work relaxing our assumptions}

Though our various simplifying assumptions have not yet been relaxed
for full calculations of overlap effects along the lines of this paper,
more general situations have long been studied by numerous authors
for either non-overlapping splittings or for certain limits of or models
of overlapping splittings.  We mention a few examples here in the context
of QCD.
For non-overlapping splittings,
the case of finite media has been of interest since the
very early work of BDMPS-Z \cite{BDMPS2,BDMPS3,Zakharov2}.
For some discussion of generalizing BDMPS-Z rate calculations
to $p_\perp$ dependence of non-overlapping splittings, see
refs.\ \cite{LOptZakharov,LOptWiedemann,LOptBlaizot,LOptApolinario}.
Ref.\ \cite{NLOptBarato} has investigated
$p_\perp$ dependence for soft emissions
overlapping harder splittings with the latter treated in antenna
approximations such as in refs.\ \cite{antenna1,antenna2}.
An example of using antenna approximations to discuss cone size dependence
of jets may be found in
refs.\ \cite{conesize1,conesize2}.
In the context of our own calculations in the case of purely
gluonic showers, a partial analysis of
$1/\Nc^2$ corrections was carried out by two of us in
ref.\ \cite{1overN}.
However, similar issues arise in the somewhat different problem of
calculating $p_\perp$ dependence for \textit{non}-overlapping splitting
rates, where going beyond the large-$\Nc$ approximation has been
studied in refs.\ \cite{NSZ6j,Zakharov6j,Konrad1}
and more recently ref.\ \cite{KonradNc3}.
Though we focus on the multiple scattering $\qhat$ approximation,
which should be appropriate at high enough energy for QCD plasmas that
are thick compared to formation lengths, there is a great deal of
work (originally for single splittings) on the opacity
expansion, which for plasmas thinner than formation times, especially for
measurements that are sensitive to the Rutherford tail (higher than
typical transverse momentum exchange) of the scatterings with the
medium \cite{ZakharovFinite,LOptWiedemann,GLV,Vitev2,Vitev3}.
For a finite but otherwise homogeneous medium,
the interpolation between the thin and thick medium limits
was mapped out with numerical calculations of single splitting in
refs.\ \cite{SimoneCharles,Carlota1}, with generalization to
an evolving medium in ref.\ \cite{Carlota2}.
More recent work introduces an expansion ---
the improved opacity expansion --- that can be used to (approximately)
study analytically the interpolation between the opacity expansion
and the $\qhat$ approximation \cite{IOC1,IOC2,IOC3}.
For some examples of discussion, in various approximations, of how to marry an
initial vacuum-like cascade of virtuality with later onshell 
showering in a finite medium and fragmentation afterward,
without the full analysis of
overlap (for simplified situations) that is the subject of our paper, see
refs.\ \cite{marryKurkela,marryIancu,marryYacine1,conesize1}.
For an example of $p_\perp$ dependence of \textit{massive} quark production
in large-$\Nc$ overlapping $q \to qg \to q c \bar c$ to first order in the
opacity expansion, see ref.\ \cite{qccbar}.

Our work here, like our earlier work on gluonic showers in
refs.\ \cite{finale,finale2},
does not use the opacity expansion, antenna approximations,
soft-radiation approximations, or approximations that ignore back-reaction
effects of the second splitting on the probability of the first splitting.
It is a complete
calculation of overlap effects, to first order in $\Nf\alpha(\mu)$,
for the simplified situation we have described.
In principle, the general formalism used could also be applied
(with a great deal of effort and development) to more general situations.

% =========================================================================

\section{Review of the building blocks: splitting rates}
\label{sec:review}

\subsection {Diagrams}

\begin {figure}[t]
\begin {center}
  \includegraphics[scale=0.6]{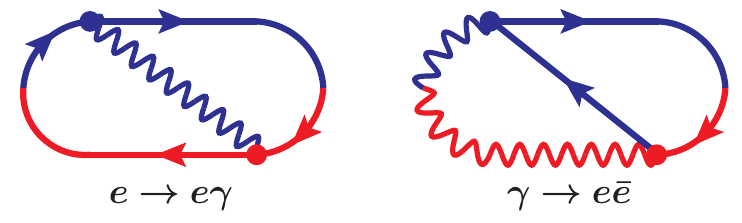}
  \caption{
     \label{fig:LOqed}
     Time-ordered interference diagrams contributing to the rate
     of $e \to e\gamma$ and $\gamma \to e\bar e$.
     Time runs from left to right.
     In both cases, all lines implicitly interact
     with the medium.  We need not follow
     particles after the emission has occurred in both the amplitude
     and conjugate amplitude because we only calculate the
     $p_\perp$-integrated rate.
     (See, for example, section 4.1 of
     ref.\ \cite{2brem} for a more explicit argument,
     although applied there to more complicated diagrams.)
     Nor need we follow them before
     the first emission because we approximate the initial particle
     as on-shell.
     Only one of the two time orderings that contribute to each
     process are shown above but both orderings can be included by
     taking $2\Re[\cdots]$.
     (Graphically, complex conjugation corresponds to flipping the diagram
     around a horizontal axis,
     exchanging the colors red and blue, and reversing the
     arrows on the fermion lines.)
  }
\end {center}
\end {figure}

In the $\qhat$ approximation, the LPM splitting rates for bremsstrahlung
and pair production are%
\footnote{
  For a translation between the $\qhat$ approximation and Migdal's
  large Coulomb logarithm approximation, see, for example, appendix C.4 of
  ref.\ \cite{qedNf}.
  The absolute value signs in (\ref{eq:LOrate0}) are unnecessary
  for the present discussion, but we include them
  to avoid confusion with the form of the formulas needed in
  ref.\ \cite{qedNf}, where (\ref{eq:LOrate0}) is sometimes evaluated for
  ``front-end transformations'' that replace $\xe$ by a negative value.
}
\begin {subequations}
\label {eq:LOrate0}
\begin {align}
  \left[ \frac{d\Gamma}{d\xe} \right]^\LO_{e\to e\gamma}
  &= \frac{\alpha}{2\pi} \, P_{e\to e}(\xe)
     \sqrt{ \frac{\qhat}{E} \left| \frac{1}{\xe} - 1 \right| }
    ,
\label {eq:LObremrate}
\\
  \left[ \frac{d\Gamma}{d\xe} \right]^\LO_{\gamma\to e\bar e}
  &= \frac{\Nf \alpha}{2\pi} \, P_{\gamma\to e}(\xe)
      \sqrt{ \frac{\qhat}{E} \left| \frac{1}{\xe} + \frac{1}{1{-}\xe} \right| }
    .
\label {eq:LOpairrate}
\end {align}
\end {subequations}
in the high energy limit.  Above, $E$ is the energy of the parent,
$\xe$ is the energy fraction of the electron daughter, and the $P(x)$
are unregulated Dokshitzer-Gribov-Lipatov-Alterelli-Parisi splitting
functions
\begin {equation}
  P_{e\to e}(x) = \frac{1+x^2}{1-x} ,
  \qquad
  P_{\gamma\to e}(x) = x^2+(1{-}x)^2 .
  \label {eq:DGLAP_Ps}
\end {equation}
We refer to (\ref{eq:LOrate0}) as the ``leading-order'' (LO) rates.
For us, leading order means leading order in the number of
high-energy splitting vertices and includes the effects of
an arbitrary number of interactions with the medium.
Adopting Zakharov's picture \cite{Zakharov1,Zakharov2}, we think of
the rate for $e{\to}e \gamma$ and $\gamma{\to}e\bar e$
as time-ordered interference diagrams, such as fig.\ \ref{fig:LOqed},
which combine
the amplitude for the splitting (blue) with the conjugate amplitude (red).
See refs.\ \cite{2brem,qedNf} for more discussion of our
graphical conventions and implementation of Zakharov's approach.

\begin {figure}[tp]
\begin {center}
  \includegraphics[scale=0.38]{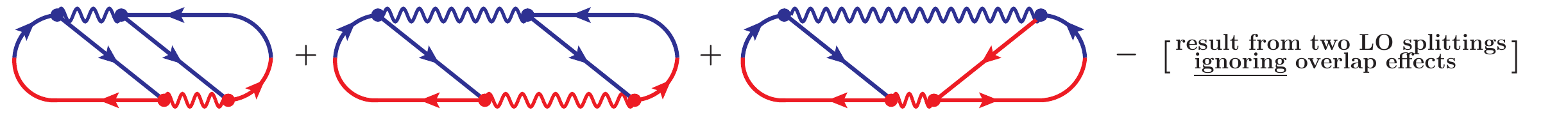}
  \caption{
     \label{fig:diagsNf1}
     Time-ordered interference diagrams for
     $e \to e e \bar e$ in large-$\Nf$ QED \cite{qedNf}.
     Here, only diagrams with transverse-polarized photons are shown.
     Complex conjugates of the above interference diagrams should also be
     included by taking $2\Re[\cdots]$ of the above.
  }
\end {center}
\end {figure}

\begin {figure}[tp]
\begin {center}
  \includegraphics[scale=0.40]{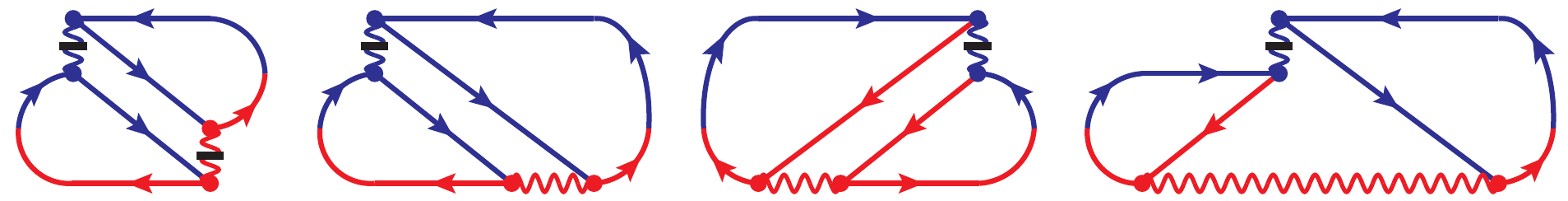}
  \caption{
     \label{fig:diagsNf2}
     More time-ordered interference diagrams for
     $e \to e e \bar e$ in large-$\Nf$ QED \cite{qedNf}.
     These involve exchange of a longitudinally-polarized
     photon in light-cone gauge,
     represented by an instantaneous (in light-cone time) vertical
     photon line crossed by a bar.
  }
\end {center}
\end {figure}

There is a factor of $\Nf$ in the pair production rate (\ref{eq:LOpairrate})
because the produced pair can have any flavor.
So, in the large-$\Nf$ limit, pair production (\ref{eq:LOpairrate}) is
parametrically faster than bremsstrahlung (\ref{eq:LObremrate}).
Correspondingly, the overlap of $e \to e\gamma$ with another splitting
process is dominated by the overlap
$e \to e\gamma \to ee\bar e$ (as opposed to $e \to e\gamma \to e\gamma\gamma$
or $\gamma\to e\bar e \to e\bar e\gamma$). 
Figs.\ \ref{fig:diagsNf1} and \ref{fig:diagsNf2}
show all of the time-ordered interference diagrams
contributing to the overlap of $e \to e\gamma \to ee\bar e$
in the large-$\Nf$ limit.  We refer to these overlap effects as one
type of next-to-leading-order (NLO) effect because these diagrams are
suppressed by one power of
high-energy $\Nf\alpha(\mu)$ compared to the leading-order
process $e \to e\gamma$.
The subtraction in fig.\ \ref{fig:diagsNf1} means that our
rates represent the {\it difference} between (i) a full calculation of
(potentially overlapping) $e \to e\gamma \to ee\bar e$ and (ii) approximating
that double splitting as two independent, consecutive single splittings
$e {\to} e\gamma$ and $\gamma \to e\bar e$
that each occur with the LO single splitting rates (\ref{eq:LOrate0}).%
\footnote{
  The key importance
  of this subtraction is explained in section 1.1 of
  ref.\ \cite{seq}.
}

\begin {figure}[tp]
\begin {center}
  \includegraphics[scale=0.40]{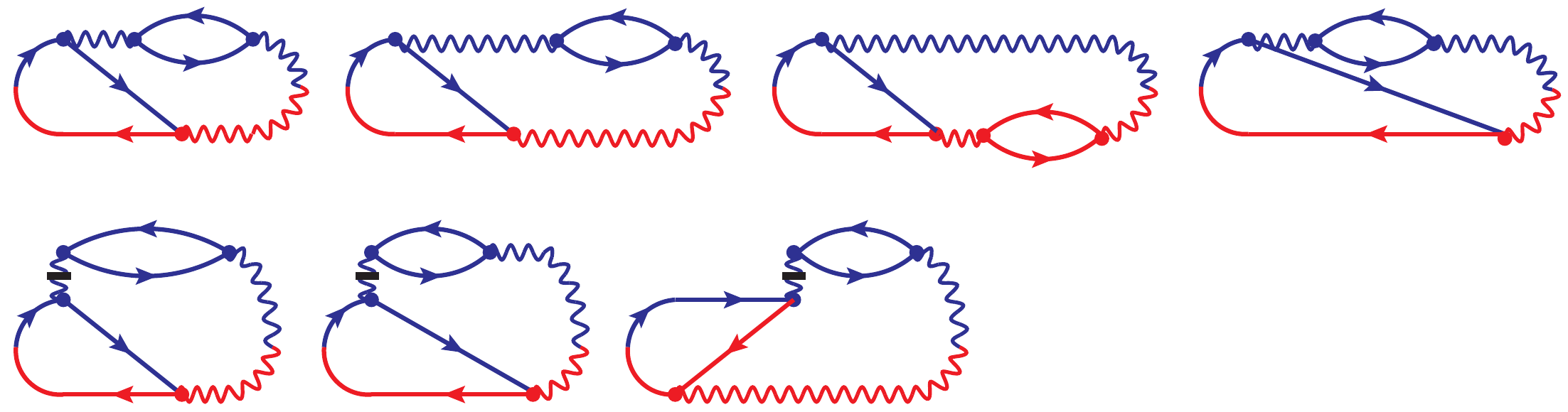}
  \caption{
     \label{fig:diagsNfVIRT}
     Time-ordered interference diagrams for the virtual correction to
     $e \to e \gamma$ in large-$\Nf$ QED \cite{qedNf}.
     Again, complex conjugates of these diagrams
     should be included by taking $2\Re[\cdots]$.
  }
\end {center}
\end {figure}

Corresponding virtual corrections to
single splitting $e \to e\gamma$, such as the interference between
$e \to e\gamma \to ee\bar e \to e\gamma$ and LO $e\to e\gamma$,
must also be accounted for.
Fig.\ \ref{fig:diagsNfVIRT} shows the relevant time-ordered
interference diagrams.%
\footnote{
   A subtraction analogous to the one in fig.\ \ref{fig:diagsNf1}
   is also made for the sum of the first three diagrams of
   fig.\ \ref{fig:diagsNfVIRT}.
   See footnote 20 of ref.\ \cite{qedNf}.
}

\begin {figure}[tp]
\begin {center}
  \includegraphics[scale=0.40]{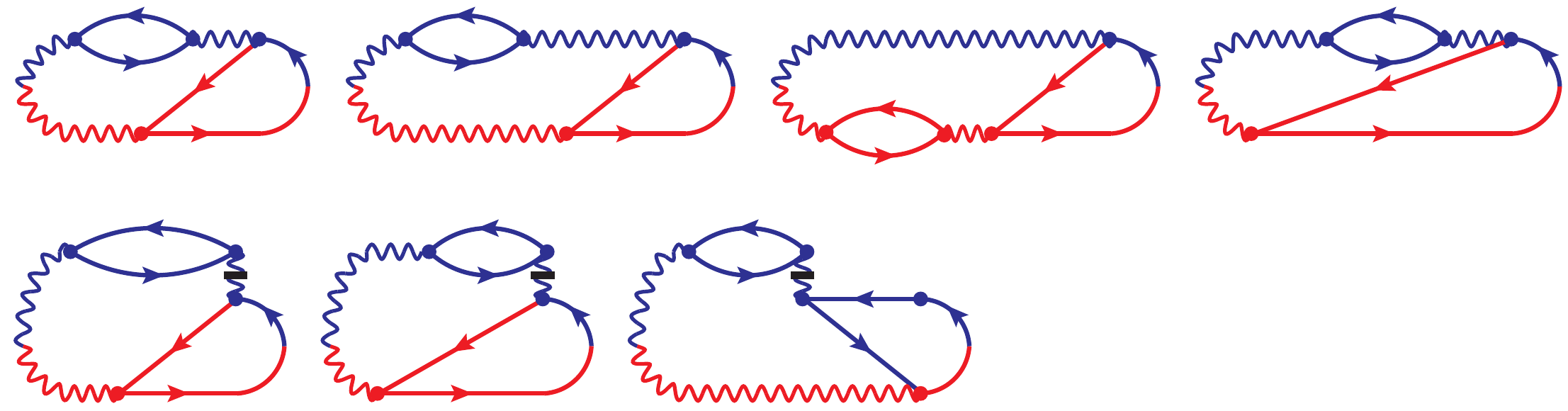}
  \caption{
     \label{fig:diagsNfVIRT2}
     Time-ordered interference diagrams for the virtual correction to
     $\gamma \to e\bar e$ in large-$\Nf$ QED \cite{qedNf}.
  }
\end {center}
\end {figure}

Finally, in the large-$\Nf$ limit, the {\it only} overlap corrections
to photon-initiated splitting $\gamma\to e\bar e$ are the virtual
corrections shown in fig. \ref{fig:diagsNfVIRT2}.
%(The non-virtual overlap of $\gamma\to e\bar e \to \gamma e\bar e$
%is suppressed by a factor of $\alpha$ compared to LO $\gamma\to e\bar e$
%and so vanishes in the large-$\Nf$ limit, which holds $\Nf\alpha$ fixed
%as $\Nf \to \infty$.)

% -------------------------------------------------------------------------

\subsection {Notation for Rates}

\begin {figure}[t]
\begin {center}
  \includegraphics[scale=0.70]{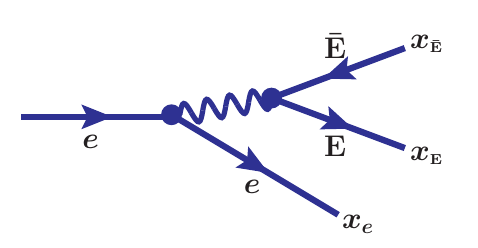}
  \caption{
     \label{fig:eEnotation}
     Our notation (\ref{eq:eEE})
     for distinguishing pair-produced electrons from the
     original electron in $e \to e\gamma \to ee\bar e$
     in the large-$\Nf$ limit.  The $x$'s are the energy fractions of
     the original electron, and $\xEbar = 1 - \xe - \xE$.
  }
\end {center}
\end {figure}

Consider overlapping bremsstrahlung followed by pair production,
$e \to e\gamma \to e e\bar e$, whose amplitude is depicted in
fig.\ \ref{fig:eEnotation}.  Here, the pair-produced electrons could
have any flavor.
It will simplify the rest of our discussion to note that, in the
$\Nf \to \infty$ limit, the two ``electron'' daughters in the final state
become distinguishable: The probability
that the flavor of the
pair-produced electron is the same as that of the initial electron
scales like $1/\Nf$, and so what we have been calling
$e \to e\gamma \to e e\bar e$ is actually more akin to
$e \to e\gamma \to e \mu\bar \mu$.  For now, we will emphasize this
distinguishability within an overlapping double-splitting process
by using the symbol $\E$ for pair-produced electrons
and so will write
\begin {equation}
  e \to e\gamma \to e\E\Ebar .
\label {eq:eEE}
\end {equation}
We will also write LO pair production as $\gamma \to \E\Ebar$ and the
corresponding one-loop virtual correction (the amplitude or conjugate
amplitude that has the loop in fig.\ \ref{fig:diagsNfVIRT2}) as
\begin {equation}
  \gamma \to \E'\Ebar' \to \gamma \to \E\Ebar .
\end {equation}

The basic rates that we will need from ref.\ \cite{qedNf} as our initial
building blocks are
leading-order splitting rates, their NLO corrections, and the overlap
correction to $e \to e \gamma \to e \E\Ebar$.  In this paper, we will
refer to them as
\begin {subequations}
\label {eq:rates}
\begin {align}
   \mbox{$1{\to}2$ rates:} \quad &
   \left[ \frac{d\Gamma}{d\xe} \right]_{e\to e\gamma} =
   \left[ \frac{d\Gamma}{d\xe} \right]^{\rm LO}_{e\to e\gamma} +
   \left[ \Delta \frac{d\Gamma}{d\xe} \right]_{e \to e\gamma}^{\rm NLO} ,
\label {eq:rateeeg}
\\ &
   \left[ \frac{d\Gamma}{d\xE} \right]_{\gamma\to \E\Ebar} =
   \left[ \frac{d\Gamma}{d\xE} \right]^{\rm LO}_{\gamma\to \E\Ebar} +
   \left[ \Delta \frac{d\Gamma}{d\xE} \right]_{\gamma \to \E\Ebar}^{\rm NLO} ,
\label {eq:rategEE}
\\
   \mbox{effective $1{\to}3$ rate:} \quad &
   \left[ \Delta \frac{d\Gamma}{d\xe\,d\xE} \right]_{e \to e\E\Ebar} .
\label {eq:1to3rate}
\end {align}
\end {subequations}
The symbol $\Delta$ in
$[ \Delta d\Gamma/d\xe\,d\xE ]_{e \to e\E\Ebar}$ is
a reminder that this rate represents a {\it correction}
(as in fig.\ \ref{fig:diagsNf1}) to a calculation of double
splitting as two, consecutive, independent LO splittings.%
\footnote{
  The fact that our effective $1{\to}3$ rate
  $[ \Delta d\Gamma/d\xe\,d\xE ]_{e \to e\E\Ebar}$ may
  therefore be negative will not
  cause any difficulties for the analysis of showers in this paper, where
  we treat high-energy $\Nf\alpha(\mu)$ as small and expand to first order in
  overlap effects.
}
Explicit formulas for the rates (\ref{eq:rates}) may be found in
ref.\ \cite{qedNf},%
\footnote{
  See appendix A of ref.\ \cite{qedNf} for a summary of rate formulas.
  Beware that our $\xE$ here is called $\ye$ in ref.\ \cite{qedNf}.
}
which carried out the calculations using Light Cone Perturbation Theory (LCPT).%
\footnote{
  In this paper, we are intentionally sloppy with some terminology.
  Technically, we should
  define the $x$'s by the splitting of lightcone longitudinal momentum:
  e.g.\ $P^+ \to \xe P^++\xE P^+ + (1{-}\xe{-}\xE) P^+$ for
  $e \to e\E\Ebar$ and $P^+ \to \xe P^+ + (1{-}\xe) P^+$ for
  $e \to e\gamma$.  But the splittings relevant to shower development are
  high energy and nearly collinear, and so we often refer to the
  $x$'s simply as ``energy fractions'' in our applications.
}
As discussed in refs.\ \cite{seq,finale,finale2} in the context of
gluon showers,
overlap effects of two consecutive splittings can
be accounted for by {\it classical}\/ probability analysis of a shower
that develops with these $1{\to}2$ splittings and $1{\to}3$ splittings.

% =========================================================================

\section{Net rates: definitions, numerics, and fits}
\label {sec:netrates}

\subsection{Basic net rates}

In refs.\ \cite{finale,finale2},
we showed how the NLO evolution of gluon showers could
be expressed in terms of the ``net'' rate $[d\Gamma/dx]_{\rm net}$ for
a splitting or pair of overlapping splittings to produce one daughter
of energy $xE$ (plus any other daughters) from a parent of energy $E$.
We then numerically evaluated the net rate
$[d\Gamma/dx]_{\rm net}$ for a mesh of $x$ values and then interpolated
using relatively simple fitting functions,
which are then used for calculations of shower development.
We will use the same
strategy here, except that now we have different types of particles
($\gamma$, $e$, and $\bar e$)
and so need multiple net rates depending on the type of parent and
daughter.

In the gluon case, one must be careful about final state,
identical particle combinatoric
factors when defining the net rate.  We may avoid that here,
and so simplify the discussion,
by using the large-$\Nf$ distinguishability between a pair-produced electron
and the direct heir of the original electron in $e \to e\gamma \to e\E\Ebar$.
Then every daughter in the process $e \to e\E\Ebar$ is distinguishable,
and the same is true of the other NLO or LO processes relevant in the
large-$\Nf$ limit: $e \to e\gamma$ and $\gamma \to \E\Ebar$.

We now establish notation by listing the basic net rates that
we need:
\begin {subequations}
\label {eq:netrates}
\begin {equation}
   \left[ \frac{d\Gamma}{dx} \right]^\net_{\uee}
   = 
   \left[ \frac{d\Gamma}{dx} \right]^\LO_{\uee}
   + \left[ \frac{d\Gamma}{dx} \right]^\NLO_{\uee} ,
\label {eq:netrateee}
\end {equation}
\begin {equation}
   \left[ \frac{d\Gamma}{dx} \right]^\net_{\ueE}
   = 
   \left[ \frac{d\Gamma}{dx} \right]^\NLO_{\ueE} ,
\end {equation}
\begin {equation}
   \left[ \frac{d\Gamma}{dx} \right]^\net_{\ueEbar}
   = 
   \left[ \frac{d\Gamma}{dx} \right]^\NLO_{\ueEbar} ,
\end {equation}
\begin {equation}
   \left[ \frac{d\Gamma}{dx} \right]^\net_{\ueg}
   = 
   \left[ \frac{d\Gamma}{dx} \right]^\LO_{\ueg}
   + \left[ \frac{d\Gamma}{dx} \right]^\NLO_{\ueg} ,
\end {equation}
\begin {equation}
   \left[ \frac{d\Gamma}{dx} \right]^\net_{\ugEbar}
   = 
   \left[ \frac{d\Gamma}{dx} \right]^\net_{\ugE}
   = 
   \left[ \frac{d\Gamma}{dx} \right]^\LO_{\ugE}
   + \left[ \frac{d\Gamma}{dx} \right]^\NLO_{\ugE} .
\end {equation}
\end {subequations}
Above, underlining of subscripts like $\uee$ indicate that we are
using the large-$\Nf$ limit to distinguish pair-produced electrons
$(\E)$
from other electron daughters ($e$) in overlapping splitting rates.
[This notational convention will help us differentiate basic net rates
(\ref{eq:netrates}) from combined quantities that we will introduce
later.]
The LO rates in (\ref{eq:netrates})
are given by (\ref{eq:LOrate0}) as
\begin {subequations}
\label {eq:LOrates}
\begin {align}
   \left[ \frac{d\Gamma}{dx} \right]^\LO_{\uee}
   \equiv& \left[ \frac{d\Gamma}{d\xe} \right]^\LO_{e\to e\gamma}
         &\quad\mbox{with $\xe=x$,$\phantom{1-{}}$}
\\
   \left[ \frac{d\Gamma}{dx} \right]^\LO_{\ueg}
   \equiv& \left[ \frac{d\Gamma}{d\xe} \right]^\LO_{e\to e\gamma}
         &\quad\mbox{with $\xe=1-x$,}
\\
   \left[ \frac{d\Gamma}{dx} \right]^\LO_{\ugE}
   \equiv& \left[ \frac{d\Gamma}{d\xE} \right]^\LO_{\gamma\to\E\Ebar}
         &\quad\mbox{with $\xE=x$.$\phantom{1-{}}$}
\end {align}
\end {subequations}
The $\ueE$ and $\ueEbar$ net rates do not have any leading-order (LO)
contribution,
since they only arise from the overlapping (and therefore NLO) splitting
$e \to e\gamma \to e\E\Ebar$.
The $\gamma{\to}\E$ and $\gamma{\to}\Ebar$ net rates are equal by
charge conjugation.
In terms of the building blocks (\ref{eq:rates}) whose formulas are given
in ref.\ \cite{qedNf}, the NLO net rates in (\ref{eq:netrates}) are
\begin {subequations}
\label {eq:NLOrates}
\begin {align}
   \left[ \frac{d\Gamma}{dx} \right]^\NLO_{\uee}
  &\equiv
   \left[ \Delta \frac{d\Gamma}{d\xe} \right]_{e \to e\gamma}^{\rm NLO}
   +
   \int_0^{1-\xe} d\xE \>
   \left[ \Delta \frac{d\Gamma}{d\xe\,d\xE} \right]_{e \to e\E\Ebar}
  && \mbox{with $\xe=x$,}
\label {eq:NLOee}
\\[15pt]
   \left[ \frac{d\Gamma}{dx} \right]^\NLO_{\ueE}
  &\equiv
   \int_0^{1-\xE} d\xe \>
   \left[ \Delta \frac{d\Gamma}{d\xe\,d\xE} \right]_{e \to e\E\Ebar}
  && \mbox{with $\xE=x$,}
\label{eq:NLOeE}
\\[15pt]
   \left[ \frac{d\Gamma}{dx} \right]^\NLO_{\ueEbar}
  &\equiv
   \int_0^{1-\xEbar} d\xe \>
   \biggl(
   \left[ \Delta \frac{d\Gamma}{d\xe\,d\xE} \right]_{e \to e\E\Ebar}
   \biggr)_{\!\!\xE = 1-\xe-\xEbar}
  && \mbox{with $\xEbar=x$,}
\\[15pt]
   \left[ \frac{d\Gamma}{dx} \right]^\NLO_{\ueg}
  &\equiv
   \left[ \Delta \frac{d\Gamma}{d\xe} \right]_{e \to e\gamma}^{\rm NLO}
  && \mbox{with $\xe=1-x$,}
\\[15pt]
   \left[ \frac{d\Gamma}{dx} \right]^\NLO_{\ugE}
  &\equiv
   \left[ \Delta \frac{d\Gamma}{d\xE} \right]_{\gamma \to\E\Ebar}^{\rm NLO}
  && \mbox{with $\xE=x$.}
\end {align}
\end {subequations}
\vspace{4pt}

Because all the daughters of our splitting processes $e{\to}e\gamma$,
$e{\to}e\E\Ebar$, and $\gamma{\to}\E\Ebar$ are distinguishable in large-$\Nf$,
the total rates for splitting of electrons or photons
are given in terms of net rates by simply
\begin {subequations}
\label {eq:Gamma}
\begin {align}
  \Gamma_e &= \int_0^1 dx \> \left[ \frac{d\Gamma}{dx} \right]^\net_\uee ,
\label {eq:Gammae}
\\
  \Gamma_\gamma &= \int_0^1 dx \> \left[ \frac{d\Gamma}{dx} \right]^\net_\ugE
               = \int_0^1 dx \> \left[ \frac{d\Gamma}{dx} \right]^\net_\ugEbar ,
\label {eq:Gammag}
\end {align}
\end {subequations}
without any identical-particle final state factors
such as those appearing in the analysis of $g{\to}gg$ and $g{\to}ggg$ in
refs.\ \cite{finale,finale2}.%
\footnote{
  See the discussion surrounding eqs.\ (3) and (4) of ref.\ \cite{finale}
  for comparison.
}
Regarding (\ref{eq:Gammae}), note that
$[d\Gamma/dx]_\uee$ accounts for both of the processes $e\to e\E\Ebar$ and
$e\to e\gamma$ that contribute to the effective electron splitting rate,
whereas, for example,
integrating $[d\Gamma/dx]_\ueE$ or $[d\Gamma/dx]_\ueEbar$ would
account only for $e \to e\E\Ebar$.

% -------------------------------------------------------------------------

\subsection{Numerics and Fits}

\subsubsection{Basic net rates}

Using the formulas from ref.\ \cite{qedNf} for the basic rates
(\ref{eq:netrates}), numerical integration%
\footnote{
  We managed numerical integration much more easily than reported
  for the gluonic case in appendix B.1 of ref.\ \cite{finale2}.
  Generally, the calculation of NLO contributions to net rates
  involve integration over (i) the energy fraction
  (call it $y$) of a real or virtual high-energy particle other than
  the one represented by $x$ in $[d\Gamma/dx]_{\uij}^\net$ and (ii)
  a time $\Delta t$ that is integrated over in the formulas of
  ref.\ \cite{qedNf} for the basic rates (\ref{eq:rates}).
  Here we found we could simply use Mathematica's
  \cite{Mathematica} built-in adaptive integrator
  NIntegrate to directly do 2-dimensional integrals over $(y,\Delta t)$
  to get results at the precision shown in Table \ref{tab:dGnet}.
  As in previous work, we still had to use more than machine precision when
  evaluating the very complicated integrands because of
  delicate cancellations that occur in limiting cases.
}
gives results for the NLO contributions (\ref{eq:NLOrates}) to the net rates
$[d\Gamma/dx]^\net_\uij$ as functions of $x$.
Those numerical integrations are sufficiently time consuming that,
following ref.\ \cite{finale2}, we will want to find a way to
accurately approximate the numerical results by relatively simple
analytic functions of $x$, which can then be used for numerically efficient
calculations of shower development.

In order to fit numerical results for the $[d\Gamma/dx]^\net_\uij$
to analytic forms, it is convenient to first transform the
$[d\Gamma/dx]^\net_\uij$ into smoother functions by
factoring out as much as we can determine about their singular
behavior as $x \to 0$ and $x \to 1$.
In ref.\ \cite{finale2}, which analyzed overlap effects for purely gluonic
showers in QCD, the NLO net rate for $g\to g$ had the same power-law
behavior as the leading-order rate, and so it was easier to search
for a good
analytic fit to the NLO/LO ratio 
$[d\Gamma/dx]^\NLO_{g\to g} / [d\Gamma/dx]^\LO_{g\to g}$
than to find a good fit directly to $[d\Gamma/dx]^\NLO_{g\to g}$.
In the study here of large-$\Nf$ QED, we modify
that procedure because the power-law divergences
of the NLO net rate as $x\to 0$ or $1$ do not always match that of the
corresponding leading-order rate.

We will define the smoother functions $f_\uij(x)$ in terms of
ratios $\left[ \frac{d\Gamma}{dx} \right]^\NLO_\uij / R_\uij(x)$
where the $R(x)$'s are chosen to be simple functions with the same
power-law divergences as the NLO net rates.  Specifically, we take
\begin {subequations}
\begin {align}
  R_\uee(\xe) &\equiv
    \xe^{-1/2} (1{-}\xe)^{-3/2} \,
    \frac{\Nf\alpha^2}{2\pi} \sqrt{ \frac{\qhat}{E} } \,,
\label{eq:Ree}
\\
  R_\ueE(\xE) &\equiv
    \xE^{-3/2} (1{-}\xE)^{+1/2} \,
    \frac{\Nf\alpha^2}{2\pi} \sqrt{ \frac{\qhat}{E} } \,,
\label{eq:ReE}
\\
  R_\ueg(x_\gamma) &\equiv R_\uee(1{-}x_\gamma) ,
\\
  R_\ugE(\xE) &\equiv
    \xE^{-1/2} (1{-}\xE)^{-1/2} \,
    \frac{\Nf^2\alpha^2}{2\pi} \sqrt{ \frac{\qhat}{E} } \,.
\end {align}
\end {subequations}
[Above, we've written the arguments $x$ of $R(x)$ as explicitly
$\xe$, $\xE$, etc.\ as a reminder of exactly what the
argument refers to for each type of net rate.]
We emphasize that there is nothing fundamental about these
exact choices of the $R(x)$'s;
they are merely the particular choices we made to simplify finding good fits.

We also found it convenient to isolate certain logarithms, associated with
the $\MSbar$ renormalization scale $\mu$.
Those logarithms appear in rates that
include loop corrections.  Specifically, we now define our
``smooth'' functions $f_\uij$ in terms of the numerically-computed NLO rates
$[d\Gamma/dx]^\NLO_\uij$ by
\begin {equation}
  \left[ \frac{d\Gamma}{dx} \right]^\NLO_\uij =
  L_\uij(x,\mu) + f_\uij(x) \, R_\uij(x) ,
\label{eq:fdef}
\end {equation}
where
\begin {subequations}
\label {eq:L}
\begin {align}
  & L_\uee(\xe,\mu) \equiv
    - \frac{\beta_0\alpha}{2}
    \left[ \frac{d\Gamma}{d\xe} \right]_\LO^{e\to e\gamma}
    \ln\left( \frac{ \mu^2 }{ \sqrt{\frac{(1{-}\xe)\qhat E}{\xe}} } \right) ,
\label{eq:Lee}
\\
  & L_\ueE \equiv L_\ueEbar \equiv 0 ,
\\
  & L_\ueg(x_\gamma,\mu) \equiv L_\uee(1{-}x_\gamma,\mu) ,
\\
  & L_\ugE(\xE,\mu) \equiv
    - \frac{\beta_0\alpha}{2}
    \left[ \frac{d\Gamma}{d\xE} \right]_\LO^{\gamma\to\E\Ebar}
    \ln\left( \frac{ \mu^2 }{ \sqrt{\qhat E} } \right) ,
\label {eq:LPH}
\end {align}
\end {subequations}
and where
\begin {equation}
   \beta_0 = \frac{2\Nf}{3\pi}
\label {eq:beta0}
\end {equation}
is the coefficient of the 1-loop renormalization group $\beta$-function for
$\alpha$.
As will be seen shortly,
the $L$'s above do not capture all of the logarithmic dependence of
the net rates on $x$ and $1{-}x$.
Our particular choice of
$x$ dependence (or lack of it) inside the logarithms of (\ref{eq:Lee})
and (\ref{eq:LPH}) is just a matter of convention for our definition
(\ref{eq:fdef}) of $f_\uij(x)$.
Readers need not ponder the logic of that choice too deeply;
mostly it is a combination of guesses we made early in our
work combined with some convenient choices for finding fits.

\begin {figure}[t]
\begin {center}
  \includegraphics[scale=0.5]{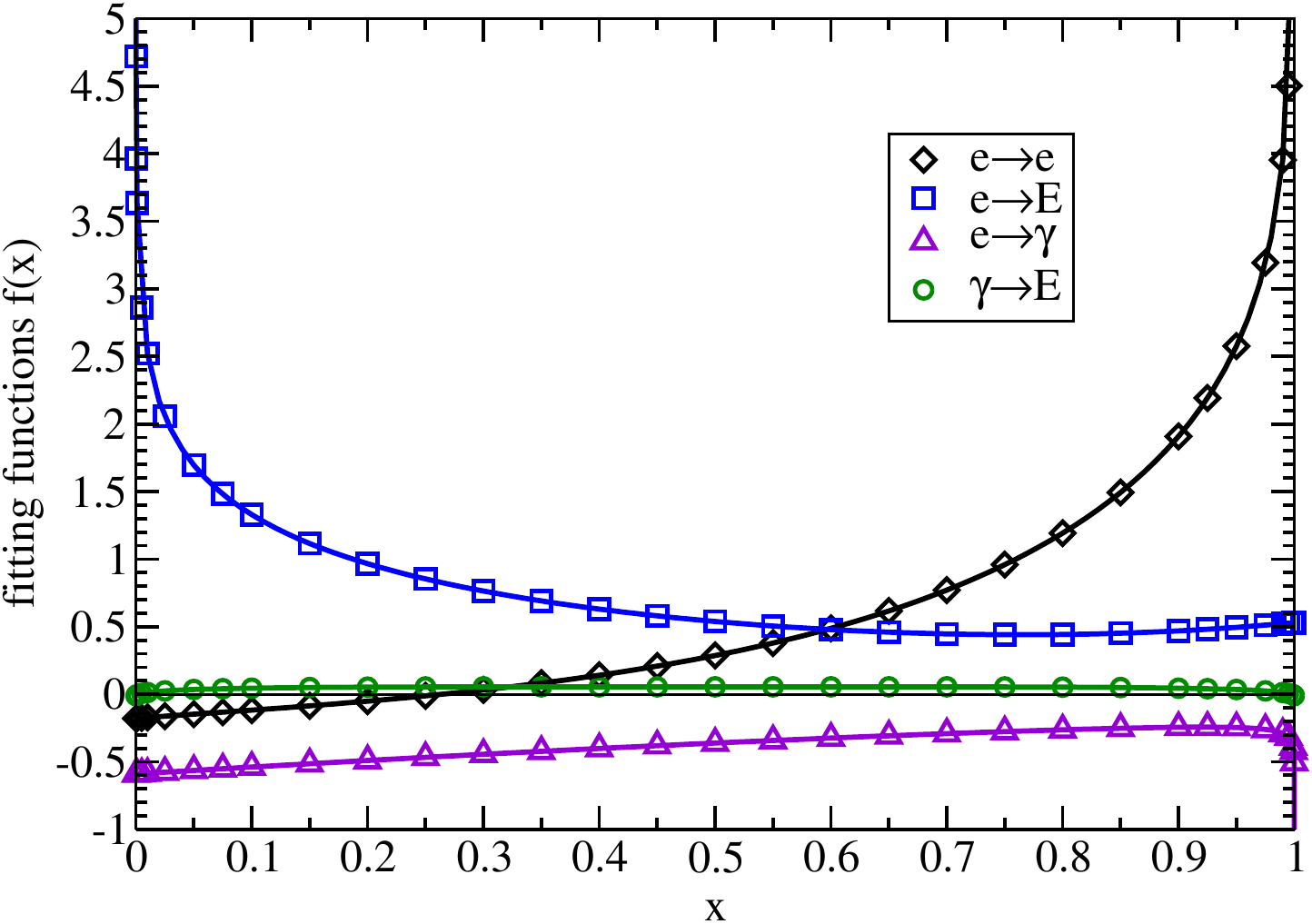}
  \caption{
     \label{fig:dGnet}
     Plots of numerically-computed data points (table \ref{tab:dGnet})
     and fits (\ref{eq:fits}) for
     the functions $f_\uij(x)$ defined by (\ref{eq:fdef}).
  }
\end {center}
\end {figure}

With these definitions,
table \ref{tab:dGnet} and the data points in fig.\ \ref{fig:dGnet}
present our numerical results for the
functions $f_\uij(x)$.  The corresponding numerical results for
our net rates (\ref{eq:netrates}) can be reconstructed using
(\ref{eq:fdef}).
There is only a single, joint column for $f_\ueE$ and $f_\ueEbar$ in
the table because they turn out to be equal:
\begin {equation}
   \left[ \frac{d\Gamma}{dx} \right]^\NLO_{\ueE} =
   \left[ \frac{d\Gamma}{dx} \right]^\NLO_{\ueEbar}
\label {eq:EequalEbar}
\end {equation}
(where, like everywhere in this paper,
the large-$\Nf$ limit is implicit).
We are not currently aware of any symmetry argument or other
high-level explanation for this equality.
Instead, we discovered numerically that the differential rate
$[\Delta\,d\Gamma/dx_e\,d\xE]_{e\to e\E\Ebar}$ appearing in
(\ref{eq:NLOrates}) is symmetric under
$\xE \to 1{-}x_e{-}\xE$ (i.e.\ $\xE \leftrightarrow \xEbar$).
See appendix \ref{app:EEbar} for some (low-level) insight into why the
formula \cite{qedNf}
for $[\Delta\,d\Gamma/dx_e\,d\xE]_{e\to e\E\Ebar}$
has this property.

\begin {table}[tp]

\setlength{\tabcolsep}{7pt}
\begin{center}
\begin{tabular}{lrrrr}
\hline
\hline
  \multicolumn{1}{c}{$x$} &
  \multicolumn{1}{c}{$f_\uee$} &
  \multicolumn{1}{c}{$ \begin{array}{c} f_\ueE\\f_\ueEbar \end{array} $} &
  \multicolumn{1}{c}{$f_\ueg$} &
  \multicolumn{1}{c}{$f_\ugE$} \\[12pt]
\hline
  % taken from my file fnew.tex produced by qednetFit.nb
  0.0001 & -0.1786 & 4.7199 & -0.5889 & -0.0099 \\
  0.0005 & -0.1777 & 3.9637 & -0.5890 & -0.0063 \\
  0.001 & -0.1774 & 3.6357 & -0.5886 & -0.0037 \\
  0.005 & -0.1745 & 2.8618 & -0.5865 & 0.0066 \\
  0.01 & -0.1712 & 2.5200 & -0.5839 & 0.0135 \\
  0.025 & -0.1620 & 2.0564 & -0.5763 & 0.0250 \\
  0.05 & -0.1472 & 1.6960 & -0.5635 & 0.0350 \\
  0.075 & -0.1324 & 1.4820 & -0.5509 & 0.0411 \\
  0.1 & -0.1173 & 1.3296 & -0.5384 & 0.0451 \\
  0.15 & -0.0856 & 1.1158 & -0.5138 & 0.0500 \\
  0.2 & -0.0508 & 0.9665 & -0.4899 & 0.0526 \\
  0.25 & -0.0118 & 0.8535 & -0.4666 & 0.0540 \\
  0.3 & 0.0323 & 0.7638 & -0.4440 & 0.0547 \\
  0.35 & 0.0828 & 0.6908 & -0.4220 & 0.0550 \\
  0.4 & 0.1410 & 0.6306 & -0.4009 & 0.0551 \\
  0.45 & 0.2085 & 0.5805 & -0.3805 & 0.0552 \\
  0.5 & 0.2871 & 0.5390 & -0.3608 & 0.0552 \\
  0.55 & 0.3792 & 0.5054 & -0.3420 & 0.0552 \\
  0.6 & 0.4877 & 0.4789 & -0.3241 & 0.0551 \\
  0.65 & 0.6166 & 0.4593 & -0.3070 & 0.0550 \\
  0.7 & 0.7711 & 0.4467 & -0.2910 & 0.0547 \\
  0.75 & 0.9590 & 0.4410 & -0.2762 & 0.0540 \\
  0.8 & 1.1927 & 0.4427 & -0.2629 & 0.0526 \\
  0.85 & 1.4937 & 0.4519 & -0.2517 & 0.0500 \\
  0.9 & 1.9083 & 0.4692 & -0.2441 & 0.0451 \\
  0.925 & 2.1920 & 0.4810 & -0.2431 & 0.0411 \\
  0.95 & 2.5759 & 0.4951 & -0.2461 & 0.0350 \\
  0.975 & 3.1929 & 0.5114 & -0.2598 & 0.0250 \\
  0.99 & 3.9530 & 0.5223 & -0.2888 & 0.0135 \\
  0.995 & 4.5022 & 0.5261 & -0.3160 & 0.0066 \\
  0.999 & 5.7373 & 0.5293 & -0.3889 & -0.0037 \\
  0.9995 & 6.2612 & 0.5297 & -0.4228 & -0.0063 \\
  0.9999 & 7.4724 & 0.5299 & -0.5051 & -0.0099 \\
\hline
\hline
\end{tabular}
\end{center}
\caption{
   \label{tab:dGnet}
   Results for the functions $f_\uij(x)$ extracted from numerical
   computation of the net rates (\ref{eq:netrates}) using the
   explicit formulas of ref.\ \cite{qedNf}.
   Numerical results for the rates were translated into numerical
   results for $f_\uij$ using the definition (\ref{eq:fdef}) of
   the $f_\uij$.
   The results for $f_\ueE$ and $f_\ueEbar$ are equal to each other.
}
\end{table}

We have found the following, reasonably good fits to the numerical rates
and will use these fits for all subsequent calculations in this paper:
\begin {subequations}
\label {eq:fits}
\begin {multline}
  f_\uee(x) =
    - \tfrac{3}{4} \ln (1{-}x)
    - 22.65461 + 43.86814\,x - 20.48818\,x^2 + 5.29318\,x^3
\\
    + 0.01427\,x^{1/2} - 1.18685\,x^{3/2} - 4.27886\,x^{5/2}
\\
    - 0.16141\,(1{-}x)^{1/2} + 12.59425\,(1{-}x)^{3/2} + 10.04309\,(1{-}x)^{5/2}
  ,
\label {eq:fee}
\end {multline}
\begin {multline}
  f_\ueE(x) = f_\ueEbar(x) =
    - \tfrac{22}{15\pi} \ln x
    + 10.45176 - 25.05713\,x + 0.71056\,x^2 - 6.76246 \,x^3
\\
    - 0.40871\,x^{1/2} + 8.94044\,x^{3/2} + 12.65584\,x^{5/2}
\\
    - 0.02067\,(1{-}x)^{1/2} - 2.06985\,(1{-}x)^{3/2} - 7.93648\,(1{-}x)^{5/2}
  ,
\label {eq:feE}
\end {multline}
\begin {multline}
  f_\ueg(x) =
    \tfrac{1}{6\pi} \ln (1{-}x)
    - 4.13754 + 8.91233\,x - 4.00731\,x^2 + 1.69990\,x^3
\\
    - 0.00903\,x^{1/2} - 0.56996\,x^{3/2} - 1.90189\,x^{5/2}
\\
    - 0.27947\,(1{-}x)^{1/2} + 1.19712\,(1{-}x)^{3/2} + 2.63102\,(1{-}x)^{5/2}
  ,
\end {multline}
\begin {multline}
  f_\ugE(x) =
    - 0.01296
    + 0.31063\,\bigl(x(1{-}x)\bigr)^{1/2}
    - 0.49837\,x(1{-}x)
\\
    + 0.44890\,\bigl(x(1{-}x)\bigr)^{3/2}
    - 0.29930\,\bigl(x(1{-}x)\bigr)^{2}
  .
\label {eq:fgE}
\end {multline}

\end {subequations}
When making fits, the coefficients of logarithms $\ln x$ and $\ln(1{-}x)$
were fixed to the exact values shown above, while all other coefficients
were allowed to float to whatever values gave the best fit.
Appendix \ref{app:logs} discusses how (either directly or indirectly) the
logarithms can be understood as arising
from vacuum-like DGLAP initial (or final)
radiation corrections to leading order (BDMPS-Z) single emission processes,
and how their coefficients may then be computed analytically.

The fits (\ref{eq:fits}) match every data point of table \ref{tab:dGnet} to
better than 0.0008 absolute error.

A convenience of our particular choices of $L_\uee(\xe,\mu)$ and
$L_\ugE(\xE,\mu)$ in (\ref{eq:L}) was the removal of a number of logarithmic
terms in the $f_\uij$.
Our choice of $L_\ugE$ removed the need for
$\ln\xE$ and symmetrically $\ln(1{-}\xE)$ terms in $f_\ugE(\xE)$.
(See appendix \ref{app:logs} for numerical evidence.)  
Our choice of $L_\uee$ did the same for $\ln\xe$
in $f_\uee(\xe)$.
[However, our choice of $\ln(1{-}\xe)$ dependence in $L_\ee$ has
no bearing on the $\ln(1{-}\xe)$ term in our fit for $f_\uee(\xe)$
and was chosen for historical reasons.%
\footnote{
  The historical reason for our choice of $x$ dependence for the logarithm in
  (\ref{eq:Lee}) comes from eq.\ (A.41) of ref.\ \cite{qedNf}, using also
  eqs.\ (A.5) and (A.7) of that reference.
  The parametric scale appearing in the denominator of the logarithm
  $\ln(\mu^2/\cdots)$ in $L_\uee(\xe)$
  happens to match a physical scale that will be
  discussed later in section \ref{sec:muAlt}, but there is no such
  correspondence in our choice of $L_\ugE(\xE)$.
  The $\ln(1{-}\xe)$ dependence of
  $L_\uee(\xe)$ is unrelated to the $\ln(1{-}\xe)$ term in (\ref{eq:fee})
  because $L_\uee(\xe)$ is suppressed compared to $R_\uee(\xe) \, f_\uee(\xe)$
  by a power of $1{-}\xe$ in the limit $\xe\to 1$.
}%
]

% ----------------------------------------------------------------------------

\subsubsection{Decomposition of \boldmath$[d\Gamma/dx]^\NLO_\ee$
                into real and virtual parts}

The NLO net rate $[d\Gamma/dx]^\NLO_\uee$ gets contributions from
both
(i) virtual corrections to single splitting $e \to e\gamma$ and
(ii) real double splitting $e\to e\gamma \to e\E\Ebar$,
respectively corresponding to the two terms in (\ref{eq:NLOee}).
Though not necessary for our final numerical
results, it will sometimes be insightful to look at these two
contributions separately.
For that purpose, let's correspondingly break down $f_\uee$ into
\begin {equation}
   f_\uee(x) = f_\ee^{\rm virt}(x) + f_\ee^{\rm real}(x) .
\label {eq:fdecompose0}
\end {equation}
By virtue of (\ref{eq:NLOrates}),
these two pieces of $f_\uee$ can be reconstructed from the data points in
table \ref{tab:dGnet}, or from the fits of (\ref{eq:fits}), as
\begin {equation}
   f_\ee^{\rm virt}(x) = f_\ueg(1-x)
   \quad \mbox{and} \quad
   f_\ee^{\rm real}(x) = f_\uee(x) - f_\ueg(1-x) .
\label {eq:fdecompose}
\end {equation}
The corresponding contributions to the net rate $[d\Gamma/dx]^\NLO_\uee$
are respectively%
\footnote{
  A similar separation of numerical results for $f_{g\to g}(x)$ into real
  and virtual contributions would not have been possible in the purely gluonic
  case of ref.\ \cite{finale2} because of the need to subtract
  infrared (IR) divergences.
  In that case,
  not only was there a double-log divergence for the net rate
  (which {\it was} subtracted),
  but the separate real and virtual contributions
  contained {\it power-law} IR divergences, which canceled only
  when those contributions were added together.
  See the discussion in section 1 of ref.\ \cite{qcd}
  and appendix E of ref.\ \cite{qcd}.
}
\begin {equation}
  \left[ \Delta \frac{d\Gamma}{d\xe} \right]_{e \to e\gamma}^{\rm NLO}
    =
    L_\uee(x,\mu) + f_\ee^{\rm virt}(x) \, R_\uee(x)
\end {equation}
and
\begin {equation}
  \int_0^{1-\xe} d\xE \>
    \left[ \Delta \frac{d\Gamma}{d\xe\,d\xE} \right]_{e \to e\E\Ebar}
    = 
    f_\ee^{\rm real}(x) \, R_\uee(x) .
\label {eq:feeRealDef}
\end {equation}

From (\ref{eq:fits}) and (\ref{eq:fdecompose}), our fits to
the real and virtual contributions for $f_\uee$ are
\begin {subequations}
\label {eq:fitrealvirt}
\begin {multline}
  f_\ee^{\rm real}(x) =
    - \tfrac{1}{6\pi} \ln x
    - \tfrac{3}{4} \ln (1{-}x)
    - 25.12199 + 49.86555\,x - 21.58057\,x^2 + 6.99308\,x^3
\\
    + 0.29374\,x^{1/2} - 2.38397\,x^{3/2} - 6.90988\,x^{5/2}
\\
    - 0.15238\,(1{-}x)^{1/2} + 13.16421\,(1{-}x)^{3/2} + 11.94498\,(1{-}x)^{5/2}
  ,
\label {eq:feeReal}
\end {multline}
\begin {multline}
  f_\ee^{\rm virt}(x) =
    \tfrac{1}{6\pi} \ln x
    + 2.46738 - 5.99741\,x + 1.09239\,x^2 - 1.69990\,x^3
\\
    - 0.27947\,x^{1/2} + 1.19712\,x^{3/2} + 2.63102\,x^{5/2}
\\
    - 0.00903\,(1{-}x)^{1/2} - 0.56996\,(1{-}x)^{3/2} - 1.90189\,(1{-}x)^{5/2}
  .
\label {eq:feeVirt}
\end {multline}
\end {subequations}

% ===========================================================================

\section{Choices of Renormalization Scale \boldmath$\mu$}
\label {sec:mu}

\subsection{QED versions of earlier scale choices}

In the context of purely-gluonic showers in QCD,
refs.\ \cite{finale,finale2} discussed three different choices of
an infrared (IR) factorization scale $\Lambda_\fac$ (introduced
to factorize out soft-radiation double logs arising in QCD in the
$\qhat$ approximation \cite{LMW}), which were used to also
set scales for the ultraviolet (UV) renormalization scale $\mu$
[as $\mu = (\qhatA \Lambda_\fac)^{1/4}$].
A soft-radiation factorization scale $\Lambda_\fac$ is unnecessary
in the QED case since the $\qhat$ approximation in QED is not afflicted
by soft-radiation double logarithms that appear in QCD.
So, we need focus only on $\mu$ in this paper.
One way to characterize the choice of $\mu$ made in
refs.\ \cite{finale,finale2} is that it is the scale of the total
transverse momentum kick that the medium gives to the
high-energy particles during a typical formation time $t_\form$,
which in the $\qhat$ approximation is
\begin {equation}
   \mu \sim \Delta p_\perp \sim \sqrt{ \qhat t_{\rm form} } \,.
\label {eq:muconcept}
\end {equation}

In the BDMPS-Z formalism for single splitting processes, the calculation
of splitting rates in the $\qhat$ approximation is formally related to
a two-dimensional non-relativistic harmonic oscillator quantum mechanics
problem with a complex frequency of oscillation given by%
\footnote{
  For more information (in the notation used here) see, for example,
  the short review in section 2 of ref.\ \cite{2brem},
  leading to eq.\ (1.5b) of ref.\ \cite{2brem} for QCD and
  eqs.\ (A.5) and (A.9) of ref.\ \cite{qedNf} for QED.
  See also sec.\ 2.1.1 of ref.\ \cite{qedNf}.
}
\begin {subequations}
\label {eq:Omegas}
\begin {align}
  \mbox{QCD $~g{\to}gg$}\,:& \qquad
    \Omega = \sqrt{ -\frac{i\qhatA}{2E}
                    \left( -1 + \frac{1}{x} + \frac{1}{1{-}x} \right)
             } \, ,
\\
  \mbox{QED $~e{\to}e\gamma$}\,:& \qquad
    \Omega = \sqrt{ -\frac{i\qhat}{2E}
                    \left( -1 + \frac{1}{\xe} + 0 \right)
                  } \, ,
\\
  \mbox{QED $\gamma{\to}\E\Ebar$}:& \qquad
    \Omega = \sqrt{ -\frac{i\qhat}{2E}
                    \left( 0 + \frac{1}{\xE} + \frac{1}{1{-}\xE} \right)
                  } \, .
\label {eq:OmegagEEbar}
\end {align}
\end {subequations}
The formation time is characterized by the time scale $1/|\Omega|$.
Focusing only on parametric behavior,
the scale choice (\ref{eq:muconcept}) would be
\begin {subequations}
\label {eq:mux}
\begin {align}
  \mbox{QCD $~g{\to}gg$}\,:& \qquad
    \mu \sim  \bigl(x(1{-}x)\qhat E\bigr)^{1/4} ,
\\
  \mbox{QED $~e{\to}e\gamma$}\,:& \qquad
    \mu \sim \left( \frac{\xe \qhat E}{x_\gamma} \right)^{1/4} ,
\\
  \mbox{QED $\gamma{\to}\E\Ebar$}:& \qquad
    \mu \sim (\xE\xEbar \qhat E)^{1/4} .
\end {align}
\end {subequations}
In the $g{\to}gg$ case, this was our preferred choice of $\mu$ in
refs.\ \cite{finale,finale2}.  The QED version was used for the results
in the last column of our table \ref{tab:chi2}.

Refs.\ \cite{finale,finale2} noted that soft emissions
($x$ or $1{-}x \ll 1$)
do not contribute significantly to the shape of energy deposition
in the case of an infinite medium, and so one could
ignore the $x$ dependence
in (\ref{eq:mux}) and instead simply choose $\mu \sim (\qhat E)^{1/4}$
as in the next-to-last column of table \ref{tab:chi2}.
Refs.\ \cite{finale,finale2} also noted that splittings
with $E \ll E_0$, in a shower that started with energy $E_0$,
do not significantly affect where energy is deposited,
and so one could alternatively use the constant value
$\mu \sim (\qhat E_0)^{1/4}$
for all the splittings in the shower,
as in the third column of table \ref{tab:chi2}.%
\footnote{
  In refs.\ \cite{finale,finale2} analysis of gluonic showers,
  the three choices of $\mu$ discussed
  above were written indirectly as $\mu = (\qhatA \Lambda_\fac)^{1/4}$
  with $\Lambda_\fac \propto x(1{-}x)E$ or $\Lambda_\fac \propto E$
  or $\Lambda_\fac \propto E_0$.
}

To summarize for future reference,
the three choices of renormalization scale for QED just discussed are
\begin {subequations}
\label {eq:muchoices}
\begin {align}
  \mu &\propto (\qhat E_0)^{1/4} ,
\label {eq:muE0}
\\
  \mu &\propto (\qhat E)^{1/4} ,
\label {eq:muE}
\\
  [\mu_{e\to e\gamma},\mu_{\gamma\to\E\Ebar}]
  &\propto
  \bigl[
%    \left( \frac{\xe \qhat E}{x_\gamma} \right)^{1/4}
    (\xe \qhat E/x_\gamma)^{1/4}
    \,,\,
    (\xE\xEbar \qhat E)^{1/4}
  \bigr] .
\label {eq:muEx}
\end {align}
\end {subequations}
The choice of overall proportionality constant represented by
proportionality signs above will not affect our results
for $\sigma/\lstop$ nor, more generally,
any aspect of the shapes $S_\rho(Z)$ and $S_\eps(Z)$ of charge
deposition $\rho(z)$ or energy deposition $\eps(z)$.%
\footnote{
  This is because a common rescaling $\mu \to \lambda\mu$
  of all $\mu$'s
  by a constant $\lambda$ changes NLO rates by an amount proportional
  to the corresponding LO rates
  and so can be absorbed by a (constant) change in the
  value of $\qhat$.  It therefore cannot affect quantities like
  $\sigma/\lstop$ and the
  shapes $S(Z)$ which are (by design) insensitive to the value
  of $\qhat$.
}

We will use (\ref{eq:muchoices}) [and especially the last two cases] to
test how sensitive our results are to different choices of
renormalization scale.  One may already see from table \ref{tab:chi2}
that different reasonable choices do not make
much of a difference,
and so it is not necessary to obsess over
which choice is best motivated.

The analogous discussion of
refs.\ \cite{finale,finale2} considered using
(\ref{eq:muconcept}) and retaining the $x$ dependence
[as in (\ref{eq:muEx}) here] to be the most physically motivated choice.
We now find that assessment somewhat less compelling because there
is a different choice of $x$-dependent scale one might consider,
which also has a reasonable physical motivation.

% --------------------------------------------------------------------------

\subsection{A different choice}
\label {sec:muAlt}

Instead of setting the renormalization scale to the transverse momentum
scale of (\ref{eq:muconcept}), one might consider setting $\mu^2$ to be
of order the
combined invariant mass $p^\mu p_\mu \equiv (p_2{+}p_3)^\mu (p_2{+}p_3)_\mu$
of the two daughters (labeled here as particles ``2'' and ``3'') of a
single BDMPS-Z splitting such as $g{\to}gg$, $e\to e\gamma$
or $\gamma\to\E\Ebar$.
This is equivalent to $\mu^2 \sim |\vec p_{\rm COM}|^2$,
where $\pm\vec p_{\rm COM}$ are the 3-momenta of the daughters in their
center-of-momentum frame.
We leave the details of the parametric analysis to appendix \ref{app:pCOM},
but the result is
\begin {equation}
   \mu^2 \sim (p_2{+}p_3)^\mu (p_2{+}p_3)_\mu \sim \frac{E}{t_{\rm form}} \,,
\label {eq:muconceptAlt}
\end {equation}
which differs from (\ref{eq:muconcept}) when one of the daughters is soft.
In particular, (\ref{eq:Omegas}) and $t_{\rm form} \sim 1/|\Omega|$ now give
\begin {subequations}
\label {eq:muxAlt}
\begin {align}
  \mbox{QCD $~g{\to}gg$}\,:& \qquad
    \mu \sim  \left( \frac{\qhat E}{x(1{-}x)} \right)^{1/4} ,
\\
  \mbox{QED $~e{\to}e\gamma$}\,:& \qquad
    \mu \sim \left( \frac{x_\gamma \qhat E}{\xe} \right)^{1/4} ,
\\
  \mbox{QED $\gamma{\to}\E\Ebar$}:& \qquad
    \mu \sim \left( \frac{\qhat E}{\xE\xEbar} \right)^{1/4}
\end {align}
\end {subequations}
in contrast to (\ref{eq:mux}).

Table \ref{tab:chi2more} shows an expanded version of table \ref{tab:chi2}
in which we have added a column for the new renormalization scale
choice (\ref{eq:muxAlt}).  The uncertainty about
whether the most sensible choice of renormalization scale should be
(\ref{eq:muconcept}) or (\ref{eq:muconceptAlt}) symmetrically brackets
(within round-off error) the $\mu{\propto}(\qhat E)^{1/4}$ result.
This is because the $x$-dependencies of $\mu$ in
(\ref{eq:muconcept}) and (\ref{eq:muconceptAlt}) are inverse to each other
(while the $\qhat$ and $E$ dependence is the same), and so their difference
from $\mu \propto (\qhat E)^{1/4}$ will generate opposite changes to the
logarithms $L_\uij$ of (\ref{eq:L}) and so opposite changes to NLO
quantities.

\begin {table}[tp]

\setlength{\tabcolsep}{7pt}
\begin {center}
\begin{tabular}{rcccc}
\hline
\hline
  && \multicolumn{3}{c}{overlap correction to $\sigma/\lstop$}
\\
\cline{2-5}
  & $\mu\propto(\qhat E_0)^{1/4}$
  & $\begin {matrix}
       \scriptstyle{\mu_{e\to e\gamma}{\propto}(x_\gamma \qhat E/\xe)^{1/4}} \\
       \scriptstyle{
          \mu_{\gamma\to e\bar e}{\propto}\bigl(\qhat E\bigr/\xe x_\gamma)^{1/4}}
     \end{matrix}
    $
  & $\mu\propto(\qhat E)^{1/4}$
  & $\begin {matrix}
       \scriptstyle{\mu_{e\to e\gamma}{\propto}(\xe \qhat E/x_\gamma)^{1/4}} \\
       \scriptstyle{
            \mu_{\gamma\to e\bar e}{\propto}\bigl(\xe x_\gamma \qhat E\bigr)^{1/4}}
    \end{matrix}$
\\
\hline
  charge, $e$ & $-87.0\%\times\Nf\alpha$
               & $~{-}88.7\%\times\Nf\alpha$
               & $~{-}84.6\%\times\Nf\alpha$
               & $~{-}80.4\%\times\Nf\alpha$
\\
  energy, $e$ &
               & $+113.6\%\times\Nf\alpha$
               & $+113.1\%\times\Nf\alpha$
               & $+112.5\%\times\Nf\alpha$
\\
  energy, $\gamma$ &
               & $~{+}99.4\%\times\Nf\alpha$
               & $~{+}98.6\%\times\Nf\alpha$
               & $~{+}97.9\%\times\Nf\alpha$
\\
\hline
\hline
\end{tabular}
\end {center}
\caption{%
\label{tab:chi2more}%
  Like table \ref{tab:chi2} except that we have inserted a new column
  (the second column of numbers) for the choice (\ref{eq:muxAlt}) of
  renormalization scale, and we have included an extra significant digit
  in our results to make clear that the values in the second and fourth
  columns of numbers symmetrically bracket those in the third column
  (within round-off error).
}
\end{table}

Because of this reflection symmetry about the $\mu{\sim}(\qhat E)^{1/4}$ result,
we will not bother explicitly showing (\ref{eq:muconceptAlt}) for
results in this paper other than $\lstop/\sigma$.  We will just
show results for (\ref{eq:muE}) and (\ref{eq:muEx}) to give a
sense of the variation of results for different reasonable choices of
renormalization scale.

% =========================================================================

\section{Charge stopping revisited}
\label {sec:charge}

For large-$\Nf$ QED, ref.\ \cite{qedNfstop} analyzed
the size of overlap corrections to the ``$\qhat$-independent'' ratio
$\sigma/\lstop$ of the width $\sigma$ of the charge-stopping distribution
$\rho(z)$ compared to the
stopping length $\lstop \equiv \langle z \rangle_\rho$.
In this section, we review that analysis in preparation for later discussion
of energy deposition in large-$\Nf$ QED.
Here we will update the notation and renormalization scales choices
of ref.\ \cite{qedNfstop}
to be closer to the related analysis of QCD energy deposition in
refs.\ \cite{finale,finale2},
and we will use the fit to the relevant NLO net rate (\ref{eq:NLOee}) to
greatly simplify numerics.%
\footnote{
  In particular, we avoid the need to puzzle through the very obscure changes
  of variables that were made in appendix D of ref.\ \cite{qedNfstop},
  which would be a headache to generalize to the case of energy
  deposition.
}

% ----------------------------------------------------------------------------

\subsection{Basic equation}

Because of the equality (\ref{eq:EequalEbar}) of the net rates
$[d\Gamma/dx]_\ueE$ and $[d\Gamma/dx]_\ueEbar$, the $\E$ and $\Ebar$
in $e \to e\E\Ebar$ will (statistically) be produced with the same
distribution in energy, and so they will subsequently deposit charge
in exactly the same way (statistically) except for sign,
and so their contributions to the total charge deposition $\rho(z)$
will exactly cancel.
Only the fate of the $e$ daughter of (large-$\Nf$)
$e\to e\E\Ebar$ will matter for charge deposition.
The only rate we need from (\ref{eq:netrates}) is therefore
$[d\Gamma/dx]_\uee$.
(The situation will be more complicated when
we later discuss energy deposition.)

Following ref.\ \cite{qedNfstop}, our starting equation is
\begin {equation}
  \rho(E,z+\Delta z)
  =
  [1 - \Gamma_e(E)\,\Delta z] \, \rho(E,z)
  +
  \int_0^1 dx \> \left[\frac{d\Gamma}{dx}(E,x)\right]^\net_\uee \Delta z \,
  \rho(x E,z)
\label {eq:rhoEvolve0}
\end {equation}
for infinitesimal $\Delta z$.
This can be understood by breaking up the distance traveled
$z+\Delta z$ on the left-hand side into
first traveling $\Delta z$ followed by traveling distance $z$.
In the first $\Delta z$ of distance,
the particle has a chance $1-\Gamma(E)\,\Delta z$ of
not splitting at all, and then
the charge density deposited after
traveling the remaining distance $z$ will just be
$\rho(E,z)$.  This possibility is represented by the first term on the
right-hand side of (\ref{eq:rhoEvolve0}).
The second term represents the alternative possibility that
the particle {\it does} split in
the first $\Delta z$.  In this case, the daughter $e$ will have energy
$xE$ and so deposit charge density $\rho(xE,z)$
after traveling the remaining distance $z$.
Eq.\ (\ref{eq:rhoEvolve0}) may be re-expressed as the differential equation
\begin {equation}
  \frac{\partial \rho(E,z)}{\partial z}
  =
  - \Gamma_e(E) \, \rho(E,z)
  +
  \int_0^1 dx \> \left[\frac{d\Gamma}{dx}(E,x) \right]^\net_\uee
  \rho(x E,z)
\label {eq:rhoEvolve1}
\end {equation}
and then rewritten using (\ref{eq:Gammae}) as
\begin {equation}
  \frac{\partial \rho(E,z)}{\partial z}
  =
  \int_0^1 dx \> \left[\frac{d\Gamma}{dx}(E,x) \right]^\net_\uee
  \bigl\{ \rho(xE,z) - \rho(E,z) \bigr \} .
\label {eq:rhoEvolve}
\end {equation}

% ----------------------------------------------------------------------------

\subsection{Scaled equation (for appropriate choices of \boldmath$\mu$)}
\label{sec:rhoscale}

The equation (\ref{eq:rhoEvolve}) can be nicely simplified if the rate
$d\Gamma/dx$ scales with energy as $E^{-1/2}$.
However, the NLO rates (\ref{eq:fdef}) also have additional logarithmic
dependence (\ref{eq:L})
on $E$ if the renormalization scale $\mu$ is held fixed.
Ref.\ \cite{qedNfstop} presents a method for dealing with that in the
QED case, but in this paper we will primarily
follow the gluon shower analysis of refs.\ \cite{finale,finale2}
and so mainly focus on the choices (\ref{eq:muE}) and
(\ref{eq:muEx}) of renormalization scale,
where $\mu$ scales with energy as $E^{1/4}$.
There is then no energy dependence in the logarithms of (\ref{eq:L}),
and the implicit $\mu$ dependence of $\alpha(\mu)$ in the NLO rates is
higher order and will not affect NLO calculations.%
\footnote{
  Higher-order corrections are not problematically
  enhanced by large logarithms because shower deposition distributions
  are dominated by the effects of quasi-democratic
  splittings with $E \sim E_0$,
  and in QED we do not have the large infrared double logarithms that
  complicated the gluon shower discussion in refs.\ \cite{finale,finale2}.
}

There is a potential issue that using an energy-dependent $\mu$ then
moves the logarithmic energy dependence to the {\it leading}-order rates
(\ref{eq:LOrate0}) through the implicit $\mu$ dependence of
$\alpha = \alpha(\mu)$.
For reasons discussed in ref.\ \cite{finale2}, we may ignore that effect
for the purpose of ascertaining whether overlap effects are large or
small;
the {\it relative} size of overlap corrections
(as in table \ref{tab:chi2}) is only affected at yet-higher order in $\alpha$
than our results.%
\footnote{
  See in particular the beginning of section 4 of ref.\ \cite{finale2}.
  Because we do not have the large logarithms of the gluon shower case
  (see the preceding footnote), the situation here is described simply
  by eq.\ (4.3) of ref.\ \cite{finale2}.
  [This argument assumes that we have chosen the renormalization scale so that
  $\mu \sim (\qhat E_0)^{1/4}$ for the special case of
  quasi-democratic splittings (neither $x$ nor $1{-}x$ small) with
  energy $E \sim E_0$, which is true for all of the choices discussed
  in section \ref{sec:mu}.]
}

So we will treat $[d\Gamma/dx]^\net_\uee$ in (\ref{eq:rhoEvolve}) as
scaling with energy exactly as $E^{-1/2}$ and introduce rescaled variables
$d\tilde\Gamma$, $\tilde z$, and $\tilde\rho$ by
\begin {equation}
    \left[ \frac{d\Gamma}{dx}(E,x) \right]_\uij^\net
    = E^{-1/2} \left[ \frac{d\tilde\Gamma}{dx}(x) \right]_\uij^\net ,
  \qquad
  z = E^{1/2} \tilde z ,
  \qquad
  \rho(E,z) = Q_0 E^{-1/2} \, \tilde\rho(\tilde z) ,
\label {eq:tilde}
\end {equation}
where $Q_0$ is the charge of the particle that
initiated the shower.  For a shower initiated by a particle
with energy $E_0$, (\ref{eq:rhoEvolve}) becomes
\begin {equation}
  \frac{\partial \tilde\rho(\tilde z)}{\partial\tilde z}
  =
  \int_0^1 dx \> \biggl[\frac{d\tilde\Gamma}{dx}(x) \biggr]_\uee^\net
  \bigl\{ x^{-1/2}\,\tilde\rho(x^{-1/2} \tilde z)
          - \tilde\rho(\tilde z) \bigr\} .
\label {eq:rhoEvolve9}
\end {equation}
Now that the variable $\tilde z$ has served its purpose, we may use
(\ref{eq:tilde}) with $E=E_0$ to convert the simplified
equation (\ref{eq:rhoEvolve9}) back to
the original unscaled variables:
\begin {equation}
  \frac{\partial \rho(z)}{\partial z}
  =
  \int_0^1 dx \> \biggl[\frac{d\Gamma}{dx}(E_0,x) \biggr]_\uee^\net
  \bigl\{ x^{-1/2}\,\rho(x^{-1/2} z)
          - \rho(z) \bigr\}
\label {eq:rhoeq}
\end {equation}
for $\rho(z) \equiv \rho(E_0,z)$.
This form of (\ref{eq:rhoEvolve}) is only valid if the rates can
be taken to scale exactly as $E^{-1/2}$ for the desired choices of $\mu$.

% ----------------------------------------------------------------------------

\subsection{Moments \boldmath$\langle z^n \rangle$
           (for appropriate choices of $\mu$)}
\label {sec:momentsrho}

Multiplying both sides of (\ref{eq:rhoeq}) by $z^n$
and integrating over $z$ gives
\begin {equation}
  -n \langle z^{n-1}\rangle_\rho
  =
  \int_0^1 dx \> \biggl[\frac{d\Gamma}{dx}(E_0,x)\biggr]^\net_\uee
  \bigl\{ x^{n/2} \langle z^n \rangle_\rho
         - \langle z^n \rangle_\rho \bigr\}
\end {equation}
and so the recursion relation
\begin {subequations}
\label {eq:znMasterRho}
\begin {equation}
   \langle z^n \rangle_\rho =
   \frac{ n \langle z^{\,n-1} \rangle_\rho }
        { \Avg_\uee[ 1{-}x^{n/2} ] }
   \,,
\label {eq:znrho}
\end {equation}
where we use the short-hand notation%
\footnote{
  ``$\Avg$'' (average) is a misnomer because our ``weight''
  $[d\Gamma/dx]_\uee^\net$ is not normalized.  As a result,
  $\Avg_\uee[1]$ equals $\Gamma_e$ instead of $1$.
  [See (\ref{eq:Gammae}).]
  We stick with the notation for the sake of consistency with
  refs.\ \cite{qedNfstop,finale2}.
}
\begin {equation}
   \Avg_{i\to j}[g(x)] \equiv
   \int_0^1 dx \> \biggl[\frac{d\Gamma}{dx}(E_0,x)\biggr]^\net_{i\to j} g(x) .
\end {equation}
\end {subequations}
We will normalize our definition of the moments as
\begin {equation}
  \langle z^n \rangle_\rho
  \equiv \frac{1}{Q_0} \int_0^\infty dz \> z^n \, \rho(E_0,z) ,
\end {equation}
where $Q_0$ is the charge of the (charged) particle that initiated the
shower, so that $\langle 1 \rangle_\rho = 1$.

Similar to the discussion in ref.\ \cite{finale2}, we then expand
moments as
\begin {equation}
   \langle z^n \rangle \simeq
   \langle z^n \rangle^\LO + \delta \langle z^n \rangle \,,
\label {eq:znexpand}
\end {equation}
where $\langle z^n \rangle^\LO$ represents the result obtained
using only leading-order rates,
and $\delta \langle z^n \rangle$ represents the
NLO (i.e.\ overlap) correction, expanded to first order.
The recursion relation (\ref{eq:znrho}) expands to
\begin {equation}
   \langle z^n \rangle_\rho^\LO =
   \frac{ n \langle z^{\,n-1} \rangle_\rho^\LO }
        { \Avg_\uee^\LO[ 1{-}x^{n/2} ] }
\label {eq:znLOrho}
\end {equation}
and
\begin {equation}
  \delta \langle z^n\rangle_\rho
  = \langle z^n \rangle_\rho^\LO
    \left[
       \frac{ \delta\langle z^{n-1} \rangle_\rho }
            { \langle z^{n-1} \rangle_\rho^\LO }
       -
       \frac{ \dAvg_\uee[ 1{-}x^{n/2} ] }
            { \Avg_\uee^\LO[ 1{-}x^{n/2} ] }
    \right] ,
\label {eq:dznrho}
\end {equation}
where
\begin {subequations}
\label {eq:Avg}
\begin {align}
   \Avg_{i\to j}^\LO[g(x)] &\equiv
   \int_0^1 dx \> \biggl[\frac{d\Gamma}{dx}(E_0,x)\biggr]_{i\to j}^\LO g(x) ,
\label {eq:AvgLO}
\\
\intertext{and where}
   \dAvg_{i\to j}[g(x)]\, &\equiv
   \int_0^1 dx \> \biggl[\frac{d\Gamma}{dx}(E_0,x)\biggr]_{i\to j}^{\NLO} g(x)
\end {align}
is the NLO correction.
\end {subequations}

% ----------------------------------------------------------------------------

\subsection{Numerical Results}
\label {sec:numericsrho}

For reference, our numerical results for the expansions
(\ref{eq:znexpand}) of various moments $\langle z^n \rangle_\rho$
are shown in table \ref{tab:momentsrho} in units of $\ell_0^n$
where
\begin {equation}
   \ell_0 \equiv \frac{1}{\alpha} \sqrt{ \frac{E_0}{\qhat} } \,.
\label {eq:ell0}
\end {equation}
We have shown results for both the $x$-independent choice (\ref{eq:muE})
and $x$-dependent choice $(\ref{eq:muEx})$ of
renormalization scale $\mu$.

Our real goal is to look at quantities, like the shape
$S_\rho(Z)$ of the charge deposition distribution $\rho(z)$,
that are insensitive to
any physics that can be absorbed into the value of $\qhat$.
Table \ref{tab:shaperho} presents values related
to the shape functions' moments $\langle Z^n \rangle$;
reduced moments
\begin {equation}
   \mu_{n,S} \equiv \bigl\langle (Z - \langle Z \rangle)^n \bigr\rangle ;
\end {equation}
and cumulants $k_{n,S}$, which are the same as $\mu_{n,S}$ for $n\le 3$
but differ for
\begin {equation}
   k_{4,S} \equiv \mu_{4,S} - 3 \mu_{2,S}^2 .
\label {eq:k4}
\end {equation}
However, for the sake of comparing apples to apples, we have followed
ref.\ \cite{finale2} by first converting all of these quantities into
corresponding lengths:
$\langle Z^n \rangle_{\vphantom{S}}^{1/n}$, $\mu_{n,S}^{1/n}$,
and $k_{n,S}^{1/n}$.  For each such quantity $Q$, the table gives
the LO value $Q_{\rm LO}$,
the NLO correction $\delta Q$ when expanded to first order, and the relative
size of overlap corrections
\begin {equation}
   \chi\alpha \equiv \frac{\delta Q}{Q_\LO} \,.
\end {equation}
For our purpose, the analysis of the various moments of the
shape function $S(Z)$ will be enough to
answer the question of whether or not $\qhat$-insensitive overlap
corrections
in QED are generically large or small compared to those in
QCD for comparable values of $N\alpha$.
Unlike refs.\ \cite{finale,finale2}, we will not make the additional
numerical effort to more generally compute the overlap corrections to
the full functional form of the shape function $S(Z)$.

\begin {table}[t]

\setlength{\tabcolsep}{7pt}
\begin{center}
\begin{tabular}{lccc}
\hline\hline
  \multicolumn{1}{c}{$z^n$}
  & \multicolumn{1}{c}{$\langle z^n\rangle_\rho^\LO$}
  & \multicolumn{2}{c}{$\delta\langle z^n \rangle_\rho$}
 \\
\cline{3-4}
  & &
  \multicolumn{1}{c}{$\scriptstyle \mu = (\qhat E)^{1/4}$}
  &
  \multicolumn{1}{c}{$
      \phantom{\strut}
      \mu_{e\to e\gamma} = (\xe \qhat E/x_\gamma)^{1/4}
      \atop
      \mu_{\gamma\to\E\Ebar} = (\xE\xEbar \qhat E)^{1/4}
  $} 
  \\[6pt]
\cline{2-4}
  & \multicolumn{3}{c}{in units of $\ell_0^{\kern1pt n}$} \\
\hline
$\langle z\rangle$   & 5.0144 & -6.8237\,$\Nf\alpha$ & -7.3990\,$\Nf\alpha$ \\
$\langle z^2\rangle$ & 35.658 & -114.83\,$\Nf\alpha$ & -122.14\,$\Nf\alpha$ \\
$\langle z^3\rangle$ & 324.38 & -1795.0\,$\Nf\alpha$ & -1886.4\,$\Nf\alpha$ \\
$\langle z^4\rangle$ & 3571.7 & -29530\phantom{.}\,$\Nf\alpha$
                                           & -30774\phantom{.}\,$\Nf\alpha$ \\
\hline\hline
\end{tabular}
\end{center}
\caption{
   \label{tab:momentsrho}
   Expansions (\ref{eq:znexpand}) of moments $\langle z^n \rangle_\rho$
   of the charge deposition distribution $\rho(z)$
   for renormalization scale choices (\ref{eq:muE}) and (\ref{eq:muEx}).
   The unit $\ell_0$ is defined by (\ref{eq:ell0}).
}
\end{table}

\begin {table}[t]
\begin {center}

\setlength{\tabcolsep}{7pt}
%\resizebox{\textwidth}{!}{%
\begin{tabular}{lcccccc}
\hline\hline
  \multicolumn{1}{l}{quantity $Q$}
  & \multicolumn{1}{c}{$Q_\rho^\LO$}
  & \multicolumn{1}{c}{$\delta Q_\rho$}
  & \multicolumn{1}{c}{$\chi\alpha$}
  & 
  & \multicolumn{1}{c}{$\delta Q_\rho$}
  & \multicolumn{1}{c}{$\chi\alpha$}
 \\
\cline{3-4}\cline{6-7}
  &&
  \multicolumn{2}{c}{$\scriptstyle \mu \propto (\qhat E)^{1/4}$}
  &&
  \multicolumn{2}{c}{$
      \phantom{\strut}
      \mu_{e\to e\gamma} \propto (\xe \qhat E/x_\gamma)^{1/4}
      \atop
      \mu_{\gamma\to\E\Ebar} \propto (\xE\xEbar \qhat E)^{1/4}
  $} 
  \\[6pt]
\hline
$\langle Z \rangle$
   & $1$ \\
$\langle Z^2 \rangle^{1/2}$
   & $1.1909$
   & $-0.2969\,\Nf\alpha$
   & $-0.2494\,\Nf\alpha$
   &
   & $-0.2825\,\Nf\alpha$
   & $-0.2372\,\Nf\alpha$ \\
$\langle Z^3 \rangle^{1/3}$
   & $1.3702$
   & $-0.6629\,\Nf\alpha$
   & $-0.4838\,\Nf\alpha$
   &
   & $-0.6343\,\Nf\alpha$
   & $-0.4629\,\Nf\alpha$ \\
$\langle Z^4 \rangle^{1/4}$
   & $1.5417$
   & $-1.0886\,\Nf\alpha$
   & $-0.7061\,\Nf\alpha$
   &
   & $-1.0461\,\Nf\alpha$
   & $-0.6785\,\Nf\alpha$ \\[1pt]
\hline
$\mu_{2,S}^{1/2} {=} k_{2,{\rm S}}^{1/2} {=} \sigma_S$
   & $0.6466$
   & $-0.5469\,\Nf\alpha$
   & $-0.8457\,\Nf\alpha$
   &
   & $-0.5202\,\Nf\alpha$
   & $-0.8044\,\Nf\alpha$ \\
$\mu_{3,S}^{1/3} {=} k_{3,{\rm S}}^{1/3}$
   & $0.6828$
   & $-1.1526\,\Nf\alpha$
   & $-1.6881\,\Nf\alpha$
   &
   & $-1.1114\,\Nf\alpha$
   & $-1.6277\,\Nf\alpha$ \\
$\mu_{4,S}^{1/4}$
   & $0.9650$
   & $-1.4646\,\Nf\alpha$
   & $-1.5177\,\Nf\alpha$
   &
   & $-1.4128\,\Nf\alpha$
   & $-1.4641\,\Nf\alpha$ \\[2pt]
%\hline
$k_{4,S}^{1/4}$
   & $0.7651$
   & $-1.9483\,\Nf\alpha$
   & $-2.5465\,\Nf\alpha$
   &
   & $-1.8927\,\Nf\alpha$
   & $-2.4738\,\Nf\alpha$ \\[2pt]
\hline\hline
\end{tabular}
%} %resizebox
\end{center}
\caption{
   \label{tab:shaperho}
   Expansions involving moments $\langle Z^n \rangle$, reduced moments
   $\mu_{n,S}$, and cumulants $k_{n,S}$ of the charge deposition
   shape function $S_\rho(Z)$,
   for renormalization scale choices (\ref{eq:muE}) and (\ref{eq:muEx}).
   There are no NLO entries for $\langle Z \rangle$ because
   $\langle Z \rangle = 1$ and $\langle Z \rangle_\LO = 1$
   by definition of $Z \equiv z/\langle z\rangle$.
}
\end{table}

The results in table \ref{tab:shaperho} for $\chi\alpha$ of
$\mu_{2,S}^{1/2}=\sigma_S$
are the numbers that were previewed in the last two columns of the
first row of table \ref{tab:chi2}, where we summarized sizes
of overlap corrections to $\sigma/\lstop$.

As in ref.\ \cite{finale2}, we should give a clarification about numerical
accuracy in tables \ref{tab:momentsrho} and \ref{tab:shaperho}
and later tables.  We implicitly pretend that
our fits (\ref{eq:fits}) to the functions $f_\uij(x)$ are exactly
correct.  In reality, though our fit is good, it is only an approximation.
As a check, however, we will now verify
that we reproduce to 3-digit accuracy
the earlier charge deposition result of ref.\ \cite{qedNfstop}
(whose numerics were handled in a completely different way)
for the relative size of overlap effects on $\sigma/\lstop$.

% -------------------------------------------------------------------------

\subsection{Check against earlier result for charge deposition
            \boldmath$\sigma/\lstop$}

Ref.\ \cite{qedNfstop} previously computed $\sigma_S = \sigma/\lstop$
for charge deposition and found that the relative size of the overlap
correction to $\sigma/\lstop$ was
\begin {equation}
   \chi\alpha = -0.870\,\Nf\alpha
   \qquad \mbox{(from ref.\ \cite{qedNfstop})}
\label {eq:oldchi}
\end {equation}
for fixed renormalization scale choice $\mu = (\qhat E_0)^{1/4}$.
This provides a good check of the effects of interpolation error
(\ref{eq:fits}) in our calculations because the two calculations make use
of interpolation in very different ways.%
\footnote{
  In total, our calculations involve
  three-dimensional numerical integration of exact analytic
  formulas presented in
  ref.\ \cite{qedNf}: (i) a time integral ($\Delta t$) described in that
  reference to get rates like the $\Delta\,d\Gamma/d\xe\,d\xE$ of
  our (\ref{eq:1to3rate}), (ii) its integral over $\xE$
  to get the net rates $[d\Gamma/dx]^\NLO_\uee$ in
  (\ref{eq:NLOee}), and (iii) the integral of that net rate over
  $x=\xe$ in the recursion relation (\ref{eq:znMasterRho}) for the moments
  $\langle z^n \rangle$.  In our paper, we have used adaptive integration
  to accurately integrate over $\Delta t$ and $\xe$, then fit the resulting
  function of $x$, and then integrated the fit over $x$.
  In contrast, ref.\ \cite{qedNfstop} used adaptive integration to integrate
  over $\Delta t$, then
  performed a very complicated 2-dimensional interpolation of
  the $(\xe,\xE)$ dependence of $\Delta\,d\Gamma/d\xe\,d\xE$, and
  then integrated that interpolation over $(\xe,\xE)$ to get results.
  There's no reason why the interpolation errors introduced by these
  two different methods would be the same.
}
So we now discuss how to convert our $\mu = (\qhat E)^{1/4}$ result
in table \ref{tab:shaperho} to $\mu = (\qhat E_0)^{1/4}$.
Ref.\ \cite{qedNfstop} devised a trick for including single-log
energy dependence, such as from our (\ref{eq:Lee}) when $\mu$ is fixed,
into the recursion relation for the moments $\langle z^n \rangle$.
We won't review the method here but will merely summarize the result,
which is that the relative size $\chi\alpha$ of overlap corrections
to $\sigma/\lstop$ is changed by%
\footnote{
  This is equivalent to eq.\ (2.26) of ref.\ \cite{qedNfstop}.
}
\begin {multline}
  \chi\alpha[\mu\,{\propto}\,(\qhat E_0)^{1/4}]
  =
  \chi\alpha[\mu\,{\propto}\,(\qhat E)^{1/4}]
  + \frac{\beta_0\alpha}{4} \left(
      \frac{ \Avg[(\sqrt{x}-x)\ln x]^\LO_\uee }
           { \Avg[(1-\sqrt{x})^2]^\LO_\uee }
    - \frac{ \Avg[x\ln x]^\LO_\uee }
           { \Avg[1-x]^\LO_\uee }
    \right)
\\
  \mbox{[for charge deposition $\sigma/\lstop$ only],}
\label {eq:E0conversion}
\end {multline}
with $\beta_0$ given by (\ref{eq:beta0}).
If we take $\chi\alpha[\mu\,{\propto}\,(\qhat E)^{1/4}]$ from
the $\mu_{2,S}^{1/2}=\sigma_S$ row of table \ref{tab:shaperho}, then
(\ref{eq:E0conversion}) gives
$\chi\alpha[\mu\,{\propto}\,(\qhat E_0)^{1/4}] = -0.8706\,\Nf\alpha$,
which agrees with (\ref{eq:oldchi}) to within 1 part in $10^3$.

% =========================================================================

\section{Energy stopping}
\label {sec:energy}

Now we reach the real goal of this paper, which is to similarly analyze
energy deposition.

% -------------------------------------------------------------------------

\subsection {Basic equations}

Like the analysis of energy deposition by purely gluonic showers in
refs.\ \cite{finale,finale2}, the energy deposition equation must track
the energy deposited by all daughters of every splitting.  The difference
with the purely gluonic case is that here the daughters are not identical
particles.  The distribution $\eps_e(E,z)$ of energy deposited by
a shower initiated by an electron will be different than the
$\eps_\gamma(E,z)$ for a shower initiated by a photon.%
\footnote{
  As in refs.\ \cite{finale,finale2}, our $\eps(E,z)$ is normalized so
  that $\int_0^\infty dz \> \eps(E,z) = E$.
  This is different than
  the normalization of appendix A of ref.\ \cite{qedNfstop}, where
  the integral of $\eps$ was normalized to 1.
}
By charge
conjugation invariance, however,
\begin {equation}
  \eps_{\bar e}(E,z) = \eps_e(E,z) .
\end {equation}

The starting point analogous to (\ref{eq:rhoEvolve1}) is now a
system of coupled equations,
\begin {subequations}
\label {eq:epsEvolve1}
\begin {align}
  \frac{\partial \eps_e(E,z)}{\partial z}
  =&~
  - \Gamma_e(E) \, \eps_e(E,z)
  + \int_0^1 dx \> \left[\frac{d\Gamma}{dx}(E,x)\right]^\net_\uee \,
     \eps_e(x E,z)
\nonumber\\ &
  + \int_0^1 dx \> \left[\frac{d\Gamma}{dx}(E,x)\right]^\net_\ueg \,
     \eps_\gamma(x E,z)
  + \int_0^1 dx \> \left[\frac{d\Gamma}{dx}(E,x)\right]^\net_\ueE \,
     \eps_e(x E,z)
\nonumber\\ &
  + \int_0^1 dx \> \left[\frac{d\Gamma}{dx}(E,x)\right]^\net_\ueEbar \,
     \eps_e(x E,z) ,
\\
  \frac{\partial \eps_\gamma(E,z)}{\partial z}
  =&~
  - \Gamma_\gamma(E) \, \eps_\gamma(E,z)
  + \int_0^1 dx \> \left[\frac{d\Gamma}{dx}(E,x)\right]^\net_\ugE \,
     \eps_e(x E,z)
\nonumber\\ &
  + \int_0^1 dx \> \left[\frac{d\Gamma}{dx}(E,x)\right]^\net_\ugEbar \,
     \eps_e(x E,z) .
\label {eq:gammastart}
\end {align}
\end {subequations}
It will be convenient to write the total rates $\Gamma_e$ and $\Gamma_\gamma$
in a particular way.  First, note from
eqs.\ (\ref{eq:netrates}--\ref{eq:Gamma})
that
\begin {align}
  \Gamma_e
  &= \int dx \> \left[ \frac{d\Gamma}{dx} \right]^\net_\uee
   =  \int d\xe \> \left[ \frac{d\Gamma}{d\xe} \right]_{e \to e\gamma}
    + \int d\xe\,d\xE \>
        \left[ \Delta \frac{d\Gamma}{d\xe\,d\xE} \right]_{e \to e\E\Ebar}
\nonumber\\
  &=  \int d\xe \> (\xe{+}x_\gamma)
        \left[ \frac{d\Gamma}{d\xe} \right]_{e \to e\gamma}
    + \int d\xe\,d\xE \> (\xe{+}\xE{+}\xEbar)
        \left[ \Delta \frac{d\Gamma}{d\xe\,d\xE} \right]_{e \to e\E\Ebar}
\end {align}
and so%
\footnote{
  Eqs.\ (\ref{eq:GammaAlt}) are the distinguishable-daughters versions
  of eq.\ (3.2) [with (3.1)]
  of ref.\ \cite{finale2}, which was for $(g{\to}gg)+(g{\to}ggg)$.
}
\begin {subequations}
\label {eq:GammaAlt}
\begin {equation}
  \Gamma_e =
  \int dx \> x
      \biggl(
        \left[ \frac{d\Gamma}{dx} \right]_\uee
        + \left[ \frac{d\Gamma}{dx} \right]_\ueg
        + \left[ \frac{d\Gamma}{dx} \right]_\ueE
        + \left[ \frac{d\Gamma}{dx} \right]_\ueEbar
      \biggr) .
\end {equation}
Similarly, we may rewrite (\ref{eq:Gammag}) as
\begin {equation}
  \Gamma_\gamma =
  \int dx \> x
      \biggl(
        \left[ \frac{d\Gamma}{dx} \right]_\ugE
        + \left[ \frac{d\Gamma}{dx} \right]_\ugEbar
      \biggr) .
\end {equation}
\end {subequations}
Using (\ref{eq:GammaAlt}), now rewrite (\ref{eq:epsEvolve1}) as
\begin {subequations}
\label {eq:epsEvolve}
\begin {align}
  \frac{\partial \eps_e(E,z)}{\partial z}
  =&~
  \int_0^1 dx \> \left[\frac{d\Gamma}{dx}(E,x)\right]^\net_{e\to e^\pm} \,
     \left\{ \eps_e(x E,z) - x \eps_e(E,z) \right\}
\nonumber\\ &
  + \int_0^1 dx \> \left[\frac{d\Gamma}{dx}(E,x)\right]^\net_\ueg \,
     \left\{ \eps_\gamma(x E,z) - x \eps_e(E,z) \right\} \,,
\label {eq:epsEvolvee}
\\
  \frac{\partial \eps_\gamma(E,z)}{\partial z}
  =&~
  \int_0^1 dx \> \left[\frac{d\Gamma}{dx}(E,x)\right]^\net_{\gamma\to e^\pm} \,
     \left\{ \eps_e(x E,z) - x \eps_\gamma(E,z) \right\} \,,
\end {align}
\end {subequations}
where we use the notation $i{\to}e^\pm$ to indicate the sum of net rates
to produce {\it any} flavor of electron or positron from particle $i$:
\begin {subequations}
\begin {align}
   \left[ \frac{d\Gamma}{dx} \right]^\net_{e\to e^\pm}
   &\equiv
   \left[ \frac{d\Gamma}{dx} \right]^\net_\uee
   +
   \left[ \frac{d\Gamma}{dx} \right]^\net_\ueE
   +
   \left[ \frac{d\Gamma}{dx} \right]^\net_\ueEbar \,,
\label {eq:dGneteepm}
\\
   \left[ \frac{d\Gamma}{dx} \right]^\net_{\gamma\to e^\pm}
   &\equiv
   \left[ \frac{d\Gamma}{dx} \right]^\net_\ugE
   +
   \left[ \frac{d\Gamma}{dx} \right]^\net_\ugEbar \,.
\label {eq:dGnetgam}
\end {align}
\end {subequations}
Eqs.\ (\ref{eq:epsEvolve}) are the energy deposition analog of
(\ref{eq:rhoEvolve}).

% -------------------------------------------------------------------------

\subsection {Scaled equations}

As long as we again choose the renormalization scale(s) $\mu$ to scale with
energy as $E^{1/4}$, we may make the same rescaling arguments as in
section \ref{sec:rhoscale}, with one modification.
For large-$\Nf$ charge deposition, we followed a particular electron
through shower development from start to finish.
Since we never needed to follow a positron or photon, the charge $Q$
of the particle followed was always
the charge $Q_0$ of the electron that initiated
the shower, which was reflected in how we rescaled $\rho(E,z)$ in
(\ref{eq:tilde}).
In contrast, in the case of energy deposition, the energy $E$ of an individual
particle in the shower is not the energy $E_0$ of the particle
that initiated the shower.  The appropriate rescaling of $\eps(E,z)$
corresponds to replacing $Q_0$ by the current particle energy $E$ in
(\ref{eq:tilde}),
\begin {equation}
    \left[ \frac{d\Gamma}{dx}(E,x) \right]_\uij^\net
    = E^{-1/2} \left[ \frac{d\tilde\Gamma}{dx}(x) \right]_\uij^\net ,
  \qquad
  z = E^{1/2} \tilde z ,
  \qquad
  \eps(E,z) = E^{1/2} \, \tilde\eps(\tilde z) ,
\label {eq:epstilde}
\end {equation}
so that (as in refs.\ \cite{finale,finale2})
the normalization of $\tilde\eps$ is independent of energy:
\begin {equation}
   \int_0^\infty d\tilde z \> \tilde\eps(\tilde z) = 1 .
\end {equation}
Using (\ref{eq:epstilde}), we may then follow the same steps as
before to obtain the following analog, for
$\eps_i(z) \equiv \eps_i(E_0,z)$, of
(\ref{eq:rhoeq}):%
\footnote{
   Eqs.\ (\ref{eq:epseq}) are analogous to eq.\ (5.15) of ref.\ \cite{finale2},
   which was for purely gluonic showers.
}
\begin {subequations}
\label {eq:epseq}
\begin {align}
  \frac{\partial \eps_e(z)}{\partial z}
  =&~
  \int_0^1 dx \> x \left[\frac{d\Gamma}{dx}(E_0,x)\right]^\net_{e\to e^\pm} \,
     \left\{ x^{-1/2} \eps_e(x^{-1/2} z) - \eps_e(z) \right\}
\nonumber\\ &
  + \int_0^1 dx \> x \left[\frac{d\Gamma}{dx}(E_0,x)\right]^\net_\ueg \,
     \left\{ x^{-1/2} \eps_\gamma(x^{-1/2} z) - \eps_e(z) \right\} \,,
\\
  \frac{\partial \eps_\gamma(z)}{\partial z}
  =&~
  \int_0^1 dx \> x \left[\frac{d\Gamma}{dx}(E_0,x)\right]^\net_{\gamma\to e^\pm} \,
     \left\{ x^{-1/2} \eps_e(x^{-1/2} z) - \eps_\gamma(z) \right\} \,.
\end {align}
\end {subequations}

% -------------------------------------------------------------------------

\subsection {Moments \boldmath$\langle z^n\rangle$}

Similar to section \ref{sec:momentsrho}, we may obtain a relation
between moments
\begin {equation}
  \langle z^n \rangle_{\eps,i}
  \equiv \frac{1}{E_0} \int_0^\infty dz \> z^n \, \eps_i(E_0,z)
\end {equation}
of the energy deposition distributions by
multiplying both sides of
(\ref{eq:epseq}) by $z^n$ and integrating over $z$ to get
\begin {equation}
  -n \langle z^{n-1} \rangle_\eps = -M_{(n)} \, \langle z^n \rangle_\eps ,
\label {eq:matrixversion}
\end {equation}
where
\begin {equation}
  \langle z^n \rangle_\eps \equiv
  \begin{pmatrix}
     \langle z^n \rangle_{\eps,e} \\ \langle z^n \rangle_{\eps,\gamma}
  \end{pmatrix}
\label {eq:znvec}
\end {equation}
and
\begin {align}
  M_{(n)} &\equiv
  \begin{pmatrix}
    M_{(n),ee} & M_{(n),e\gamma} \\ M_{(n),\gamma e} & M_{(n),\gamma\gamma}
  \end{pmatrix}
\nonumber\\
  &=
  \Avg_{\,e\to e^\pm}
    \begin{pmatrix}
       x{-}x^{1+\frac{n}{2}} & 0 \\ 0 & 0
    \end{pmatrix}
  +
  \Avg_\ueg
    \begin{pmatrix}
       x & - x^{1+\frac{n}{2}} \\ 0 & 0
    \end{pmatrix}
  +
  \Avg_{\,\gamma\to e^\pm}
    \begin{pmatrix}
       0 & 0 \\ - x^{1+\frac{n}{2}} & x
    \end{pmatrix}
   .
\label {eq:M}
\end {align}
Inverting (\ref{eq:matrixversion}) gives the recursion relation
\begin {equation}
  \langle z^n \rangle_\eps = n M_{(n)}^{-1} \, \langle z^{n-1} \rangle_\eps \,,
\label {eq:recursionM}
\end {equation}
where $M_{(n)}^{-1}$ is the matrix inverse of $M_{(n)}$.

The recursion relation may be further simplified because of the
large-$\Nf$ limit that we took to simplify our analysis.
The leading-order $e{\to}e\gamma$ rate (\ref{eq:LObremrate}) is
$O(\alpha) = O(\Nf^{-1})$, and the leading-order $\gamma{\to}e\bar e$
rate (\ref{eq:LOpairrate}) is $O(\Nf\alpha) = O(\Nf^0)$.
In both cases, NLO corrections are suppressed by relative factors of
$O(\Nf\alpha) = O(\Nf^0)$ as far as $\Nf$ counting is concerned.%
\footnote{
  The analysis of this paper, as well as refs.\ \cite{qedNf,qedNfstop},
  formally assumes $\Nf^{-1} \ll \Nf\alpha \ll 1$ with
  $\Nf\alpha$ held fixed in the $\Nf{\to}\infty$ limit
}
That means that, in terms of powers of $\Nf$,
\begin {subequations}
\label {eq:Nfhierarchy}
\begin {equation}
   \left[\frac{d\Gamma}{dx}(E,x)\right]^\net_{e\to e^\pm}
   ~\mbox{and}~
   \left[\frac{d\Gamma}{dx}(E,x)\right]^\net_\ueg
   = O(\Nf^{-1})
\end {equation}
\begin {equation}
   \mbox{while}~
   \left[\frac{d\Gamma}{dx}(E,x)\right]^\net_{\gamma\to e^\pm} = O(\Nf^0) ,
\end {equation}
\end {subequations}
and so
\begin {equation}
  \begin{pmatrix}
    M_{(n),ee} & M_{(n),e\gamma} \\ M_{(n),\gamma e} & M_{(n),\gamma\gamma}
  \end{pmatrix}
  =
  \begin{pmatrix}
    O(\Nf^{-1}) & O(\Nf^{-1}) \\ O(\Nf^0) & O(\Nf^0)
  \end{pmatrix} .
\end {equation}
In our $\Nf{\to}\infty$ limit, the inverse of this matrix becomes
\begin {equation}
  M_{(n)}^{-1} \to
  \frac{1}{\det M_{(n)}}
  \begin{pmatrix}
    M_{(n),\gamma\gamma} & 0 \\ -M_{(n),\gamma e} & 0
  \end{pmatrix} .
\end {equation}
The coupled recursion relation (\ref{eq:recursionM}) then reduces
to an uncoupled recursion relation for the moments
$\langle z^n \rangle_{\eps,e}$ of electron-initiated showers,
\begin {subequations}
\label {eq:znRecursion}
\begin {equation}
  \langle z^n \rangle_{\eps,e}
  =
  \frac{n M_{(n),\gamma\gamma}}{\det M_{(n)}}
  \,
  \langle z^{n-1} \rangle_{\eps,e} \,,
\label {eq:zneRecursion}
\end {equation}
and a dependent result for moments of photon-initiated showers,
\begin {equation}
  \langle z^n \rangle_{\eps,\gamma}
  =
  - \frac{M_{(n),\gamma e}}{M_{(n),\gamma\gamma}}
  \,
  \langle z^n \rangle_{\eps,e} \,.
\label {eq:zngRecursion}
\end {equation}
\end {subequations}
Appendix \ref{app:znRecursion} outlines
an alternative way to derive (\ref{eq:znRecursion}) by baking
in the large-$\Nf$ hierarchy (\ref{eq:Nfhierarchy}) much earlier.

% -------------------------------------------------------------------------

\subsection {Numerical Results}

For reference, table \ref{tab:momentseps} shows the expansions of the
raw moments $\langle z^n\rangle_{\eps,i}$ of electron-initiated
and photon-initiated energy deposition.  These moments
depend on the value of $\qhat$.
Similar to the discussion in section \ref{sec:numericsrho}, our interest
is in moments of the corresponding shapes, given in table \ref{tab:shapeeps},
which are insensitive to physics that can be absorbed into $\qhat$.

\begin {table}[t]

\setlength{\tabcolsep}{7pt}
\begin{center}
\begin{tabular}{lccc}
\hline\hline
  \multicolumn{1}{c}{$z^n$}
  & \multicolumn{1}{c}{$\langle z^n\rangle_{\eps,i}^\LO$}
  & \multicolumn{2}{c}{$\delta\langle z^n \rangle_{\eps,i}$}
 \\
\cline{3-4}
  & &
  \multicolumn{1}{c}{$\scriptstyle \mu = (\qhat E)^{1/4}$}
  &
  \multicolumn{1}{c}{$
      \phantom{\strut}
      \mu_{e\to e\gamma} = (\xe \qhat E/x_\gamma)^{1/4}
      \atop
      \mu_{\gamma\to\E\Ebar} = (\xE\xEbar \qhat E)^{1/4}
  $} 
  \\[6pt]
\cline{2-4}
  & \multicolumn{3}{c}{in units of $\ell_0^{\kern1pt n}$} \\
\hline
\multicolumn{4}{l}{\textbf{electron initiated ($i=e$):}}\\
$\langle z\rangle$   & 7.7744 & -39.525\,$\Nf\alpha$ & -39.531\,$\Nf\alpha$ \\
$\langle z^2\rangle$ & 75.639 & -734.74\,$\Nf\alpha$ & -735.02\,$\Nf\alpha$ \\
$\langle z^3\rangle$ & 879.41& -12614\phantom{.}\,$\Nf\alpha$
                                           & -12621\phantom{.}\,$\Nf\alpha$ \\
$\langle z^4\rangle$ & 11854\phantom{.}& -2.2669$\times10^5\,\Nf\alpha$
                                           & -2.2683$\times10^5\,\Nf\alpha$ \\
\hline
\multicolumn{4}{l}{\textbf{photon initiated ($i=\gamma$):}}\\
$\langle z\rangle$   & 6.7877 & -34.536\,$\Nf\alpha$ & -34.479\,$\Nf\alpha$ \\
$\langle z^2\rangle$ & 59.881 & -582.13\,$\Nf\alpha$ & -581.32\,$\Nf\alpha$ \\
$\langle z^3\rangle$ & 644.57 & -9252.5\,$\Nf\alpha$ & -9242.1\,$\Nf\alpha$ \\
$\langle z^4\rangle$ & 8149.4 & -1.5596$\times10^5\,\Nf\alpha$
                                           & -1.5581$\times10^5\,\Nf\alpha$ \\
\hline\hline
\end{tabular}
\end{center}
\caption{
   \label{tab:momentseps}
   Like table \ref{tab:momentsrho} but moments $\langle z^n\rangle$
   for energy deposition
   instead of electron-initiated
   charge deposition.
   The unit $\ell_0$ is defined by (\ref{eq:ell0}).
}
\end{table}

\begin {table}[t]
\begin {center}

\setlength{\tabcolsep}{7pt}
%\resizebox{\textwidth}{!}{%
\begin{tabular}{lcrrcrr}
\hline\hline
  \multicolumn{1}{l}{quantity $Q$}
  & \multicolumn{1}{c}{$Q_{\eps,i}^\LO$}
  & \multicolumn{1}{c}{$\delta Q_{\eps,i}$}
  & \multicolumn{1}{c}{$\chi\alpha$}
  & 
  & \multicolumn{1}{c}{$\delta Q_{\eps,i}$}
  & \multicolumn{1}{c}{$\chi\alpha$}
 \\
\cline{3-4}\cline{6-7}
  &&
  \multicolumn{2}{c}{$\scriptstyle \mu \propto (\qhat E)^{1/4}$}
  &&
  \multicolumn{2}{c}{$
      \phantom{\strut}
      \mu_{e\to e\gamma} \propto (\xe \qhat E/x_\gamma)^{1/4}
      \atop
      \mu_{\gamma\to\E\Ebar} \propto (\xE\xEbar \qhat E)^{1/4}
  $} 
  \\[6pt]
\hline
\multicolumn{4}{l}{\textbf{electron initiated ($i=e$):}}\\
$\langle Z \rangle$
   & $1$ \\
$\langle Z^2 \rangle^{1/2}$
   & $1.1187$
   & $0.2541\,\Nf\alpha$
   & $0.2271\,\Nf\alpha$
   &
   & $0.2529\,\Nf\alpha$
   & $0.2261\,\Nf\alpha$ \\
$\langle Z^3 \rangle^{1/3}$
   & $1.2323$
   & $0.3730\,\Nf\alpha$
   & $0.3027\,\Nf\alpha$
   &
   & $0.3708\,\Nf\alpha$
   & $0.3009\,\Nf\alpha$ \\
$\langle Z^4 \rangle^{1/4}$
   & $1.3421$
   & $0.4067\,\Nf\alpha$
   & $0.3031\,\Nf\alpha$
   &
   & $0.4037\,\Nf\alpha$
   & $0.3008\,\Nf\alpha$ \\[1pt]
\hline
$\mu_{2,S}^{1/2} {=} k_{2,{\rm S}}^{1/2} {=} \sigma_S$
   & $0.5014$
   & $0.5669\,\Nf\alpha$
   & $1.1305\,\Nf\alpha$
   &
   & $0.5642\,\Nf\alpha$
   & $1.1252\,\Nf\alpha$ \\
$\mu_{3,S}^{1/3} {=} k_{3,{\rm S}}^{1/3}$
   & $0.4893$
   & $-0.0086\,\Nf\alpha$
   & $-0.0175\,\Nf\alpha$
   &
   & $-0.0114\,\Nf\alpha$
   & $-0.0233\,\Nf\alpha$ \\
$\mu_{4,S}^{1/4}$
   & $0.7191$
   & $0.3677\,\Nf\alpha$
   & $0.5112\,\Nf\alpha$
   &
   & $0.3639\,\Nf\alpha$
   & $0.5061\,\Nf\alpha$ \\[2pt]
%\hline
$k_{4,S}^{1/4}$
   & $0.5281$
   & $-0.5273\,\Nf\alpha$
   & $-0.9984\,\Nf\alpha$
   &
   & $-0.5298\,\Nf\alpha$
   & $-1.0032\,\Nf\alpha$ \\[2pt]
\hline\hline
\multicolumn{4}{l}{\textbf{photon initiated ($i=\gamma$):}}\\
$\langle Z \rangle$
   & $1$ \\
$\langle Z^2 \rangle^{1/2}$
   & $1.1400$
   & $0.2592\,\Nf\alpha$
   & $0.2274\,\Nf\alpha$
   &
   & $0.2572\,\Nf\alpha$
   & $0.2256\,\Nf\alpha$ \\
$\langle Z^3 \rangle^{1/3}$
   & $1.2726$
   & $0.3859\,\Nf\alpha$
   & $0.3032\,\Nf\alpha$
   &
   & $0.3819\,\Nf\alpha$
   & $0.3001\,\Nf\alpha$ \\
$\langle Z^4 \rangle^{1/4}$
   & $1.3998$
   & $0.4253\,\Nf\alpha$
   & $0.3038\,\Nf\alpha$
   &
   & $0.4195\,\Nf\alpha$
   & $0.2997\,\Nf\alpha$ \\[1pt]
\hline
$\mu_{2,S}^{1/2} {=} k_{2,{\rm S}}^{1/2} {=} \sigma_S$
   & $0.5475$
   & $0.5399\,\Nf\alpha$
   & $0.9861\,\Nf\alpha$
   &
   & $0.5357\,\Nf\alpha$
   & $0.9785\,\Nf\alpha$ \\
$\mu_{3,S}^{1/3} {=} k_{3,{\rm S}}^{1/3}$
   & $0.5451$
   & $0.1139\,\Nf\alpha$
   & $0.2089\,\Nf\alpha$
   &
   & $0.1077\,\Nf\alpha$
   & $0.1975\,\Nf\alpha$ \\
$\mu_{4,S}^{1/4}$
   & $0.7918$
   & $0.3592\,\Nf\alpha$
   & $0.4537\,\Nf\alpha$
   &
   & $0.3521\,\Nf\alpha$
   & $0.4447\,\Nf\alpha$ \\[2pt]
%\hline
$k_{4,S}^{1/4}$
   & $0.5928$
   & $-0.4196\,\Nf\alpha$
   & $-0.7078\,\Nf\alpha$
   &
   & $-0.4268\,\Nf\alpha$
   & $-0.7199\,\Nf\alpha$ \\[2pt]
\hline\hline
\end{tabular}
%} %resizebox
\end{center}
\caption{
   \label{tab:shapeeps}
   Expansions of moments $\langle Z^n \rangle$, reduced moments $\mu_{n,S}$,
   and cumulants $k_{n,S}$ of the energy deposition shape functions
   $S_{\eps,e}(Z)$ and $S_{\eps,\gamma}(Z)$ for electron-initiated and
   photon-initiated showers, respectively. 
   Like table \ref{tab:shaperho} but for energy deposition
   instead of electron-initiated
   charge deposition.
}
\end{table}

The results in table \ref{tab:shapeeps} for $\chi\alpha$ of
$\mu_{2,S}^{1/2}=\sigma_S$
are the numbers previewed in the last
two rows of table \ref{tab:chi2}, where we summarized the sizes
of overlap corrections to $\sigma/\lstop$.

Before moving on, we mention that it is possible to write
{\it exact}\/ analytic results for all of our {\it leading}-order
(LO) entries appearing in tables \ref{tab:momentsrho}--\ref{tab:shapeeps}.
As an example, the entry for the width $\sigma_S$ of the LO shape
$S_{\eps,e}^\LO$ of electron-initiated energy deposition
in table \ref{tab:shapeeps} is the numerical value of
\begin {equation}
   (\sigma_S)_{\eps,e}^\LO =
   \left( \frac{\sigma}{\lstop} \right)_{\eps,e}^\LO =
%   \left(
%     \tfrac{11}{16} - \tfrac{118}{105\pi} - \tfrac{2432}{1225\pi^2}
%   \right)^{-1}
   \sqrt{
     \tfrac{4096}{421}
       \left(
         \tfrac{11}{16} - \tfrac{118}{105\pi} - \tfrac{2432}{1225\pi^2}
       \right)
     - 1
  } .
\label {eq:exactsigmaS}
\end {equation}
See appendix \ref{app:LO}.  However, since we must do numerics anyway for
the NLO results, we find it simpler to just implement the recursion relation
(\ref{eq:recursionM}) numerically in the LO case as well.

% =========================================================================

%%\section{Cross-ratios}

% =========================================================================

\section{Conclusion}
\label {sec:conclusion}

As previewed in the introduction, our immediate conclusion is simply that
there is no important
qualitative difference between the size of $\qhat$-insensitive
overlap effects in
charge vs.\ energy deposition for large-$\Nf$ QED.
{\it Both} are
very large compared to the size of such corrections to large-$\Nc$
gluonic showers \cite{finale,finale2},
for comparable values of $\Nf\alpha$ and $\Nc\alphas$.

This leaves open the question of whether there is something
special or accidental about the relatively small result for
gluonic showers.  One possibility that our current analysis can help
with is to determine whether there is an important qualitative difference
due to the inescapable presence of fermions in a QED shower calculation
vs.\ the lack of fermions in previous gluon shower calculations.
The framework developed in this paper should be able to shed light
by adapting our analysis here to large-$\Nf$ QCD.
(As a first
step for judging whether including quarks in QCD medium-induced showers
will have a large qualitative impact on overlap effects,
analyzing large-$\Nf$ QCD will involve substantially less additional
work than the case of moderate $\Nf$.)
We leave such a large-$\Nf$ analysis
of QCD overlap effects for later work \cite{qcdNf}.

% =========================================================================
% =========================================================================

\begin{acknowledgments}

The work of Arnold and Elgedawy (while at U. Virginia) was supported,
in part, by the U.S. Department of Energy under Grant No.~DE-SC0007974.
Elgedawy's work at South China Normal University was supported by
Guangdong Major Project of Basic and Applied Basic Research No.\ 2020B0301030008
and by the
National Natural Science Foundation of China under Grant No.\ 12035007.
Arnold is grateful for the hospitality of the EIC Theory Institute
at Brookhaven National Lab for one of the months during which he was
working on this paper.

\end{acknowledgments}

% =========================================================================
\appendix
% =========================================================================

\section {Equality of \boldmath$e{\to}\E$ and $e{\to}\Ebar$ net rates}
\label {app:EEbar}

In this appendix, we provide a sketch of why the differential rate
$\Delta\,d\Gamma/d\xe\,d\xE$ for the
overlapping process $e\to e\E\Ebar$ is
symmetric under $\xE \leftrightarrow \xEbar$ (i.e.\ $\xE \to 1{-}\xe{-}\xE$),
which in turn is responsible for the equality of the net rates
$[d\Gamma/dx]_\ueE$ and $[d\Gamma/dx]_\ueEbar$ in large-$\Nf$ QED.
We will have to discuss some details of the machinery of the
calculation, but we will try to keep the discussion as high level as possible.
(Alternatively, one may just accept the equality as a property
of the final formulas that has been verified numerically and be
done with it.)

\begin {figure}[t]
\begin {center}
  \includegraphics[scale=0.5]{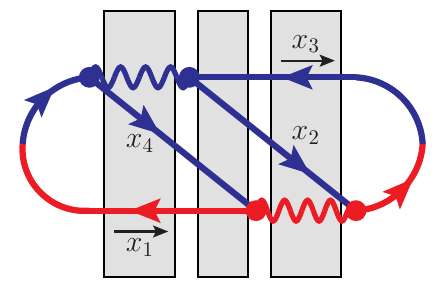}
  \caption{
     \label{fig:seqoverlap}
     The first interference
     diagram of fig.\ \ref{fig:diagsNf1}, here showing how we label
     various energy fractions as $(x_1,x_2,x_3,x_4)$ following
     the convention of refs.\ \cite{qedNf}.
     The shaded areas are discussed in the text.
  }
\end {center}
\end {figure}

To make the discussion concrete, we will focus on the particular example
of the rate diagram shown in fig.\ \ref{fig:seqoverlap}.
The diagram also shows the notation $(x_1,x_2,x_3,x_4) = (-1,\xE,\xEbar,\xe)$
used in ref.\ \cite{qedNf} for the energy fractions of various particles.
In this language, the symmetry we want to explain corresponds to
switching the values of $x_2$ and $x_3$.

In Zakharov's formalism \cite{Zakharov1,Zakharov2}, this time-ordered
diagram is evaluated by treating the gray regions as problems of
3- or 4-particle evolution in two-dimensional
quantum mechanics, with an imaginary-valued
potential that accounts for the effect of medium-averaged interactions of
the high-energy particles with the medium.  Those evolutions are then
tied together with quantum field theory calculations of the vertices in
fig.\ \ref{fig:seqoverlap}.
In the case of 4-particle evolution
(the middle gray area), the corresponding Hamiltonian is \cite{qedNf}%
\footnote{
  Our (\ref{eq:Hfour}) is the 4-particle generalization of the 3-particle
  version reviewed (using our notation) in eq.\ (2.11) of ref.\ \cite{2brem}.
  For our (\ref{eq:Vfour}), see eqs.\ (E.11--12) of ref.\ \cite{qedNf}.
}
\begin {subequations}
\begin {equation}
   \frac{(\p_1^\perp)^2}{2x_1 E} + \frac{(\p_2^\perp)^2}{2x_2 E}
                 + \frac{(\p_3^\perp)^2}{2x_3 E} + \frac{(\p_4^\perp)^2}{2x_4 E}
   + V(\b_1,\b_2,\b_3,\b_4)
\label {eq:Hfour}
\end {equation}
with potential
\begin {equation}
   V(\b_1,\b_2,\b_3,\b_4)
   = -\frac{i\qhat}{4} \left(
      b_{12}^2 + b_{23}^2 + b_{34}^2 + b_{41}^2 - b_{13}^2 - b_{24}^2
   \right)
   = -\frac{i\qhat}{4} (\b_1 - \b_2 + \b_3 - \b_4)^2 ,
\label {eq:Vfour}
\end {equation}
\end {subequations}
where $(\b_1,\b_2,\b_3,\b_4)$ are the transverse positions of the
four particles, $(\p_1^\perp,\p_2^\perp,\p_3^\perp,\p_4^\perp)$
are the corresponding transverse
momenta, and $\b_{ij} \equiv \b_i - \b_j$.  This Hamiltonian is not
symmetric under exchanging the values of $x_2$ and $x_3$.
%(One might think to additionally swap $\b_2$ and $\b_3$, but judging the
%effects of that would require detailed analysis of how
%particles 2 and 3 in the middle gray area of fig.\ \ref{fig:seqoverlap}
%connect to the rest of the diagram.  We'll be able to avoid most of that.)
However, the rate we are interested in calculating is invariant under
(i) translations in the transverse plane and (ii) rotations that (at most)
change the directions of the $z$ axis by a parametrically small amount
(preserving the high-energy approximation that $p^\perp \ll p^z$).
This is enough symmetry to allow reduction of the 4-particle problem
to an effective 2-particle problem with Hamiltonian%
\footnote{
  The kinetic terms in (\ref{eq:Htwo}) correspond to those of the
  Lagrangian of eqs.\ (5.15--17) of ref.\ \cite{2brem}, but with
  particle labels $(1,2,3,4)$ there permuted to $(2,3,4,1)$ here, and
  $\dot C_{ij}$'s converted to $P_{ij}$'s.
}
\begin {equation}
   \frac{P_{41}^2}{2 x_1 x_4(x_1{+}x_4) E}
   + \frac{P_{23}^2}{2 x_2 x_3(x_2{+}x_3) E}
   - \frac{i\qhat}{4} (x_1{+}x_4)^2 \left( \C_{41} - \C_{23} \right)^2
\label {eq:Htwo}
\end {equation}
with degrees of freedom $\C_{41}$ and $\C_{23}$ defined by
$\C_{ij} \equiv (\b_i-\b_j)/(x_i+x_j)$ with
conjugate momenta $\P_{ij} \equiv x_j\p_i^\perp - x_i\p_j^\perp$.
The reduced Hamiltonian (\ref{eq:Htwo}) {\it is} symmetric under
exchanging the values of $x_2$ and $x_3$, which turns out to be
the most critical reason that the final result will have that
property.%
\footnote{
   The full set of $\C_{ij}$ that one can write
   the 4-body potential in terms
   of have only two linearly independent degrees of freedom.
   (See, for example, eqs.\ (5.14) of ref.\ \cite{2brem}, which are
   also valid for any cyclic permutation of the indices.)
   We chose to write (\ref{eq:Htwo}) in terms of $\C_{41}$ and
   $\C_{23}$.  If we had chosen to use $\C_{34}$ and $\C_{12}$ instead,
   the Hamiltonian would {\it not}\/ have looked $x_2{\leftrightarrow}x_3$
   symmetric.  We are relying here on the fact that $(\C_{41},\C_{23})$
   turns out to be the natural choice of basis for this diagram, as we will
   see in (\ref{eq:seqform}), because of the way the lines are connected
   in the diagram.
}
The other aspects of the problem can be mostly understood in
terms of charge conjugation symmetry.

To see this, we should sketch a little more
how the elements of the calculation fit together.
The contribution from fig.\ \ref{fig:seqoverlap} to the rate
$\Delta\,d\Gamma/d\xe\,d\xE$ was originally derived from a formula
of the form \cite{qedNf,seq}%
\footnote{
  Specifically, see eq.\ (E.1) of ref.\ \cite{seq}, with the QED modifications
  described in appendices E.1 and E.2 of ref.\ \cite{qedNf}.
}
\begin {align}
   \mbox{(splitting}&~\mbox{amplitude factors from the vertices)}
   \times
   \frac{\Nf\alpha^2}{(x_1{+}x_4)^2}
   \int_{\rm times} \int_{\B',\B''}
\nonumber\\ &\times
   \grad_{\B'''}
   \langle\B''',t'''|\B'',t''\rangle
   \Bigr|_{\B'''=0}
\nonumber\\ &\times
   \grad_{\C''_{41}}
   \grad_{\C'_{23}}
   \langle\C''_{41},\C''_{23},t''|\C'_{41},\C'_{23},t'\rangle
   \Bigr|_{\C_{41}''=0=\C_{23}'; ~ \C_{23}''=\B''; ~ \C_{41}'=\B'}
\nonumber\\ &\times
   \grad_{\B}
   \langle\B',t'|\B,t\rangle
   \Bigr|_{\B=0} .
\label {eq:seqform}
\end {align}
Above, the $\langle\C''_{41},\C''_{23},t''|\C'_{41},\C'_{23},t'\rangle$
is the propagator associated with the Hamiltonian (\ref{eq:Htwo}).
The other two $\langle \cdots \rangle$'s are similar factors for
the initial and final 3-particle evolution (leftmost and rightmost
gray areas in fig.\ \ref{fig:seqoverlap}), where the same translation
and rotational symmetries
mentioned before have been used to reduce the problem from
3-particle quantum mechanics to effectively 1-particle quantum mechanics,
with a variable we conventionally call $\B$ in the 3-particle case.
Various separations $\B$ or $\C_{ij}$ are set to zero at vertices
where two lines come together (and so their separation vanishes).
The derivatives $\grad$ are position-space versions of
transverse momentum factors associated with splitting vertices.

With (\ref{eq:seqform}) in hand, we sketch the other reasons for
the $x_2{\leftrightarrow}x_3$ symmetry.  (i) The expression only cares about
the 4-particle propagator in terms of the variables $\C_{41}$ and $\C_{23}$,
which is the choice of variables for which we noted (\ref{eq:Htwo}) was
$x_2{\leftrightarrow}x_3$ symmetric.  (ii) The initial 3-particle
evolution in fig.\ \ref{fig:seqoverlap} (leftmost gray area)
is independent of the values of $x_2$ and $x_3$.
(iii) The final 3-particle evolution (of $\E$, $\Ebar$, and a
conjugate-amplitude photon) is symmetric in $x_2{\leftrightarrow}x_3$
by charge conjugation invariance, and (iv) the vertices associated
with $\gamma\to\E\Ebar$ at the start and end of the final 3-particle
evolution come with amplitudes that are also symmetric by charge
conjugation invariance.

Finally, the same style of argument can be used for the other
large-$\Nf$ diagrams, including fig.\ \ref{fig:diagsNf2}, by
labeling the particles in any 4-particle evolution
the same way [$(x_1,x_2,x_3,x_4) = (-1,\xE,\xEbar,\xe)$] and then
describing the 4-particle evolution with (\ref{eq:Htwo}).

% =========================================================================

\section {DGLAP origin of logarithms \boldmath$\ln x$ and $\ln(1{-}x)$ in
          eqs.\ (\ref{eq:fits}) for $f_{i\to j}(x)$}
\label {app:logs}

To understand the coefficients of the logarithms in (\ref{eq:fits})
for our NLO fit functions $f_\uij(x)$, we start by looking at the
piece $f_{e\to e}^{\rm real}(x)$ (\ref{eq:feeReal}) of $f_\uee(x)$
that corresponds to real, double
splitting $e \to e\E\Ebar$.

% ---------------------------------------------------------------------------

\subsection{\boldmath$e \to e\E\Ebar$}

\begin {figure}[t]
\begin {center}
  \resizebox{.7\textwidth}{!}{
  \begin{picture}(355,110)(0,0)
    \put(0,5){\includegraphics[scale=1.0]{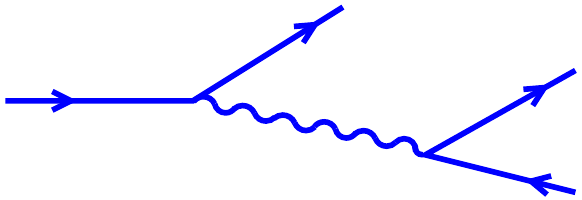}}
    \put(40,40){\large$E$}
    \put(170,100){\large$\xe$}
    \put(105,33){\rotatebox{-12}{\large$E_\gamma = (1{-}\xe)E$}}
    \put(284,73){\large$x_\E=\yfrakE (1{-}\xe)$}
    \put(284,5){\large$x_\Ebar=(1{-}\yfrakE)(1{-}\xe)$}
  %\put(0,0){.}
  %\put(0,110){.}
  %\put(355,0){.}
  %\put(340,110){.}
  \end{picture}
  } % resizebox
  \caption{
     \label {fig:eEE}
     $e \to e\E\Ebar$ process
  }
\end {center}
\end {figure}

In this appendix, it will be convenient to use the notation shown in
fig.\ \ref{fig:eEE} for the energies of particles in the
$e{\to}e\gamma \to e\E\Ebar$ double splitting process.
In particular, we introduce
$E_\gamma$ as the energy of the intermediate photon and
\begin {equation}
   \yfrakE \equiv \frac{\xE}{x_\gamma} = \frac{\xE}{1{-}\xe}
\label {eq:yfrakEdef}
\end {equation}
as the energy fraction of the pair electron $\E$ relative to its
immediate parent, the photon.

For what follows, it will be useful to have
at hand parametric formulas for the formation times for
individual, single splitting processes $e\to e\gamma$ and
$\gamma\to\E\Ebar$:
\begin {subequations}
\label {eq:tformsReal}
\begin {align}
  t_{\rm form}^{e\to e\gamma} &\sim \sqrt{ \frac{\xe E}{x_\gamma\qhat} }
  = \sqrt{ \frac{\xe E}{(1{-}\xe)\qhat} } \,,
\\
  t_{\rm form}^{\gamma\to \E\Ebar} &\sim
    \sqrt{ \frac{\yfrakE\yfrakEbar E_\gamma}{\qhat} }
  = \sqrt{ \frac{\yfrakE(1{-}\yfrakE)(1{-}\xe)E}{\qhat} } \,,
\end {align}
\end {subequations}
coming from $t_{\rm form} \sim |\Omega|^{-1}$ and (\ref{eq:Omegas})
[with $(\xE,\xEbar,E)$ in (\ref{eq:OmegagEEbar})
replaced here by $(\yfrakE,\yfrakEbar,E_\gamma)$].

% ..........................................................................

\subsubsection{\boldmath$\xe\to 1$}

In the limit $\xe \to 1$, (\ref{eq:tformsReal}) gives
\begin {equation}
   t_\form^{\gamma\to\E\Ebar} \ll t_\form^{e \to e\gamma} .
\label {eq:hierarchy1}
\end {equation}
The splitting with the smallest formation time is the one that most
disrupts the LPM effect.  Following the argument of appendix B.1 of
ref.\ \cite{seq}, we will treat the splitting with the parametrically
smaller formation time (in this case $\gamma \to \E\Ebar$)
as the ``underlying'' medium-induced splitting process, and
we will treat the other splitting (here the earlier $e \to e\gamma$)
as a vacuum-like DGLAP correction to that underlying process.
Specifically, following through to eqs.\ (B.6) and (B.7)
of ref.\ \cite{seq}, we approximate
\begin {equation}
   \left[ \Delta\, \frac{d\Gamma}{d\xe\,d\yfrakE} \right]_{e\to e\E\Ebar}
   \approx \frac{\alpha}{2\pi} \, P_{e\to e}(\xe) \,
     \ln\biggl( \frac{t_{\rm form}^{e\to e\gamma}}{t_{\rm form}^{\gamma\to\E\Ebar}} \biggr)
     \left[ \frac{d\Gamma}{d\yfrakE} \right]^\LO_{\gamma\to\E\Ebar} ,
\label {eq:DGLAPlog1}
\end {equation}
where $\approx$ indicates a \textit{leading-log} approximation.
Using (\ref{eq:tformsReal}), remembering that we are taking $\xe{\to}1$, and
then integrating both sides with respect to $\yfrakE$ gives
\begin {equation}
   \int_0^1 d\yfrakE \>
   \left[ \Delta\, \frac{d\Gamma}{d\xe\,d\yfrakE} \right]_{e\to e\E\Ebar}
   \approx \frac{\alpha}{2\pi} \, P_{e\to e}(\xe) \,
     \ln\Bigl( \frac{1}{1{-}\xe} \Bigr)
     \, \Gamma^\LO_{\gamma\to\E\Ebar} ,
\label{eq:reallog1a}
\end {equation}
where, using (\ref{eq:LOpairrate}) and (\ref{eq:DGLAP_Ps}),
\begin {equation}
  \Gamma^\LO_{\gamma\to\E\Ebar} =
  \frac{\Nf \alpha}{2\pi} \sqrt{\frac{\qhat}{E_\gamma}}
  \int_0^1 d\yfrakE \>
     \frac{ \yfrakE^2+(1{-}\yfrakE)^2 }{ \sqrt{\yfrakE(1{-}\yfrakE)} }
  =
  \frac{3\Nf \alpha}{8} \sqrt{\frac{\qhat}{E_\gamma}} .
\end {equation}
Using this in (\ref{eq:reallog1a}), and taking the $\xe{\to}1$ limit of
$P_{e\to e}(x)$ from (\ref{eq:DGLAP_Ps}),
\begin {equation}
   \int_0^1 d\yfrakE \>
   \left[ \Delta \, \frac{d\Gamma}{d\xe\,d\yfrakE} \right]_{e\to e\E\Ebar}
   \approx
     -\frac{3\Nf\alpha^2}{8\pi} \,
     \frac{\ln(1{-}\xe)}{(1{-}\xe)^{3/2}}
     \sqrt{ \frac{\qhat}{E} } \,.
\label{eq:reallog1b}
\end {equation}
The left-hand side above is equivalent (after changing variables)
to the left-hand side of (\ref{eq:feeRealDef}), and so
$f_{e\to e}^{\rm real}(\xe)$ is given by dividing (\ref{eq:reallog1b}) by
the $R_{\uee}(\xe)$ of (\ref{eq:Ree}),
\begin {equation}
  f_{e\to e}^{\rm real}(\xe) \approx -\tfrac34 \ln(1{-}\xe) .
\label {eq:feeLog1}
\end {equation}
This is the $\ln(1{-}x)$ term that we used in our fit (\ref{eq:feeReal}).

\begin {figure}[t]
\begin {center}
  \includegraphics[scale=0.3]{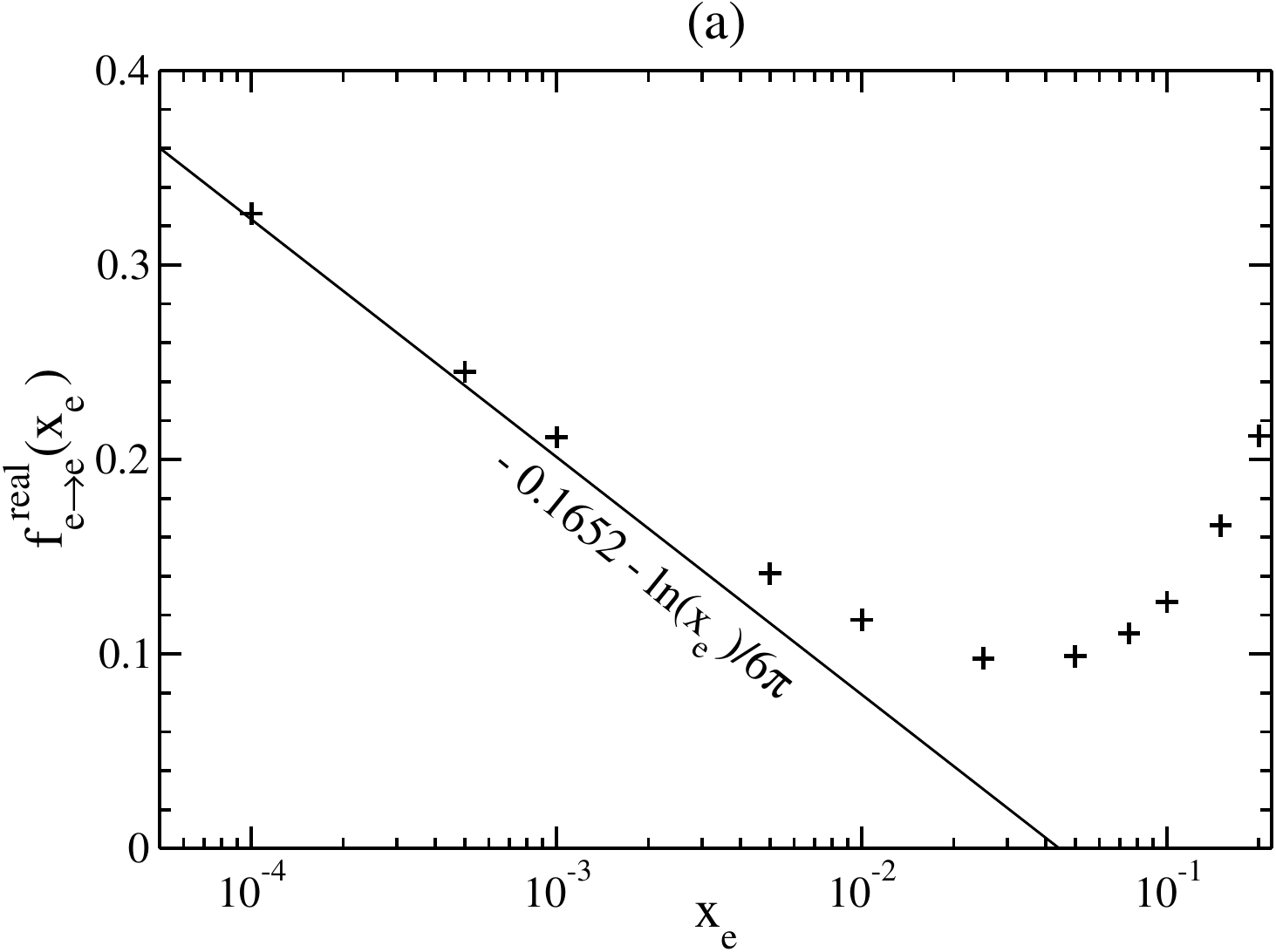}
  ~~
  \includegraphics[scale=0.3]{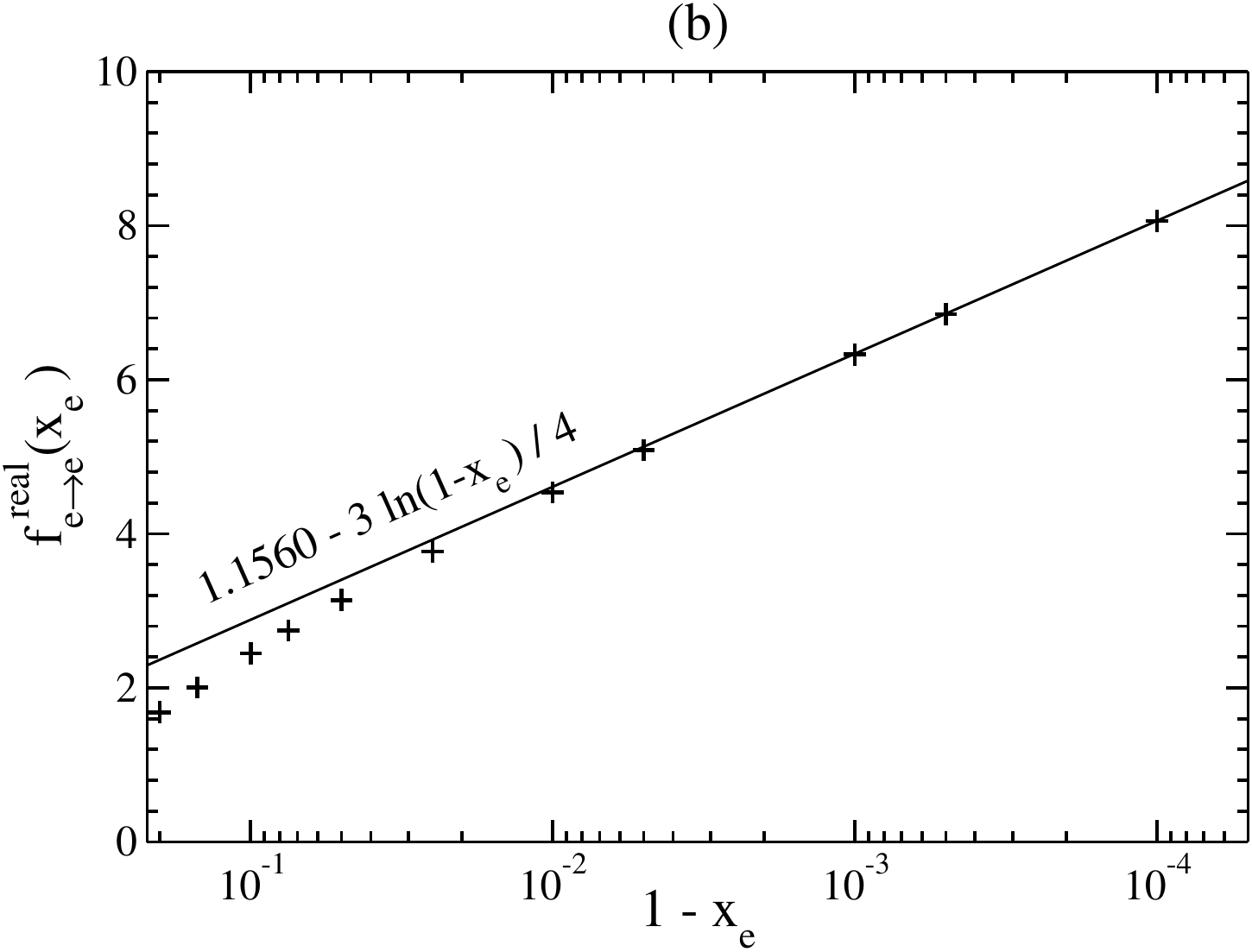}
  \caption{
     \label{fig:REALlog}
     Log-linear plots of numerical results for $f_{e\to e}^{\rm real}(\xe)$
     vs.\ (a) $\xe$ and (b) $1{-}\xe$.
     The numerical data points ($+$)
     are taken from table \ref{tab:dGnet} using
     (\ref{eq:fdecompose}).
     The points are compared to lines whose slopes on these plots
     correspond to our leading-log analytic results
     (a) $-\frac{1}{6\pi} \ln x$ and (b) $-\frac34 \ln(1{-}\xe)$ for the
     limits $\xe\to 0$ and $\xe \to 1$ respectively.
     Note that the horizontal axis in both plots is oriented so that
     $\xe\to 0$ is toward the left and $\xe\to 1$ is toward the right. 
  }
\end {center}
\end {figure}

Readers not convinced by our fast and loose argument for
(\ref{eq:DGLAPlog1}) may be reassured to see numerical evidence that
the coefficient on the logarithm in (\ref{eq:feeLog1}) has been
correctly identified.
Fig.\ \ref{fig:REALlog}b shows a log-linear plot of the numerical data
points for $f_{e\to e}^{\rm real}(x)$ vs.\ $1{-}x$, arranged
so that $x\simeq1$ corresponds to the right-hand side of the plot.
The coefficient of $\ln(1{-}x)$ is determined by the limit of the slope
of this plot as $x \to 1$.  To check the slope, we have also drawn a line
\begin {equation}
   -\tfrac34 \ln(1{-}\xe) + {\rm constant}
\end {equation}
corresponding to the second term of our fit (\ref{eq:feeReal})
plus the (constant) $\xe\to 1$ limit of all the other terms.
The slopes of that line and of the numerical data indeed match
extremely well as $x \to 1$.

% ...........................................................................

\subsubsection{\boldmath$\xe\to 0$}

In the limit $\xe \to 0$ with $\yfrakE$ held fixed,
(\ref{eq:tformsReal}) gives
\begin {equation}
   t_\form^{\gamma\to\E\Ebar} \gg t_\form^{e \to e\gamma} ,
\label {eq:thierarchy0}
\end {equation}
and so, by the previous logic, we now consider $e\to e\gamma$ to be
the underlying medium-induced splitting process, and $\gamma\to\E\Ebar$
is a vacuum-like fragmentation correction which can also be expressed
as a DGLAP-like correction.
The analog of (\ref{eq:DGLAPlog1}) is
\begin {equation}
   \left[ \Delta\, \frac{d\Gamma}{d\xe\,d\yfrakE} \right]_{e\to e\E\Ebar}
   \approx
     \frac{\alpha}{2\pi} \, \Nf P_{\gamma\to e}(\yfrakE) \,
     \ln\biggl( \frac{t_{\rm form}^{\gamma\to\E\Ebar}}{t_{\rm form}^{e\to e\gamma}}
        \biggr)
     \left[ \frac{d\Gamma}{d\xe} \right]^\LO_{e\to\gamma e} ,
\label {eq:DGLAPlog0}
\end {equation}
from which
\begin {multline}
   \int_0^1 d\yfrakE \>
   \left[ \Delta\, \frac{d\Gamma}{d\xe\,d\yfrakE} \right]_{e\to e\E\Ebar}
   \approx
     \frac{\Nf\alpha}{2\pi} \, \ln(\xe^{-1/2})
     \left[ \frac{d\Gamma}{d\xe} \right]^\LO_{e\to\gamma e}
     \int_0^1 d\yfrakE \> P_{\gamma\to e}(\yfrakE)
\\
   \simeq
     -\frac{\Nf\alpha^2}{12\pi^2} \,
     \frac{\ln \xe}{\xe^{1/2}}
     \sqrt{ \frac{\qhat}{E} } \,.
\label{eq:reallog0a}
\end {multline}
Dividing by the $R_{\uee}(\xe)$ of (\ref{eq:Ree}) gives
\begin {equation}
  f_{e\to e}^{\rm real}(\xe) \approx -\tfrac{1}{6\pi} \ln\xe ,
\label {eq:feeLog0}
\end {equation}
which is the $\ln x$ term that we used in our fit (\ref{eq:feeReal}).

The \textit{a posteriori} check that it was okay to take
$\xe{\to}0$ in (\ref{eq:thierarchy0}) while ignoring the possibility
that $\yfrake$ or $1{-}\yfrake$ was also very small is that the
$\yfrakE$ integral in (\ref{eq:reallog0a}) was convergent.

As a numerical check of (\ref{eq:feeLog0}), see fig.\ \ref{fig:REALlog}a.
The line is
\begin {equation}
   -\tfrac{1}{6\pi} \ln\xe + {\rm constant} ,
\end {equation}
corresponding to the first term of (\ref{eq:feeReal})
plus the (constant) $\xe{\to}0$ limit of all the other terms.

% ...........................................................................

\subsubsection{\boldmath$\xE\to 0$}

We now study the behavior of the NLO net rate $[d\Gamma/d\xE]_\ueE^\NLO$ of
(\ref{eq:NLOeE}) as
$\xE \to 0$.  Only the $e\to e\E\Ebar$ process contributes to that net rate.
Consider now the
limit $\xE\to0$ of the formation times (\ref{eq:tformsReal}).
Throughout this discussion, we will take that limit while assuming that
both $\yfrakE \equiv \xE/(1{-}\xe)$ and $1{-}\yfrakE$
remain $O(1)$, and so $1{-}\xe = O(\xE)$
and $\xe \to 1$.
The {\it a posteriori} justification is that
we will encounter no divergence when we
later integrate over $\yfrakE$ in this approximation.

The limit $\xE\to0$ ($\xe\to1$) of (\ref{eq:tformsReal}) therefore
gives us the same hierarchy of formation times as
(\ref{eq:hierarchy1}), and so we have the same
leading-log approximation (\ref{eq:DGLAPlog1}) for
$[\Delta\,d\Gamma/d\xe\,d\yfrakE]_{e\to e\E\Ebar}$.
The $\xe{\to}1$ limit of (\ref{eq:DGLAPlog1})
is the unintegrated version of (\ref{eq:reallog1a}):
\begin {equation}
   \left[ \Delta\, \frac{d\Gamma}{d\xe\,d\yfrakE} \right]_{e\to e\E\Ebar}
   \approx
     \frac{\alpha}{2\pi} \, P_{e\to e}(\xe) \,
     \ln\Bigl( \frac{1}{1{-}\xe} \Bigr)
     \left[ \frac{d\Gamma}{d\yfrakE} \right]^\LO_{\gamma\to\E\Ebar}
        \!\!(\yfrakE,E_\gamma)
    ,
\end {equation}
where we find it useful to now explicitly list the momentum fraction and
energy arguments appropriate for the LO $\gamma\to\E\Ebar$ rate.
The difference here will be in how we then integrate
to get $[d\Gamma/d\xE]^\NLO_\ueE$ instead of the
behavior of the real double splitting contribution
(\ref{eq:feeRealDef}) to the net rate $[d\Gamma/d\xe]^\NLO_\uee$.

\begin {figure}[t]
\begin {center}
  \includegraphics[scale=0.3]{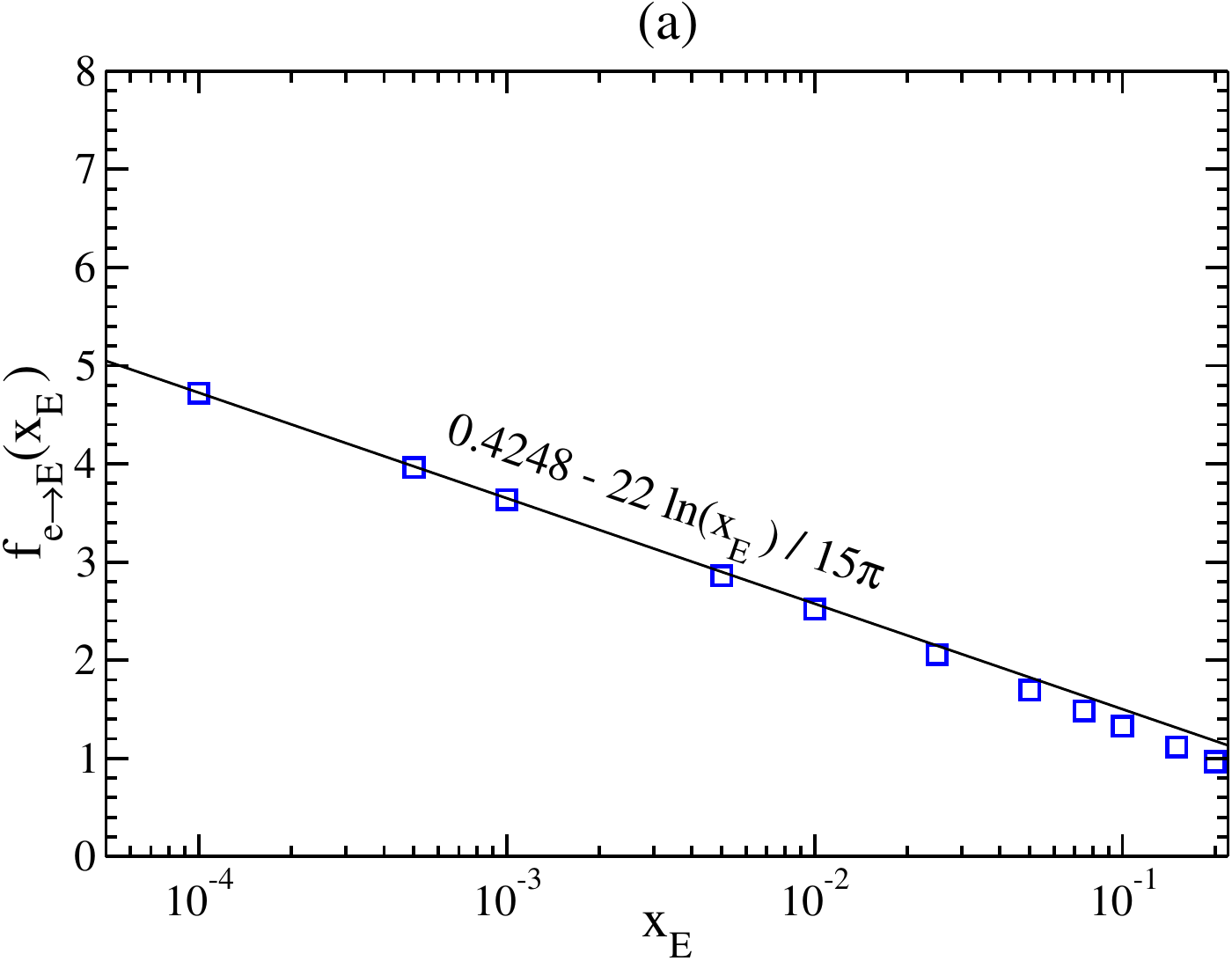}
  ~~
  \includegraphics[scale=0.3]{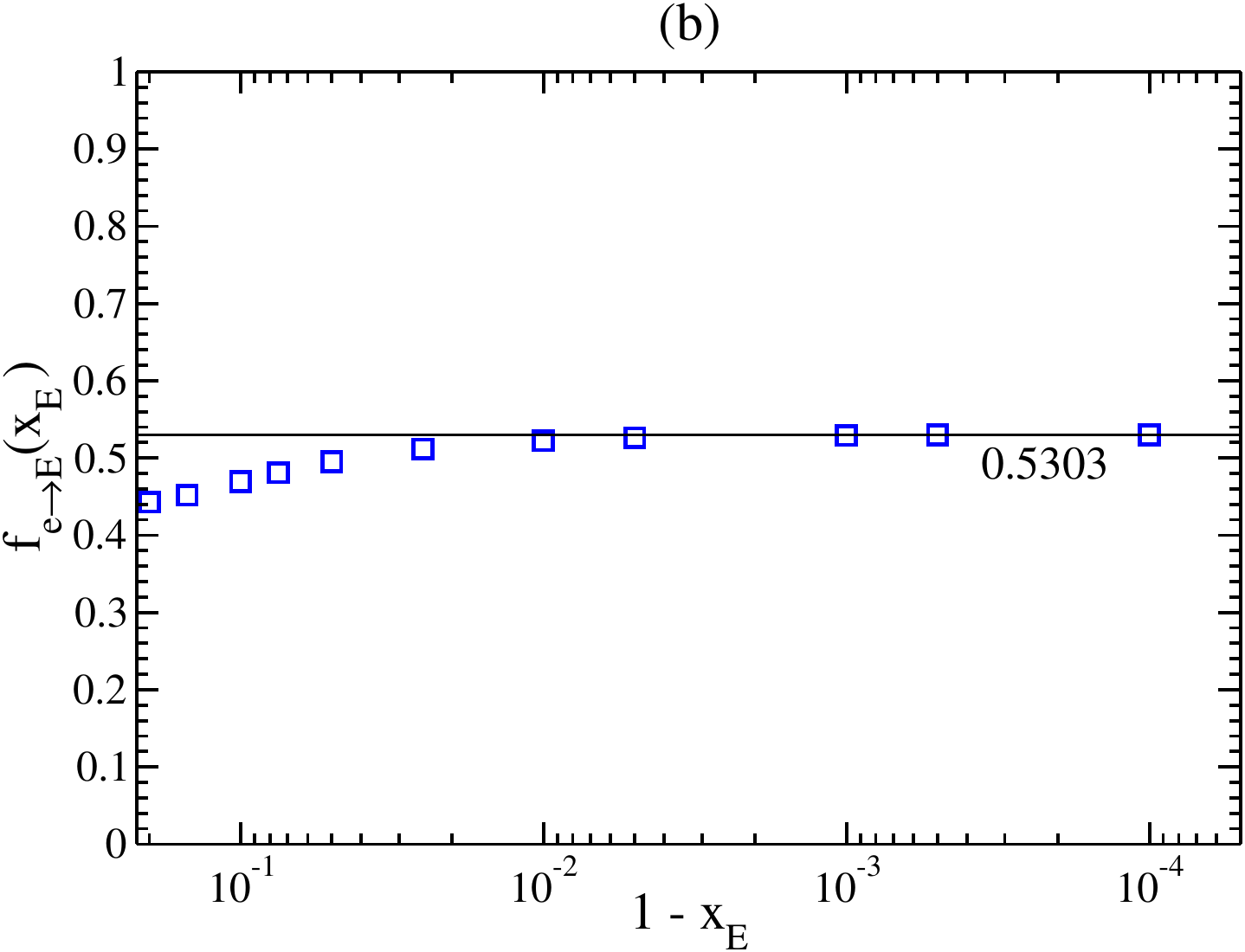}
  \caption{
     \label{fig:feElog}
     Like fig.\ \ref{fig:REALlog} but here
     log-linear plots of numerical results for $f_\ueE(\xE)$
     vs.\ (a) $\xE$ and (b) $1{-}\xE$.
  }
\end {center}
\end {figure}

Use the definition (\ref{eq:yfrakEdef}) of $\yfrakE$ to change variables from
$(\xe,\yfrakE)$ to $(\xE,\yfrakE)$,
\begin {subequations}
\label {eq:xElog1a}
\begin {equation}
   \left[ \Delta\, \frac{d\Gamma}{d\xE\,d\yfrakE} \right]_{e\to e\E\Ebar}
   \approx
     \frac{\alpha}{2\pi\yfrakE} \,
     P_{e\to e}\bigl(1 - \tfrac{\xE}{\yfrakE}\bigr) \,
     \ln\Bigl( \frac{\yfrakE}{\xE} \Bigr)
     \left[ \frac{d\Gamma}{d\yfrakE} \right]^\LO_{\gamma\to\E\Ebar}
          \!\!\bigl(\yfrakE, \tfrac{\xE}{\yfrakE} E\bigr) .
\end {equation}
With these variables, the NLO net rate for $e\to\E$ is given by the integral
\begin {equation}
  \left[ \frac{d\Gamma}{d\xE} \right]^\NLO_\ueE =
  \int_{\xE}^1 d\yfrakE \>
  \left[ \Delta\, \frac{d\Gamma}{d\xE\,d\yfrakE} \right]_{e\to e\E\Ebar} .
\end {equation}
\end {subequations}
Taking $\xE \to 0$ while treating $\yfrakE$ as $O(1)$,
(\ref{eq:xElog1a}) gives
\begin {equation}
  \left[ \frac{d\Gamma}{d\xE} \right]^\NLO_\ueE \approx
    - \frac{\Nf\alpha^2}{2\pi^2} \,
    \frac{\ln\xE}{\xE^{3/2}}
    \sqrt{\frac{\qhat}{E}}
    \int_0^1 d\yfrakE \> \frac{ \yfrakE^2+(1{-}\yfrakE)^2 }{ \sqrt{1-\yfrakE} }
  \simeq
    -\frac{11\,\Nf\alpha^2}{15\pi^2} \,
    \frac{\ln \xE}{\xE^{3/2}}
    \sqrt{ \frac{\qhat}{E} } \,.
\end {equation}
Finally, dividing by the corresponding $R_\ueE(\xE)$ of (\ref{eq:ReE}),
\begin {equation}
  f_\ueE(\xE) \approx -\tfrac{22}{15\pi} \ln\xE
\label {eq:feELog0}
\end {equation}
as in (\ref{eq:feE}).
A numerical check of this result is shown in fig.\ \ref{fig:feElog}a.

% ...........................................................................

\subsubsection{\boldmath$\xE\to 1$}

The limit $\xE\to1$ requires both $\xe\to0$ and $\xEbar\to0$.
If we assume $\xe\sim\xEbar$, then (\ref{eq:tformsReal})
does not give any hierarchy of formation times:
\begin {equation}
   t_\form^{\gamma\to\E\Ebar} \sim t_\form^{e \to e\gamma} .
\end {equation}
No hierarchy suggests no logarithmic enhancement, and so we might expect
$f_\ueE(\xE)$ to have no $\ln(1{-}\xE)$ behavior as $\xE\to 1$:
\begin {equation}
  f_\ueE(\xE) \to {\rm constant} .
\end {equation}
We verify this expectation in fig.\ \ref{fig:feElog}b, where we compare
numerical results to the constant taken from the $\xE{\to}1$ limit
of our fit (\ref{eq:feE}).

% ---------------------------------------------------------------------------

\subsection {Virtual diagrams}

We do not have a method for deducing {\it ab initio} the logarithmic
behavior of $f_{\uij}$'s that contain virtual diagrams.
Here, we will rely on numerics to identify which limits lack
{\it any} logarithmic terms $\ln x$ or $\ln(1{-}x)$.
We will then be able to
combine those cases with the previous $e\to e\E\Ebar$ results for
$f^{\rm real}_{e\to e}(\xe)$ to
determine the remaining logarithms in (\ref{eq:fits}) and
(\ref{eq:feeVirt}).

\begin {figure}[t]
\begin {center}
  \includegraphics[scale=0.3]{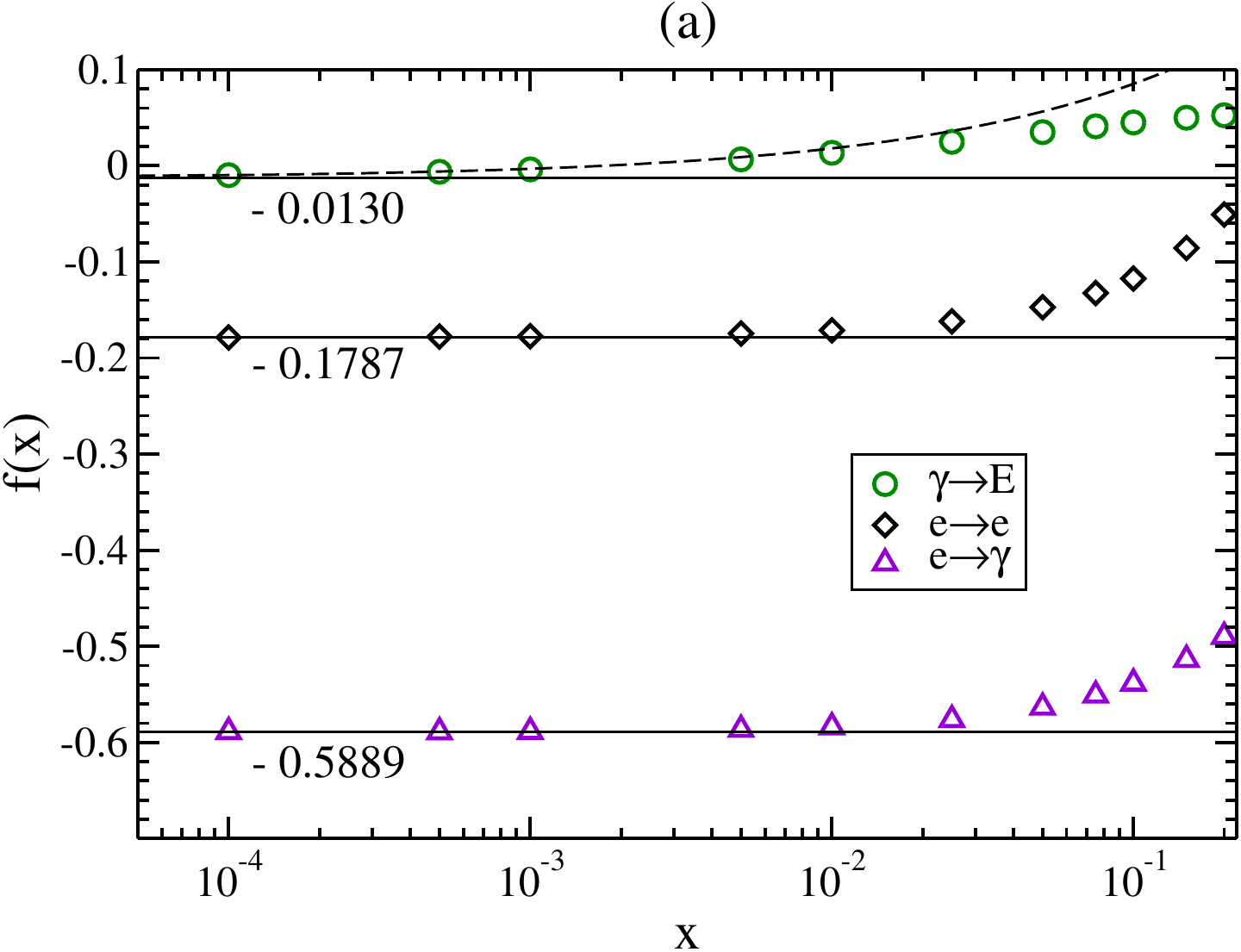}
  ~~
  \includegraphics[scale=0.3]{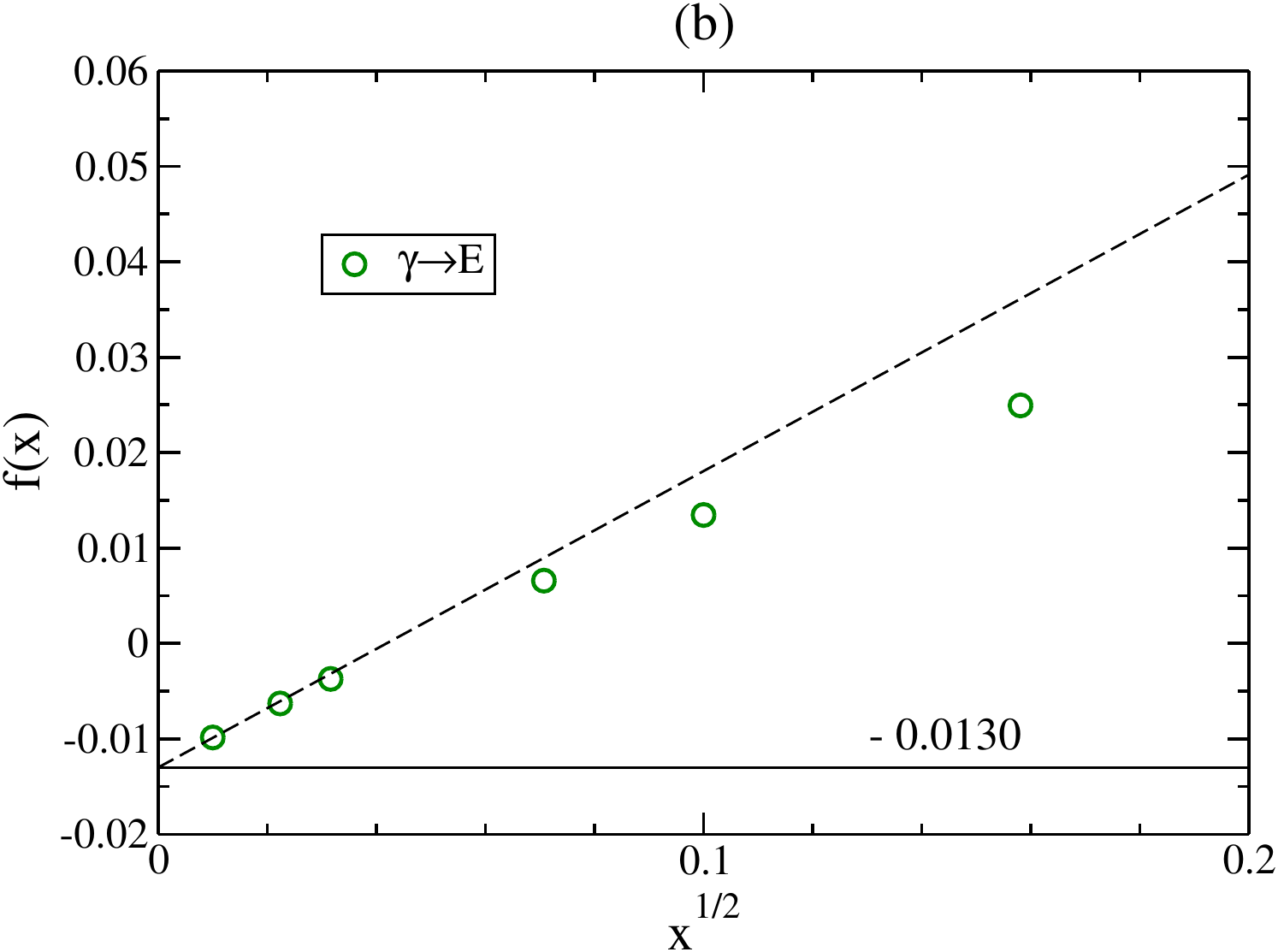}
  \caption{
     \label{fig:fotherLog}
     (a) Like fig.\ \ref{fig:REALlog}a but here for
     log-linear plots of numerical results for $f_\uee(x)$, $f_\ueg(x)$,
     and $f_\ugE(x)$ vs.\ $x$.
     The dashed curve is $-0.0130 + 0.3106 \sqrt{x}$ as in (\ref{eq:fgEsqrt}).
     (b) A linear plot of $f_\ugE(x)$ vs. $\sqrt{x}$.
  }
\end {center}
\end {figure}

Fig.\ \ref{fig:fotherLog}a shows the $x{\to}0$ behavior of our numerical
results for $f_\uee(x)$, $f_\ueg(x)$, and $f_\ugE(x)$, and the horizontal
lines show the constants given by the $x{\to}0$ limit
of the corresponding fit functions in (\ref{eq:fits}).
It is clear from the plot that the numerical data points for
$f_\uee(x)$ and $f_\ugE(x)$ approach a constant as $x \to 0$, and there
is no sign of any $\ln x$ behavior.
By (\ref{eq:fdecompose}), this also means that $f_{e\to e}^{\rm virt}(x)$
has no $\ln(1{-}x)$ behavior, as in (\ref{eq:feeVirt}).

The case of $f_\ugE(x)$ is less clear; visually, it is hard to
determine from fig.\ \ref{fig:fotherLog}a
whether the slope of the numerical data is definitely
approaching zero as $x \to 0$.
However, as noted in ref.\ \cite{finale2},%
\footnote{
   Specifically, see the brief discussion following eq.\ (3.19b) of
   ref.\ \cite{finale2}.
}
expansions can be in
$\sqrt{x}$ rather than $x$, as was reflected in the form of our
fit functions (\ref{eq:fits}).
For (\ref{eq:fgE}), the first two terms in the expansion of the fit
for small $x$ are
\begin {equation}
   f_\ugE(x_\gamma) \simeq -0.0130 + 0.3106 \sqrt{x_\gamma} ,
\label {eq:fgEsqrt}
\end {equation}
represented by the dashed curve in fig.\ \ref{fig:fotherLog}.
The
slow convergence of the $\ugE$ data points to a constant in
fig.\ \ref{fig:fotherLog}a is consistent with
the presence of a $\sqrt{x}$ correction, which is made clearer by
fig.\ \ref{fig:fotherLog}b, where $f_\ugE(x)$ is plotted vs.\ $\sqrt{x}$.
There is no evidence of $\ln x$ behavior.
Since $f_\ugE(x)$ is symmetric under $x\to 1{-}x$ by charge conjugation,
this also means that there is no $\ln(1{-}x)$ behavior.

We now have enough information to reconstruct other limits.
From (\ref{eq:fdecompose0}) and (\ref{eq:fdecompose}), we have
\begin {equation}
   f_\uee(\xe) = f_\ueg(1{-}\xe) + f_{e\to e}^{\rm real}(\xe) .
\end {equation}
Combined with our numerical result that $f_\ueg(x_\gamma)$ approaches a constant
as $x_\gamma\to0$ and that $f_{e\to e}^{\rm real}(\xe)$ behaves like
(\ref{eq:feeLog1}) as $\xe\to 1$, we get
\begin {equation}
  f_\uee(\xe) \approx f_{e\to e}^{\rm real}(\xe) \approx -\tfrac34 \ln(1{-}\xe)
  ~~\mbox{for}~~ \xe \to 1 ,
\end {equation}
which is what we used in the fit (\ref{eq:fee}).

Similarly, eq.\ (\ref{eq:fdecompose0}) plus our
numerical result that $f_\uee(\xe)$
approaches a constant as $\xe \to 0$ gives
$f_{e\to e}^{\rm virt}(\xe) \approx -f_{e\to e}^{\rm real}(\xe)$ for $\xe\to 0$,
and then (\ref{eq:feeLog0}) gives
\begin {equation}
  f_{e\to e}^{\rm virt}(\xe)
  \approx \tfrac{1}{6\pi} \ln\xe
  ~~\mbox{for}~~ \xe \to 0
\end {equation}
as in (\ref{eq:feeVirt}).

% =========================================================================

\section {Parametric estimate of \boldmath$(p_2{+}p_3)^\mu (p_2{+}p_3)_\mu$}
\label {app:pCOM}

In (\ref{eq:muconceptAlt}), we asserted that parametrically
$(p_2{+}p_3)^\mu (p_2{+}p_3)_\mu \sim E/t_{\rm form}$ for a leading-order
BDMPS-Z single splitting where the two daughters have momenta $p_2$ and
$p_3$.  A mnemonic for remembering this formula is to remember that,
in the case of a single virtual particle, its virtuality is
$P^\mu P_\mu \simeq 2 E \Delta E$
when the particle is off-shell in energy by $\Delta E \ll E$.
If we uncritically use that same formula in our case and interpret
$E$ as the energy of the parent and $\Delta E$ as the typical off-shellness
of the splitting process during the formation time, then we can use
the uncertainty relation $\Delta E \sim 1/\Delta t$ to guess
(\ref{eq:muconceptAlt}).

In this appendix, we want to be more concrete by keeping all of the
argument specific to the case of BDMPS-Z splitting.  The splitting
process is high energy and nearly collinear in the frame we usually
work in, the rest frame of the medium.  So we can approximate the
4-momenta of each on-shell daughter as
\begin {equation}
  (p_i^0,\p_i^\perp,p_i^z)
  \simeq
  \left( x_i E + \frac{(p^\perp_i)^2}{2 x_i E} \,,\, \p^\perp_i \,,\,
         x_i E \right)
  ,
\end {equation}
where for this purpose we define $x_i$ as the $p^z$ fraction relative
to the parent,
and we use $({+}{-}{-}{-})$ metric convention.
Then, in the high-energy approximation,
\begin {subequations}
\label {eq:pCOM1}
\begin {equation}
  (p_2{+}p_3)^\mu (p_2{+}p_3)_\mu = 2p_2^\mu p_{3\mu}
  \simeq
  \frac{ (x_3 \p^\perp_2 - x_2 \p^\perp_3 )^2 }{ x_2 x_3 } .
\end {equation}
The combination
\begin {equation}
   \P^\perp \equiv x_3 \p^\perp_2 - x_2 \p^\perp_3
\end {equation}
\end {subequations}
is invariant under
rotations that preserve the high-energy approximation
$p_i^\perp \ll p_i^z$.
In the $\qhat$ approximation, solving the single splitting BDMPS-Z
problem involves solving a two-dimensional non-Hermitian
harmonic oscillator problem with Hamiltonian%
\footnote{
  For a review in the notation of this paper, 
  see section 2 of ref.\ \cite{2brem}.
}
\begin {equation}
  {\cal H} = \frac{(P^\perp)^2}{2 M} + \frac12 M \Omega^2 B^2 ,
\label {eq:calH}
\end {equation}
where $\B \equiv \b_2-\b_3$ is the separation of the two daughters
in the transverse plane and is conjugate to $\P^\perp$;
$\Omega$ above is given by (\ref{eq:Omegas}); and $M \equiv x_2 x_3 E$.
In the form introduced by Zakharov \cite{Zakharov1,Zakharov2},
the
leading-order splitting rate (in an infinite medium)
is then given in terms of the
propagator $\langle\B,t|\B',t'\rangle$
of the above harmonic oscillator by
\begin {equation}
   \frac{d\Gamma}{dx}
   =
   \frac{\alpha P(x)}{M^2}
   \Re \int_0^\infty d(\Delta t) \>
   \grad_{\B} \cdot \grad_{\B'}
   \langle \B,\Delta t | \B',0 \rangle
   \Bigr|_{\B = \B' = 0} ,
\label {eq:drate123}
\end {equation}
where $P(x)$ is the relevant DGLAP vacuum splitting function, $\Delta t$ is
the duration of the splitting process, and the $\Delta t$ integral
is dominated by $\Delta t \sim t_{\rm form} \sim 1/|\Omega|$.
During the formation time, the typical size of ${\cal H}$ is parametrically
$1/t_{\rm form} \sim |{\cal H}| \sim (P^\perp)^2/2M$, and so
$(P^\perp)^2 \sim M/t_{\rm form} = x_2 x_3 E/t_{\rm form}$.  Using (\ref{eq:pCOM1})
then gives the promised parametric estimate (\ref{eq:muconceptAlt}).

% =========================================================================

\section {Another path to the large-\boldmath$\Nf$ recursion relations for
          $\langle z^n \rangle_{\eps,i}$}
\label {app:znRecursion}

In this appendix, we discuss another way that one can arrive at
the large-$\Nf$ recursion relations (\ref{eq:znRecursion}), by
taking the large-$\Nf$ limit at the {\it beginning} of the derivation.

% ---------------------------------------------------------------------------

\subsection{\boldmath$e^\pm$ evolution}

\begin {figure}[t]
\begin {center}
  \includegraphics[scale=0.5]{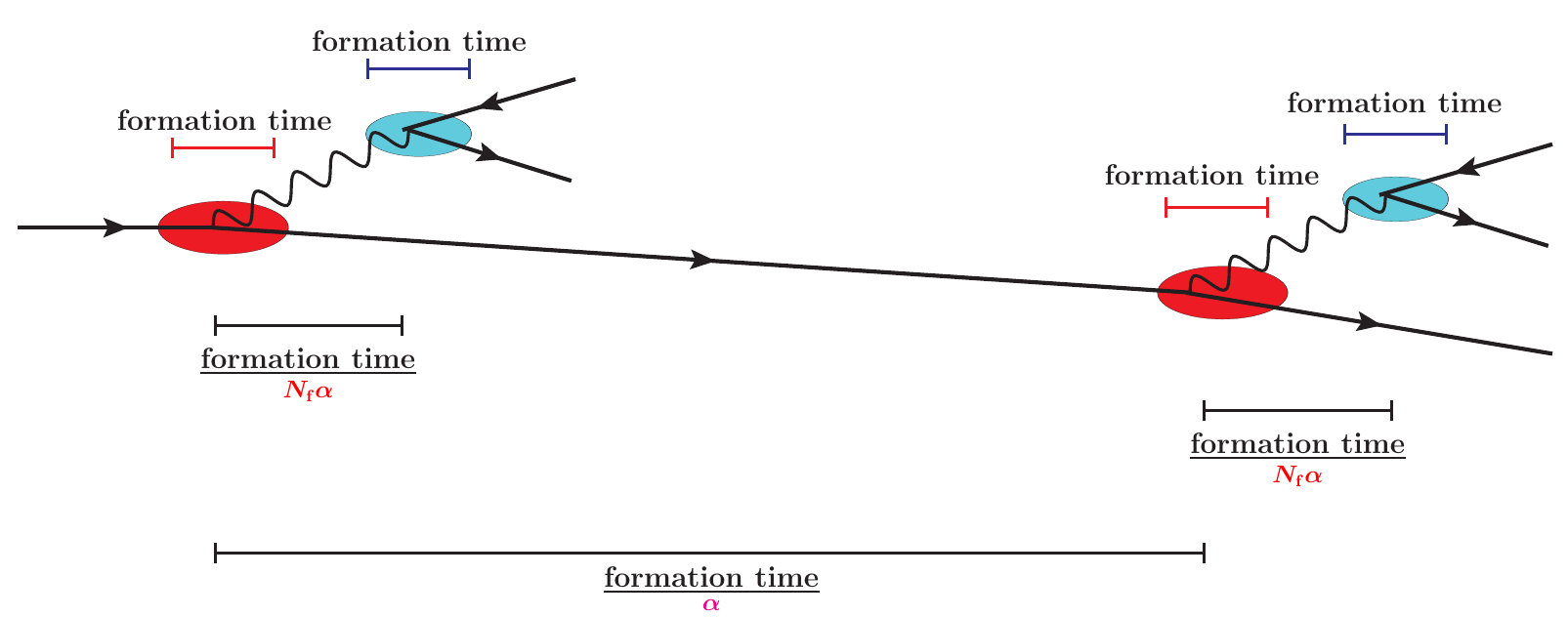}
  \caption{
     \label{fig:typicalNf}
     Hierarchy of scales in large-$\Nf$ QED \cite{qedNfstop}.
  }
\end {center}
\end {figure}

Because the $e\to e\gamma$ rate is suppressed by a factor of $\Nf^{-1}$
compared to the $\gamma\to\E\Ebar$ rate, it is the
$e\to e\gamma$ rate that will be the bottleneck to shower development
and so will parametrically determine the stopping length (and other
moments) for charge and energy deposition.
This hierarchy of scales is depicted in fig.\ \ref{fig:typicalNf}.
In the large-$\Nf$ limit, the lifetime of photons in the in-medium
shower is negligible compared to the duration of the shower, and so
we may always treat the combination $e \to e\gamma \to e\E\Ebar$ of
$e \to e\gamma$ and $\gamma\to\E\Ebar$ splittings as effectively
instantaneous, even for the case of {\it non}-overlapping splittings.
When overlap effects are ignored, the combined rate for such
a sequential splitting would be (i) the {\it rate} for the initial
$e\to e\gamma$ splitting multiplied by (ii) the {\it probability distribution}
for energy fractions of the subsequent, inevitable $\gamma\to\E\Ebar$ splitting
a moment later:
\begin {subequations}
\label {eq:dGindep}
\begin {equation}
   \left[\frac{d\Gamma}{dx_e\,d\yfrakE}(E)\right]^\indep_{e\to e\E\Ebar}
   \equiv
   \left[\frac{d\Gamma}{dx_e}(E)\right]_{e\to e\gamma}
   \!\times
   \frac{1}{\Gamma_\gamma\bigl((1{-}x_e)E\bigr)} \,
   \left[\frac{d\Gamma}{d\yfrakE}\bigl((1{-}x_e)E\bigr)\right]_{\gamma\to\E\Ebar}
   ,
\label {eq:eEEnormal}
\end {equation}
where $\yfrakE$ is the energy fraction (\ref{eq:yfrakEdef}) of the
final pair electron ($\E$) relative to its
immediate parent $\gamma$.
Note that
\begin {equation}
   \left[\frac{d\Gamma}{dx_e\,d\xE}(E)\right]^\indep_{e\to e\E\Ebar}
   =
   \frac{1}{(1{-}x_e)}
   \left[\frac{d\Gamma}{dx_e\,d\yfrakE}(E)\right]^\indep_{e\to e\E\Ebar}
   .
\end {equation}
\end {subequations}
The superscript ``$\indep$'' in (\ref{eq:eEEnormal}) means
``independent'' and indicates that
the possibility that $e\to e\gamma$ and $\gamma\to\E\Ebar$ overlap each
other has been ignored.  However, we \textit{do} include
virtual NLO corrections to each individual splitting.
That is, the single-splitting
rates appearing on the right-hand side of (\ref{eq:eEEnormal}) are each the sum
of LO+NLO single-splitting rates as in (\ref{eq:rateeeg}) and
(\ref{eq:rategEE}), and similarly for the total single-splitting rate
$\Gamma_\gamma$ in the denominator.

\begin {figure}[t]
\begin {center}
  \includegraphics[scale=0.5]{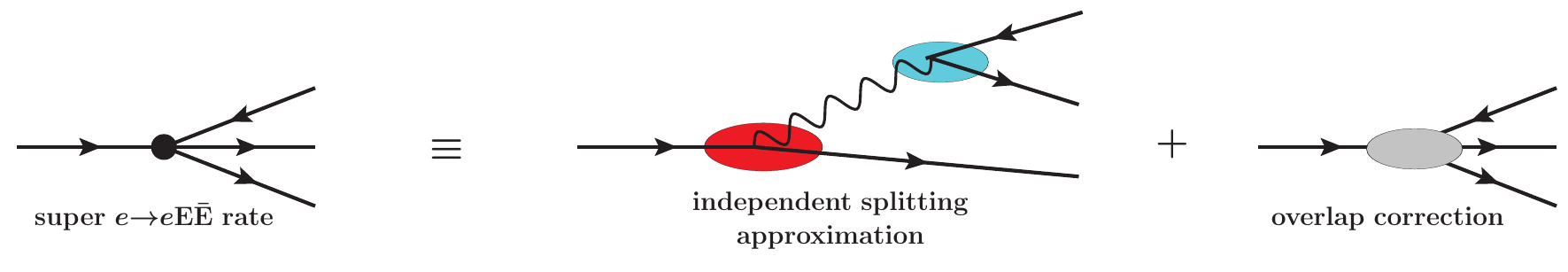}
  \caption{
     \label{fig:superrate}
     Definition (\ref{eq:super}) of the ``super'' $e{\to}e\E\Ebar$ rate,
     where all intermediate photons have been integrated out in
     $e\to e\gamma \to e\E\Ebar$, regardless of whether or not the two
     splittings $e\to e\gamma$ and $\gamma \to \E\Ebar$ overlap.
     Above, each picture of a diagram is meant to schematically
     represent its corresponding contribution to {\it rates}
     (not amplitudes), and it
     is rate contributions that are added.
  }
\end {center}
\end {figure}

We can now undo the approximation that
$e\to e\gamma$ and $\gamma\to\E\Ebar$ do not overlap
by adding in the known overlap
correction $[\Delta d\Gamma/d\xe\,d\xE]_{e\to e\E\Ebar}$.
This will allow us to describe shower development in the
large-$\Nf$ limit using just one rate,
which we will call the {\it super}-effective $1{\to}3$
rate:
\begin {equation}
   \left[\frac{d\Gamma}{d\xe\,d\xE}\right]^\super_{e\to e\E\Ebar}
   \equiv
   \left[\frac{d\Gamma}{d\xe\,d\xE}\right]^\indep_{e\to e\E\Ebar}
   +
   \left[\frac{\Delta\,d\Gamma}{d\xe\,d\xE}\right]_{e\to e\E\Ebar} \,,
\label{eq:super}
\end {equation}
depicted pictorially in fig.\ \ref{fig:superrate}.
The analog of the net $e{\to}e^\pm$ rate given by (\ref{eq:dGneteepm}) is%
\begin {equation}
  \left[ \frac{d\Gamma}{dx} \right]^\snet_{e\to e^\pm}
  \equiv
  \int_0^{1-x} dy \>
  \biggl\{
    \frac{d\Gamma^\super_{\eeEEsub}}{d\xe\,d\xE}(x,y)
    + \frac{d\Gamma^\super_{\eeEEsub}}{d\xe\,d\xE}(y,x)
%\\
    + \frac{d\Gamma^\super_{\eeEEsub}}{d\xe\,d\xE}(y,1{-}x{-}y)
  \biggr\}
  .
\label{eq:snet}
\end {equation}

The now purely-$e^\pm$ energy deposition equation is
\begin {equation}
  \frac{\partial \eps_e(E,z)}{\partial z} =
  \int_0^1 dx \>
       \left[ \frac{d\Gamma}{dx}(E,x) \right]^\snet_{e\to e^\pm}
       \left\{ \eps_e(x E, z) - x \eps_e(E, z) \right\} ,
\end {equation}
which is the analog here of (\ref{eq:epsEvolvee}).
Assuming $E^{-1/2}$ scaling of rates, this gives
\begin {equation}
  \frac{\partial \eps_e(z)}{\partial z} =
  \int_0^1 dx \> x
       \left[ \frac{d\Gamma}{dx}(E_0,x) \right]^\snet_{e\to e^\pm}
       \left\{ x^{-1/2} \eps_e(x^{-1/2} z) - \eps_e(z) \right\} .
\end {equation}
Taking moments gives the recursion relation
\begin {equation}
  -n \langle z^{n-1} \rangle_{\eps,e} =
  \int_0^1 dx \> x
       \left[ \frac{d\Gamma}{dx}(E_0,x) \right]^\snet_{e\to e^\pm}
       \left\{ x^{n/2} \langle z^n \rangle_{\eps,e}
               - \langle z^n \rangle_{\eps,e} \right\}
\end {equation}
and so
\begin {subequations}
\label{eq:zneRecursion3}
\begin {equation}
  \langle z^n \rangle_{\eps,e} = n \mu^{-1}_{(n)} \langle z^{n-1} \rangle_{\eps,e}
\end {equation}
with the number
\begin {equation}
  \mu_{(n)} \equiv
  \int_0^1 dx \> x
       \left[ \frac{d\Gamma}{dx}(E_0,x) \right]^{\snet}_{e\to e^\pm}
       \{ 1 - x^{n/2} \}
  = \Avg^{\super,\net}_{e\to e^\pm}[ x(1-x^{n/2}) ] .
\end {equation}
\end {subequations}
This must be equivalent to
the recursion relation (\ref{eq:zneRecursion}) derived for
$\langle z^n\rangle_{\eps,e}$ in the main text, which means
that
\begin {equation}
  \Avg^{\super,\net}_{e\to e^\pm}[ x(1-x^{n/2}) ] =
  \frac{\det M_{(n)}}{M_{(n),\gamma\gamma}} \,.
\label {eq:Miracle}
\end {equation}
Eq.\ (\ref{eq:Miracle}) is not obvious from the formulas for its
left-hand and right-hand sides, and so we
will show how to verify it in section \ref{app:Miracle} below.

% ---------------------------------------------------------------------------

\subsection{Photon initiated showers}

For a photon initiated shower, we may again make use of the fact that,
in the large-$\Nf$ limit, the photon pair-produces instantly relative to
the duration of the shower.  That means that rather than an evolution
equation (\ref{eq:gammastart}), we may just write an instantaneous
relation,
\begin {multline}
  \eps_\gamma(E,z)
  =
  \frac{1}{\Gamma_\gamma(E)}
  \left(
  \int_0^1 dx \> \left[\frac{d\Gamma}{dx}(E,x)\right]^\net_\ugE \,
     \eps_e(x E,z)
  + \int_0^1 dx \> \left[\frac{d\Gamma}{dx}(E,x)\right]^\net_\ugEbar \,
     \eps_e(x E,z)
  \right)
\\
  =
  \frac{1}{\Gamma_\gamma(E)}
  \int_0^1 dx \> \left[\frac{d\Gamma}{dx}(E,x)\right]^\net_{\gamma\to e^\pm} \,
     \eps_e(x E,z) .
\end{multline}
Assuming $E^{-1/2}$ scaling of the rates, this gives
\begin {equation}
  \eps_\gamma(z)
  =
  \frac{1}{\Gamma_\gamma(E_0)}
  \int_0^1 dx \> x
       \left[ \frac{d\Gamma}{dx}(E_0,x) \right]^\net_{\gamma\to e^\pm}
       \left\{ x^{-1/2} \eps_e(x^{-1/2} z) \right\} ,
\end {equation}
and taking moments yields the same relation (\ref{eq:zngRecursion})
between $\langle z^n \rangle_{\eps,\gamma}$ and $\langle z^n \rangle_{\eps,e}$
that was derived in the main text.

% ---------------------------------------------------------------------------

\subsection{Verifying eq.\ (\ref{eq:Miracle})}
\label {app:Miracle}

Now we discuss how to directly verify the relation (\ref{eq:Miracle})
rather than merely asserting that it must be true.
We find
it convenient to rewrite the relation as
\begin {equation}
  M_{ee} - \frac{M_{e\gamma} M_{\gamma e}}{M_{\gamma\gamma}}
  \overset{?}{=} \Avg^{\super,\net}_{e\to e^\pm}[ x(1-x^{n/2}) ] ,
\label {eq:verify1}
\end {equation}
where the question mark over an equality indicates an assertion that
is being checked.  To reduce notational
clutter, in this section we abbreviate $M_{(n)}$ as $M$.

Consider the overlap correction
$[\Delta\,d\Gamma/d\xe\,\xE]_{e\to e\E\Ebar}$ to double splitting
$e \to e\gamma \to e\E\Ebar$.
On the left-hand side of (\ref{eq:verify1}), that overlap correction
contributes only to the first term of
\begin {equation}
  M_{ee} \equiv
  \Avg_{e\to e^\pm}\bigl[ x{-}x^{1+\frac{n}{2}} \bigr]
  +
  \Avg_{e\to\gamma}\bigl[ x ]
\label {eq:Mee2}
\end {equation}
[see (\ref{eq:M})] and not to any of the other $M_{ij}$.
Given the definitions%
\footnote{
  See (\ref{eq:dGneteepm}) and (\ref{eq:NLOrates}) for
  $[d\Gamma/dx]^\net_{e\to e^\pm}$
  versus (\ref{eq:snet}) and (\ref{eq:super}) for
  $[d\Gamma/dx]^{\super,\net}_{e\to e^\pm}$.
}
of $[d\Gamma/dx]^\net_{e\to e^\pm}$
and $[d\Gamma/dx]^{\super,\net}_{e\to e^\pm}$, that very same contribution
also appears on the right-hand side of (\ref{eq:verify1}).
So what remains is that we need to check the equality (\ref{eq:verify1})
for everything that's \textit{not} a double-splitting overlap correction.
That's
\begin {equation}
    M_{ee}^{1\to 2}
    - \frac{M_{e\gamma} M_{\gamma e}}{M_{\gamma\gamma}}
  \overset{?}{=}
  \int_0^1 dx \> \left[ \frac{d\Gamma}{dx} \right]^{\indep,\net}_{e\to e^\pm}
    x(1-x^{n/2}) ,
\label {eq:verify3}
\end {equation}
where
\begin {equation}
  M^{1\to 2}_{ee} \equiv
  \Avg_{e\to e\gamma}\bigl[\xe-\xe^{1+\frac{n}{2}}\bigr]
  +
  \Avg_{e\to e\gamma}\bigl[x_\gamma] ,
\label {eq:Mee1to2}
\end {equation}
which represents the
contribution to the original $M_{ee}$
(\ref{eq:Mee2})
from \textit{just} the LO+NLO $1{\to}2$ process
$e{\to}e\gamma$ but \textit{not}
from the effective $1{\to}3$ rate
$[\Delta\,d\Gamma/d\xe\,\xE]_{e\to e\E\Ebar}$ representing the overlap
correction to $e \to e\gamma \to e\E\Ebar$.
Note that every $M_{ij}$ on the left-hand side of (\ref{eq:verify3})
now only involves LO+NLO $1{\to}2$ processes.
On the right-hand side of (\ref{eq:verify3}),
\begin {equation}
  \left[ \frac{d\Gamma}{dx} \right]^\inet_{e\to e^\pm}
  \equiv
  \int_0^{1-x} dy \>
  \biggl\{
    \frac{d\Gamma^\indep_{\eeEEsub}}{d\xe\,d\xE}(x,y)
    + \frac{d\Gamma^\indep_{\eeEEsub}}{d\xe\,d\xE}(y,x)
    + \frac{d\Gamma^\indep_{\eeEEsub}}{d\xe\,d\xE}(y,1{-}x{-}y)
  \biggr\} ,
\label{eq:inet}
\end {equation}
analogous to (\ref{eq:snet}), and recall that the ``independent splitting''
rate (\ref{eq:dGindep}) was also defined exclusively in terms of
LO+NLO $1{\to}2$ splitting processes.

Our next step is to realize that $\xe{+}x_\gamma = 1$ for $e\to e\gamma$,
and so (\ref{eq:Mee1to2}) is
\begin {equation}
  M^{1\to 2}_{ee}
  = \Avg_{e\to e\gamma}\bigl[\xe{+}x_\gamma\bigr]
    - \Avg_{e\to e\gamma}\bigl[\xe^{1+\frac{n}{2}}\bigr]
  = \Gamma_{e\to e\gamma}
    - \Avg_{e\to e\gamma}\bigl[\xe^{1+\frac{n}{2}}\bigr] .
\label {eq:Mee1to2b}
\end {equation}
There is a similar $\Gamma_{e\to e\gamma}$ term hiding in the
$n$-independent term on the right-hand side of (\ref{eq:verify3}).
Specifically,
\begin {align}
 \int_0^1 dx \> x & \left[ \frac{d\Gamma}{dx} \right]^{\indep,\net}_{e\to e^\pm}
\nonumber\\
  &=
  \int_0^1 dx \> \int_0^{1-x} dy \> x
  \biggl\{
    \frac{d\Gamma^\indep_{\eeEEsub}}{dx_e dy_\ssE}(x,y)
    + \frac{d\Gamma^\indep_{\eeEEsub}}{dx_e dy_\ssE}(y,x)
    + \frac{d\Gamma^\indep_{\eeEEsub}}{dx_e dy_\ssE}(y,1{-}x{-}y)
  \biggr\}
\nonumber\\
  &=
  \int_0^1 dx_e \> \int_0^{1-x_e} dx_\ssE \>
  (x_e + x_\ssE + x_\ssEbar) \,
  \frac{d\Gamma^\indep_{\eeEEsub}}{dx_e dx_\ssE}(x_e,x_\ssE)
\nonumber\\
  &=
  \Gamma^\indep_{\eeEEsub}
\nonumber\\
  &=
  \int_0^1 dx_e \> \int_0^1 d\yfrakE
  \left[ \frac{d\Gamma}{dx_e}(E) \right]_{e\to e\gamma} \,
  \frac{1}{\Gamma_\gamma\bigl((1{-}x_e)E\bigr)} \,
  \left[ \frac{d\Gamma}{d\yfrak_\ssE}\bigl((1{-}x_e)E\bigr)
       \right]_{\gamma\to\E\Ebar}
\nonumber\\
  &=
  \int_0^1 dx_e \>
  \left[ \frac{d\Gamma}{dx_e}(E) \right]_{e\to e\gamma} \,
  \frac{1}{\Gamma_\gamma\bigl((1{-}x_e)E\bigr)}
  \int_0^1 d\yfrakE
  \left[ \frac{d\Gamma}{d\yfrakE}\bigl((1{-}x_e)E\bigr) \right]_{\gamma\to\E\Ebar}
\nonumber\\
  &=
  \int_0^1 dx_e \>
  \left[ \frac{d\Gamma}{dx_e}(E) \right]_{e\to e\gamma}
\nonumber\\
  &=
  \Gamma_{e\to e\gamma} .
\label {eq:A0}
\end {align}
Eq.\ (\ref{eq:verify3}) then reduces to
\begin {equation}
  \Avg_{e\to e\gamma}[ \xe^{1+\frac{n}{2}} ]
    + \frac{M_{e\gamma} M_{\gamma e}}{M_{\gamma\gamma}}
  \overset{?}{=}
  \int_0^1 dx \> \left[ \frac{d\Gamma}{dx} \right]^{\indep,\net}_{e\to e^\pm}
    x^{1+\frac{n}{2}} .
\label {eq:verify4}
\end {equation}
Using (\ref{eq:inet}), this may be rewritten as
\begin {equation}
  \Avg_{e\to e\gamma}[ \xe^{1+\frac{n}{2}} ]
    + \frac{M_{e\gamma} M_{\gamma e}}{M_{\gamma\gamma}}
  \overset{?}{=}
  \int_0^1 d\xe\int_0^{1-\xe}d\xE \>
    \left[ \frac{d\Gamma}{d\xe\,d\xE} \right]^{\indep}_{e\to e\E\Ebar}
    \bigl( \xe^{1+\frac{n}{2}} + \xE^{1+\frac{n}{2}} + \xEbar^{1+\frac{n}{2}} \bigr).
\label {eq:verify4b}
\end {equation}
The \textit{first} term on the right-hand side is equal
(after changing integration variable $\xE$ to $\yfrakE$) to
\begin {align}
  \int_0^1 dx_e \> \int_0^1 d\yfrakE \>&
  \left[ \frac{d\Gamma}{dx_e}(E)\right]_{e\to e\gamma} \,
  \frac{1}{\Gamma_\gamma\bigl((1{-}x_e)E\bigr)} \,
  \left[ \frac{d\Gamma}{d\yfrakE}\bigl((1{-}x_e)E\bigr) \right]_{\gamma\to\E\Ebar}
  \,\xe^{1+\frac{n}{2}}
\nonumber\\
  &=
  \int_0^1 dx_e
  \left[ \frac{d\Gamma}{dx_e}(E) \right]_{e\to e\gamma}
  \xe^{1+\frac{n}{2}}
  =
  \Avg_{e\to e\gamma}[ \xe^{1+\frac{n}{2}} ]
\end {align}
and so cancels against the same term on the left-hand side of
(\ref{eq:verify4b}), leaving
\begin {equation}
  \frac{M_{e\gamma} M_{\gamma e}}{M_{\gamma\gamma}}
  \overset{?}{=}
  \int_0^1 d\xe\int_0^{1-\xe}d\xE \>
    \left[ \frac{d\Gamma}{d\xe\,d\xE} \right]^{\indep}_{e\to e\E\Ebar}
    \bigl( \xE^{1+\frac{n}{2}} + \xEbar^{1+\frac{n}{2}} \bigr).
\label {eq:verify5}
\end {equation}

Now look at the first term on the right-hand side of (\ref{eq:verify5}),
which is
\begin {subequations}
\begin {align}
  \int_0^1 & d\xe \int_0^{1-\xe}d\xE \>
    \left[ \frac{d\Gamma}{d\xe\,d\xE} \right]^{\indep}_{e\to e\E\Ebar}
    \xE^{1+\frac{n}{2}}
\nonumber\\
  &=
  \int_0^1 dx_e \> \int_0^1 d\yfrakE \>
  \left[ \frac{d\Gamma}{dx_e}(E) \right]_{e\to e\gamma} \,
  \frac{1}{\Gamma_\gamma\bigl((1{-}x_e)E\bigr)} \,
  \left[ \frac{d\Gamma}{d\yfrakE}\bigl((1{-}x_e)E\bigr) \right]_{\gamma\to\E\Ebar}
  \, [\yfrakE (1{-}\xe)]^{1+\frac{n}{2}}
\nonumber\\
  &=
  \int_0^1 dx_e \> \left[ \frac{d\Gamma}{dx_e}(E) \right]_{e\to e\gamma}
    (1{-}\xe)^{1+\frac{n}{2}}
\nonumber\\ & \hspace{8em}
  \times
    \frac{1}{\Gamma_\gamma\bigl((1{-}x_e)E\bigr)}
    \int_0^1 d\yfrakE \>
    \left[ \frac{d\Gamma}{d\yfrakE}\bigl((1{-}x_e)E\bigr) \right]_{\gamma\to\E\Ebar}
    \, \yfrakE^{1+\frac{n}{2}} .
\label{eq:1stterm0}
\\
\intertext{
  Since our analysis in this appendix (and most of the paper) has
  assumed that rates scale
  as a power of energy (specifically $E^{-1/2}$), the energy dependence cancels
  in the combination $1/\Gamma \times d\Gamma/d\yfrakE$,
  and so we may rewrite (\ref{eq:1stterm0})  as
}
  &= 
  \int_0^1 dx_e \> \left[ \frac{d\Gamma}{dx_e}(E) \right]_{e\to e\gamma}
    (1{-}\xe)^{1+\frac{n}{2}}
  \times
    \frac{1}{\Gamma_\gamma(E)}
    \int_0^1 d\xE \>
    \left[ \frac{d\Gamma}{d\xE}(E) \right]_{\gamma\to\E\Ebar}
    \, \xE^{1+\frac{n}{2}}
\nonumber\\
   &=
  \Avg_{e\to e\gamma}[x_\gamma^{1+\frac{n}{2}}]
  \times \frac{1}{\Gamma_\gamma(E)}
  \times \Avg_{\gamma\to\E\Ebar}[\xE^{1+\frac{n}{2}}] ,
\end {align}
\end{subequations}
where we have also renamed the integration variable
$\yfrakE\vphantom{\yfrak_{E_s}}$
to
``$\xE$'' to aid the comparison we will shortly make between the two sides
of (\ref{eq:verify5}).
The $\xEbar^{1+\frac{n}{2}}$ term on the right-hand side of (\ref{eq:verify5})
gives a similar result [and in fact an exactly equal result
because of the charge conjugation
symmetry of (independent) LO+NLO $\gamma \to \E\Ebar$ splitting].
So (\ref{eq:verify5}) becomes
\begin {equation}
  \frac{M_{e\gamma} M_{\gamma e}}{M_{\gamma\gamma}}
  \overset{?}{=}
  \frac{
    \Avg_{e\to e\gamma}[x_\gamma^{1+\frac{n}{2}}] \,
    \Avg_{\gamma\to\E\Ebar}[\xE^{1+\frac{n}{2}} {+} \xEbar^{1+\frac{n}{2}}]
  }
  {
    \Gamma_\gamma(E)
  }
  \,.
\label {eq:verify6}
\end {equation}
The numerators match up by (\ref{eq:M}) and (\ref{eq:dGnetgam}).
The denominators match up because
\begin {equation}
  M_{\gamma\gamma}
  = \Avg_{\gamma\to e^\pm}[x]
  = \Avg_\ugE[x] + \Avg_\ugEbar[x]
  = \Avg_{\gamma\to\E\Ebar}[\xE{+}\xEbar]
  = \Gamma_\gamma ,
\end {equation}
where the last equality follows because $\xE{+}\xEbar = 1$ for
LO+NLO single splitting $\gamma \to \E\Ebar$.
That completes our verification of (\ref{eq:Miracle}).

% ============================================================================

\section{Analytic LO results}
\label {app:LO}

For QED, the integrals in (\ref{eq:znrho}) and
(\ref{eq:M}) may be carried out
analytically for the LO contribution.  The results are
\begin {multline}
  \Avg^\LO_\uee[1-\xe^{n/2}]
  = \frac{ B(\tfrac12,\tfrac12) + B(\tfrac52,\tfrac12) }{ 2\pi\ell_0 }
  - \frac{ B(\tfrac{n+1}2,\tfrac12)
           + B(\tfrac{n+5}2,\tfrac12) }{ 2\pi\ell_0 }
\\
  = \frac{11}{16\ell_0}
  - \frac{ B(\tfrac{n+1}2,\tfrac12)
           + B(\tfrac{n+5}2,\tfrac12) }{ 2\pi\ell_0 }
\end {multline}
for charge deposition and
\begin {subequations}
\begin {align}
  M^\LO_{(n),ee} &=
  \frac{11}{16\,\ell_0}
  - \frac{ B(\tfrac{n+3}2,\tfrac12)
           + B(\tfrac{n+7}2,\tfrac12) }
         { 2\pi\ell_0 } \,,
\\
  M^\LO_{(n),e\gamma} &=
  - \frac{ B(\tfrac{n+3}2,\tfrac12)
           + B(\tfrac{n+3}2,\tfrac52) }
         { 2\pi\ell_0 } \,,
\\
  M^\LO_{(n),\gamma e} &=
  - \frac{ B(\tfrac{n+7}2,\tfrac12)
              + B(\tfrac{n+3}2,\tfrac52) }
            { \pi\ell_0 } \, \Nf,
\\
  M^\LO_{(n),\gamma\gamma} &=
  \frac{3\Nf}{8\,\ell_0}
\end {align}
\end {subequations}
for energy deposition, where
$B(x,y) \equiv \Gamma(x)\,\Gamma(y)/\Gamma(x{+}y)$
is the Euler beta function
and $\ell_0$ is defined by (\ref{eq:ell0}).
Table \ref{tab:M} shows the simple results for $n\le 4$.
Such analytic LO results are special to QED;
we do not know how to do the analogous integral
$\Avg^\LO_{g\to g}[x{-}x^{1+\frac{n}{2}}]$ analytically for the gluonic showers
of ref.\ \cite{finale2}.

\begin {table}[tp]

\setlength{\tabcolsep}{7pt}
\begin {center}
\begin{tabular}{cccccc}
\hline
\hline
  $n$ & $\Avg^\LO_\uee[1{-}\xe^{n/2}]$
      & $M^\LO_{(n),ee}$ & $M^\LO_{(n),e\gamma}$
      & $M^\LO_{(n),\gamma e}$ & $M^\LO_{(n),\gamma\gamma}$
\\
\cline{2-6}
& \multicolumn{5}{c}{in units of $1/\ell_0$}
\\
\hline
  1 & $\tfrac{11}{16}{-}\tfrac{23}{15\pi}$
    & $\tfrac{11}{16}{-}\tfrac{118}{105\pi}$ & $-\tfrac{76}{105\pi}$
    & $-\tfrac{36}{35\pi}\Nf$ & $\tfrac38\Nf$ \\
  2 & $\tfrac{9}{32}$
    & $\tfrac{93}{256}$ & $-\tfrac{51}{256}$
    & $-\tfrac{19}{64}\Nf$ & $\tfrac38\Nf$ \\
  3 & $\tfrac{11}{16}{-}\tfrac{118}{105\pi}$
    & $\tfrac{11}{16}{-}\tfrac{296}{315\pi}$ & $-\tfrac{176}{315\pi}$
    & $-\tfrac{272}{315\pi}\Nf$ & $\tfrac38\Nf$ \\
  4 & $\tfrac{93}{256}$
    & $\tfrac{209}{512}$ & $-\tfrac{83}{512}$
    & $-\tfrac{33}{128}\Nf$ & $\tfrac38\Nf$
\\
\hline
\hline
\end{tabular}
\end {center}
\caption{%
\label{tab:M}%
  Exact values for the parameters that appear in the leading-order version
  of the recursion relations (\ref{eq:znrho}) and (\ref{eq:recursionM}).
}
\end{table}

The values of the above coefficients may be used in the recursion
relations for $\langle z^n \rangle_\rho$, $\langle z^n \rangle_{\eps,e}$,
and $\langle z^n \rangle_{\eps,e}$ to obtain exact values for those moments
and thence exact values for the various moments, reduced moments, and
cumulants of the corresponding shape functions $S(Z)$, such as
the example given in (\ref{eq:exactsigmaS}).  However, the recursion causes most
of those formulas to look very messy; so we content ourselves with
the one example and will not explicitly write out any others.

%%%%%%%%%%%%%%%%%%%%%%%%%%%%%%%%%%%%%%%%%%%%%%%%%%%%%%%%%%%%%%%%%%%%%%%%%%%%%%


\begin{thebibliography}{}

\bibitem{LP1}
  L.~D.~Landau and I.~Pomeranchuk,
  ``Limits of applicability of the theory of bremsstrahlung electrons and
  pair production at high-energies,''
  Dokl.\ Akad.\ Nauk Ser.\ Fiz.\  {\bf 92} (1953) 535.

\bibitem{LP2}
  L.~D.~Landau and I.~Pomeranchuk,
  ``Electron cascade process at very high energies,''
  Dokl.\ Akad.\ Nauk Ser.\ Fiz.\  {\bf 92} (1953) 735.

\bibitem{Migdal}
  A.~B.~Migdal,
  ``Bremsstrahlung and pair production in condensed media at high-energies,''
   Phys.\ Rev.\  {\bf 103}, 1811 (1956);

\bibitem{LPenglish}
  L. Landau,
  {\sl The Collected Papers of L.D. Landau}\/
  (Pergamon Press, New York, 1965).

\bibitem{BDMPS1}
  R.~Baier, Y.~L.~Dokshitzer, A.~H.~Mueller, S.~Peigne and D.~Schiff,
  ``The Landau-Pomeranchuk-Migdal effect in QED,''
  Nucl.\ Phys.\  B {\bf 478}, 577 (1996)
  [arXiv:hep-ph/9604327];

\bibitem{BDMPS2}
  R.~Baier, Y.~L.~Dokshitzer, A.~H.~Mueller, S.~Peigne and D.~Schiff,
  ``Radiative energy loss of high-energy quarks and gluons in a
    finite volume quark-gluon plasma,''
  Nucl.\ Phys.\  B {\bf 483}, 291 (1997) [arXiv:hep-ph/9607355].
  %%CITATION = NUPHA,B483,291;%%

\bibitem{BDMPS3}
  R.~Baier, Y.~L.~Dokshitzer, A.~H.~Mueller, S.~Peigne and D.~Schiff,
  ``Radiative energy loss and $p_\perp$-broadening of high energy partons in
    nuclei,''
  {\it ibid.}\ {\bf 484} (1997)
  [arXiv:hep-ph/9608322].
  %%CITATION = NUPHA,B484,265;%%

\bibitem{Zakharov1}
 B.~G.~Zakharov,
 ``Fully quantum treatment of the Landau-Pomeranchuk-Migdal effect in
   QED and QCD,''
 JETP Lett.\  {\bf 63}, 952 (1996)
 [Pis'ma Zh.\ \'Eksp.\ Teor.\ Fiz.\  {\bf 63}, 906 (1996)]
 [arXiv:hep-ph/9607440].

\bibitem{Zakharov2}
 B.~G.~Zakharov,
 ``Radiative energy loss of high-energy quarks in finite size nuclear matter
   and quark-gluon plasma,''
 JETP Lett.\  {\bf 65}, 615 (1997)
 [Pis'ma Zh.\ \'Eksp.\ Teor.\ Fiz.\  {\bf 65}, 585 (1997)]
 [arXiv:hep-ph/9704255].
 %%CITATION = JTPLA,63,952.%%

\bibitem{Blaizot}
  J.~P.~Blaizot and Y.~Mehtar-Tani,
  ``Renormalization of the jet-quenching parameter,''
  Nucl.\ Phys.\ A {\bf 929}, 202 (2014)
  [arXiv:1403.2323 [hep-ph]].
  %%CITATION = ARXIV:1403.2323;%%

\bibitem{Iancu}
  E.~Iancu,
  ``The non-linear evolution of jet quenching,''
  JHEP \textbf{10}, 95 (2014)
  [arXiv:1403.1996 [hep-ph]].
  %%CITATION = ARXIV:1403.1996;%%

\bibitem{Wu}
  B.~Wu,
  ``Radiative energy loss and radiative $p_{\bot}$-broadening of
    high-energy partons in QCD matter,''
  JHEP \textbf{12}, 081 (2014)
  [arXiv:1408.5459 [hep-ph]].
  %%CITATION = ARXIV:1408.5459;%%

\bibitem{qedNfstop}
  P.~Arnold, S.~Iqbal and T.~Rase,
  ``Strong- vs. weak-coupling pictures of jet quenching: a dry run using QED,''
  JHEP \textbf{05}, 004 (2019)
  %doi:10.1007/JHEP05(2019)004
  [arXiv:1810.06578 [hep-ph]].

\bibitem{finale}
  P.~Arnold, O.~Elgedawy and S.~Iqbal,
  ``Are gluon showers inside a quark-gluon plasma strongly coupled?
    a theorist's test,''
  Phys. Rev. Lett. \textbf{131}, no.16, 162302 (2023)
  %doi:10.1103/PhysRevLett.131.162302
  [arXiv:2212.08086 [hep-ph]].

\bibitem{finale2}
  P.~Arnold, O.~Elgedawy and S.~Iqbal,
  ``The LPM effect in sequential bremsstrahlung: gluon shower development,''
  Phys. Rev. D \textbf{108}, no.7, 074015 (2023)
  %doi:10.1103/PhysRevD.108.074015
  [arXiv:2302.10215 [hep-ph]].

\bibitem{qedNf}
  P.~Arnold and S.~Iqbal,
  ``In-medium loop corrections and longitudinally polarized gauge bosons
    in high-energy showers,''
  JHEP \textbf{12}, 120 (2018)
  %doi:10.1007/JHEP12(2018)120
  [erratum: JHEP \textbf{12}, 098 (2023)]
  [arXiv:1806.08796 [hep-ph]].

\bibitem{qcd}
  P.~Arnold, T.~Gorda and S.~Iqbal,
  ``The LPM effect in sequential bremsstrahlung:
    nearly complete results for QCD,''
  JHEP \textbf{11}, 053 (2020)
  [{\it erratum} JHEP \textbf{05}, 114 (2022)]
  %doi:10.1007/JHEP11(2020)053, 10.1007/JHEP05(2022)114
  [arXiv:2007.15018 [hep-ph]].

\bibitem{LOptZakharov}
  B.~G.~Zakharov,
  ``Transverse spectra of radiation processes in-medium,''
  JETP Lett. \textbf{70}, 176-182 (1999)
  %doi:10.1134/1.568149
  arXiv:hep-ph/9906536 [hep-ph]].
  
\bibitem{LOptWiedemann}
  U.~A.~Wiedemann and M.~Gyulassy,
  ``Transverse momentum dependence of the Landau-Pomeranchuk-Migdal effect,''
  Nucl.\ Phys.\ B \textbf{560}, 345-382 (1999)
  %doi:10.1016/S0550-3213(99)00458-7
  [arXiv:hep-ph/9906257 [hep-ph]];
  U.~A.~Wiedemann,
  ``Gluon radiation off hard quarks in a nuclear environment:
     Opacity expansion,''
  Nucl.\ Phys.\ B \textbf{588}, 303-344 (2000)
  %doi:10.1016/S0550-3213(00)00457-0
  [arXiv:hep-ph/0005129 [hep-ph]].

\bibitem{LOptBlaizot}
  J.~P.~Blaizot, F.~Dominguez, E.~Iancu and Y.~Mehtar-Tani,
  ``Medium-induced gluon branching,''
  JHEP \textbf{01}, 143 (2013)
  %doi:10.1007/JHEP01(2013)143
  [arXiv:1209.4585 [hep-ph]].

\bibitem{LOptApolinario}
  L.~Apolin\'ario, N.~Armesto, J.~G.~Milhano and C.~A.~Salgado,
  ``Medium-induced gluon radiation and colour decoherence beyond
    the soft approximation,''
  JHEP \textbf{02}, 119 (2015)
  %doi:10.1007/JHEP02(2015)119
  [arXiv:1407.0599 [hep-ph]].

%\bibitem{LOptBeyond1/N}
%  J.~H.~Isaksen and K.~Tywoniuk,
% ``Precise description of medium-induced emissions,''
% [arXiv:2303.12119 [hep-ph]].

\bibitem{NLOptBarato}
  J.~Barata, F.~Dom\'\i{}nguez, C.~A.~Salgado and V.~Vila,
  ``A modified in-medium evolution equation with color coherence,''
  JHEP \textbf{05}, 148 (2021)
  %doi:10.1007/JHEP05(2021)148
  [arXiv:2101.12135 [hep-ph]].

\bibitem{antenna1}
  J.~Casalderrey-Solana, Y.~Mehtar-Tani, C.~A.~Salgado and K.~Tywoniuk,
  ``New picture of jet quenching dictated by color coherence,''
  Phys. Lett. B \textbf{725}, 357-360 (2013)
  %doi:10.1016/j.physletb.2013.07.046
  [arXiv:1210.7765 [hep-ph]].

\bibitem{antenna2}
  Y.~Mehtar-Tani, C.~A.~Salgado and K.~Tywoniuk,
  ``The Radiation pattern of a QCD antenna in a dense medium,''
  JHEP \textbf{10}, 197 (2012)
  %doi:10.1007/JHEP10(2012)197
  [arXiv:1205.5739 [hep-ph]].

\bibitem{conesize1}
  Y.~Mehtar-Tani, D.~Pablos and K.~Tywoniuk,
  ``Cone-Size Dependence of Jet Suppression in Heavy-Ion Collisions,''
  Phys. Rev. Lett. \textbf{127}, no.25, 252301 (2021)
  %doi:10.1103/PhysRevLett.127.252301
  [arXiv:2101.01742 [hep-ph]].

\bibitem{conesize2}
Y.~Mehtar-Tani, D.~Pablos and K.~Tywoniuk,
  ``Jet suppression and azimuthal anisotropy from RHIC to LHC,''
  Phys. Rev. D \textbf{110}, no.1, 014009 (2024)
  %doi:10.1103/PhysRevD.110.014009
  [arXiv:2402.07869 [hep-ph]].

\bibitem{1overN}
  P.~Arnold and O.~Elgedawy,
  ``The LPM effect in sequential bremsstrahlung:
    1/$ {\mathrm{N}}_{\mathrm{c}}^2 $ corrections,''
  JHEP \textbf{08}, 194 (2022)
  %doi:10.1007/JHEP08(2022)194
  [arXiv:2202.04662 [hep-ph]].

\bibitem{NSZ6j}
  N.~N.~Nikolaev, W.~Schafer and B.~G.~Zakharov,
  ``Nonlinear $k_\perp$ factorization for gluon-gluon
    dijets produced off nuclear targets,''
  Phys.\ Rev.\ D {\bf 72}, 114018 (2005)
  %doi:10.1103/PhysRevD.72.114018
  [hep-ph/0508310].
  %%CITATION = doi:10.1103/PhysRevD.72.114018;%%
  %27 citations counted in INSPIRE as of 14 May 2019

\bibitem{Zakharov6j}
  B.~G.~Zakharov,
  ``Color randomization of fast gluon-gluon pairs in the quark-gluon plasma,''
  J. Exp. Theor. Phys. \textbf{128}, no.2, 243-258 (2019)
  [arXiv:1806.04723 [hep-ph]].
  %%CITATION = ARXIV:1806.04723;%%

\bibitem{Konrad1}
  J.~H.~Isaksen and K.~Tywoniuk,
  ``Wilson line correlators beyond the large-N$_{c}$,''
  JHEP \textbf{11}, 125 (2021)
  %doi:10.1007/JHEP11(2021)125
  [arXiv:2107.02542 [hep-ph]].

\bibitem{KonradNc3}
  J.~H.~Isaksen and K.~Tywoniuk,
  ``Precise description of medium-induced emissions,''
  JHEP \textbf{09}, 049 (2023)
  %doi:10.1007/JHEP09(2023)049
  [arXiv:2303.12119 [hep-ph]].

\bibitem{ZakharovFinite}
B.~G.~Zakharov,
  ``On the energy loss of high-energy quarks in a finite size
    quark-gluon plasma,''
  JETP Lett. \textbf{73}, 49-52 (2001)
  %doi:10.1134/1.1358417
  [arXiv:hep-ph/0012360 [hep-ph]].

\bibitem{GLV}
  M.~Gyulassy, P.~Levai and I.~Vitev,
  ``Reaction operator approach to nonAbelian energy loss,''
  Nucl. Phys. B \textbf{594}, 371-419 (2001)
  %doi:10.1016/S0550-3213(00)00652-0
  [arXiv:nucl-th/0006010 [nucl-th]].

\bibitem{Vitev2}
  M.~Fickinger, G.~Ovanesyan and I.~Vitev,
  ``Angular distributions of higher order splitting functions
    in the vacuum and in dense QCD matter,''
  JHEP \textbf{07}, 059 (2013)
  %doi:10.1007/JHEP07(2013)059
  [arXiv:1304.3497 [hep-ph]].

\bibitem{Vitev3}
  M.~D.~Sievert, I.~Vitev and B.~Yoon,
  ``A complete set of in-medium splitting functions to any order in opacity,''
  Phys. Lett. B \textbf{795}, 502-510 (2019)
  %doi:10.1016/j.physletb.2019.06.019
  [arXiv:1903.06170 [hep-ph]].

\bibitem{SimoneCharles}
S.~Caron-Huot and C.~Gale,
  ``Finite-size effects on the radiative energy loss of a fast parton
    in hot and dense strongly interacting matter,''
  Phys. Rev. C \textbf{82}, 064902 (2010)
  %doi:10.1103/PhysRevC.82.064902
  [arXiv:1006.2379 [hep-ph]].

\bibitem{Carlota1}
  C.~Andres, L.~Apolin\'ario and F.~Dominguez,
  ``Medium-induced gluon radiation with full resummation of
    multiple scatterings for realistic parton-medium interactions,''
  JHEP \textbf{07}, 114 (2020)
  %doi:10.1007/JHEP07(2020)114
  [arXiv:2002.01517 [hep-ph]].

\bibitem{Carlota2}
  C.~Andres, L.~Apolin\'ario, F.~Dominguez and M.~G.~Martinez,
  ``In-medium gluon radiation spectrum with all-order
    resummation of multiple scatterings in longitudinally evolving media,''
  [arXiv:2307.06226 [hep-ph]].

\bibitem{IOC1}
Y.~Mehtar-Tani,
  ``Gluon bremsstrahlung in finite media beyond multiple soft
  scattering approximation,''
  JHEP \textbf{07}, 057 (2019)
  %doi:10.1007/JHEP07(2019)057
  [arXiv:1903.00506 [hep-ph]].

\bibitem{IOC2}
J.~Barata, Y.~Mehtar-Tani, A.~Soto-Ontoso and K.~Tywoniuk,
  ``Medium-induced radiative kernel with the Improved Opacity Expansion,''
  JHEP \textbf{09}, 153 (2021)
  %doi:10.1007/JHEP09(2021)153
  [arXiv:2106.07402 [hep-ph]].

\bibitem{IOC3}
J.~Barata and Y.~Mehtar-Tani,
  ``Improved opacity expansion at NNLO for medium induced gluon radiation,''
  JHEP \textbf{10}, 176 (2020)
  %doi:10.1007/JHEP10(2020)176
  [arXiv:2004.02323 [hep-ph]].

\bibitem{marryIancu}
  P.~Caucal, E.~Iancu, A.~H.~Mueller and G.~Soyez,
  ``Vacuum-like jet fragmentation in a dense QCD medium,''
  Phys. Rev. Lett. \textbf{120}, 232001 (2018)
  %doi:10.1103/PhysRevLett.120.232001
  [arXiv:1801.09703 [hep-ph]].

\bibitem{marryKurkela}
  A.~Kurkela and U.~A.~Wiedemann,
  ``Picturing perturbative parton cascades in QCD matter,''
  Phys. Lett. B \textbf{740}, 172-178 (2015)
  %doi:10.1016/j.physletb.2014.11.054
  [arXiv:1407.0293 [hep-ph]].

\bibitem{marryYacine1}
  Y.~Mehtar-Tani and K.~Tywoniuk,
    ``Sudakov suppression of jets in QCD media,''
    Phys. Rev. D \textbf{98}, no.5, 051501 (2018)
    %doi:10.1103/PhysRevD.98.051501
    [arXiv:1707.07361 [hep-ph]].

%\bibitem{marryYacine2}
%  Y.~Mehtar-Tani, D.~Pablos and K.~Tywoniuk,
%    ``Cone-Size Dependence of Jet Suppression in Heavy-Ion Collisions,''
%    Phys. Rev. Lett. \textbf{127}, no.25, 252301 (2021)
%    %doi:10.1103/PhysRevLett.127.252301
%    [arXiv:2101.01742 [hep-ph]].

\bibitem{qccbar}
  M.~Attems, J.~Brewer, G.~M.~Innocenti, A.~Mazeliauskas, S.~Park,
  W.~van der Schee and U.~A.~Wiedemann,
  ``The medium-modified $ g\to c\overline{c} $ splitting function
    in the BDMPS-Z formalism,''
  JHEP \textbf{01}, 080 (2023)
  %doi:10.1007/JHEP01(2023)080
  [arXiv:2203.11241 [hep-ph]].

\bibitem{2brem}
  P.~Arnold and S.~Iqbal,
  ``The LPM effect in sequential bremsstrahlung,''
  JHEP \textbf{04}, 070 (2015)
  [{\it erratum} JHEP \textbf{09}, 072 (2016)]
  %doi:10.1007/JHEP09(2016)072, 10.1007/JHEP04(2015)070
  [arXiv:1501.04964 [hep-ph]].
  %%CITATION = doi:10.1007/JHEP09(2016)072, 10.1007/JHEP04(2015)070;%%
  %9 citations counted in INSPIRE as of 20 Sep 2016

\bibitem{seq}
  P.~Arnold, H.~C.~Chang and S.~Iqbal,
  ``The LPM effect in sequential bremsstrahlung 2: factorization,''
  JHEP \textbf{09}, 078 (2016)
  [arXiv:1605.07624 [hep-ph]].
  %%CITATION = ARXIV:1605.07624;%%

\bibitem{Mathematica}
  Wolfram Research, Inc., Mathematica (various versions),
  Champaign, IL (2018--2021).

\bibitem{LMW}
  T.~Liou, A.~H.~Mueller and B.~Wu,
  ``Radiative $p_\bot$-broadening of high-energy quarks and gluons in
    QCD matter,''
  Nucl.\ Phys.\ A {\bf 916}, 102 (2013)
  [arXiv:1304.7677 [hep-ph]].
  %%CITATION = ARXIV:1304.7677;%%

\bibitem{qcdNf}
P.~Arnold, O.~Elgedawy and S.~Iqbal,
  ``Are in-medium quark-gluon showers strongly coupled?
    Results in the large-$N_f$ limit,''
  arXiv:2408.07129 [hep-ph].

\end{thebibliography}
\end {document}